\documentclass[12pt]{article}
\usepackage{axodraw,epsf}
\newcommand{\be}{\begin{eqnarray*}}
\newcommand{\ee}{\end{eqnarray*}}
\newcommand{\bea}{\begin{eqnarray}}
\newcommand{\eea}{\end{eqnarray}}
\newcommand{\al}{\alpha}

\newcommand{\msb}{\overline{\rm MS}}
\newcommand{\ep}{\varepsilon}

\newcommand{\Li}[2]{{\mbox{Li}}_{#1}\left(#2\right)}
\newcommand{\Cl}[2]{{\mbox{Cl}}_{#1}\left(#2\right)}
\newcommand{\Ls}[2]{{\mbox{Ls}}_{#1}\left(#2\right)}
\newcommand{\LS}[3]{{\mbox{Ls}}_{#1}^{(#2)}\left(#3\right)}
\newcommand{\Lsc}[2]{{\mbox{Lsc}}_{#1\!}\left(#2\right)}
\newcommand{\tfrac}[2]{{\textstyle{\frac{#1}{#2}}}}

\newcommand{\Snp}[2]{{\mbox{S}}_{#1\!}\left(#2\right)}
\newcommand{\MIN}{\mbox{i}^{1-d}}

\newcommand{\U}{A_0(m_u^2)}
\newcommand{\h}{A_0(m_H^2)}
\newcommand{\W}{A_0(m_W^2)}
\newcommand{\Z}{A_0(m_Z^2)}
\newcommand{\X}{\Delta(0,m_u,m_W)}
\newcommand{\g}{\Delta(m_W,m_W,m_H)}
\newcommand{\Y}{B_0(0,m_u^2;m_W^2)}
\newcommand{\J}{B_0(m_H^2,m_W^2;m_W^2)}
\newcommand{\K}{B_0(m_Z^2,m_W^2;m_W^2)}
\newcommand{\M}{B_0(0,0;m_W^2)}

\newcommand{\R}{\Delta(m_u,m_u,m_Z)}
\newcommand{\Q}{\Delta(m_Z,m_Z,m_H)}
\newcommand{\T}{B_0(m_u^2,m_u^2;m_Z^2)}
\newcommand{\I}{B_0(0,0;m_Z^2)}
\newcommand{\D}{B_0(m_H^2,m_Z^2;m_Z^2)}

\newcommand{\MSb}{$\overline{\rm MS}$ }
\newcommand{\EP}{\; .}
\newcommand{\EC}{\; ,}
\newcommand{\Item}[1]{ \begin{itemize}
\item #1
\end{itemize} }
\newcommand{\crn}{\nonumber \\ }
\newcommand{\sinw}{\sin \theta_W}
\newcommand{\sinW}{\sin^2 \theta_W}
\newcommand{\mv}{\mbox{MeV}}
\newcommand{\gv}{\mbox{GeV}}
\newcommand{\ba}{\begin{eqnarray}}
\newcommand{\ea}{\end{eqnarray}}
\newcommand{\epm}{e^+e^-}

\newcommand{\MSs}{$\overline{\mathrm{MS}}$ scheme}

\newcommand{\alms}{\alpha_{\overline{\rm MS}}}

\newcommand{\alos}{\alpha_{\rm OS}}

\begin{document}

\begin{flushright}
{DESY 02-156}\\
{hep-ph/0212319} \\[3mm]
\end{flushright}
 \begin{center}
 {\large \bf
 $\overline{{\rm MS}}$ vs. Pole Masses of Gauge Bosons II: \\
 Two-Loop Electroweak Fermion  Corrections }\footnote{Supported by DFG
under Contract SFB/TR 9-03}
 \end{center}
 \vspace{1cm}
 \begin{center}
{\large
F.~Jegerlehner%
\footnote{~E-mail: fred.jegerlehner@desy.de},
M.~Yu.~Kalmykov%
\footnote{~E-mail: kalmykov@ifh.de}
\footnote{
~On leave of absence from BLTP, JINR, 141980, Dubna (Moscow Region),
Russia},~
}
\vspace{5mm}

{\it ~DESY Zeuthen, Platanenallee 6, D-15738, Zeuthen, Germany}

\vspace{5mm}

{\large
O.~Veretin%
\footnote{~E-mail: veretin@particle.uni-karlsruhe.de}
}

\vspace{5mm}
{\it Institut f\"ur Theoretische Teilchenphysik, Universit\"at Karlsruhe, D-76128, Karlsruhe, Germany}
\end{center}
\begin{abstract}
We have calculated the fermion contributions to the shift of the
position of the poles of the massive gauge boson propagators at
two--loop order in the Standard Model. Together with the bosonic
contributions calculated previously the full two--loop corrections are
available. This allows us to investigate the full correction in the
relationship between $\overline{\rm MS}$ and pole masses of the vector
bosons $Z$ and $W$. Two--loop renormalization and the corresponding
renormalization group equations are discussed. Analytical results for
the master--integrals appearing in the massless fermion contributions
are given. A new approach of summing multiple binomial sums has been
developed.
\end{abstract}
\section{Introduction}

In our previous paper~\cite{I}, referred to as I in the following, we
presented the two--loop calculation of the bosonic contributions to
the $W$-- and $Z$--self--energies on the mass shell and discussed
the main properties and possible applications. In the present paper
II we extend this calculation to a full Standard Model (SM)
calculation by including the missing fermion--loop and mixed
contributions. Throughout the paper, we shall adopt the terminology
calling ``bosonic corrections'' the one's represented by diagrams
without any fermions and ``fermionic corrections'' the remaining one's
given by diagrams exhibiting at least one fermion loop.

In I, in particular, we have shown by an explicit calculation that up to
two-loop order the purely bosonic corrections to the position of the
pole of the gauge-boson propagators are real for arbitrary values of the
Higgs-boson mass.  In contrast, since the $W$-- and $Z$--bosons decay,
at leading order, into light fermion pairs, the light (massless) fermion
corrections give rise to a non--zero imaginary part at the one loop
level already. The problem of gauge (in)dependence of the complex pole
has been extensively discussed in literature~\cite{pole}.  Only
recently, two important results have been proven to all orders in
perturbation theory: i) the position of the complex pole is a gauge
independent quantity~\cite{pole:SM}; ii) the branching ratios and
partial widths associated with the pole residues are gauge
independent~\cite{width}.  Moreover, it has been shown that the pole
mass of the W-boson is an infrared finite quantity with respect to
massless photonic corrections.  An alternative proof of the infrared
finiteness of the two-loop bosonic contributions to the pole of the
gauge bosons was presented in I by explicite calculation. 
It is based on the fact that within dimensional
regularization~\cite{dimreg}, in $d=4-2\ep$ space-time dimensions, 
the singular $1/\ep$ terms, which regularize both ultraviolet (UV) and
infrared (IR) singularities, are absent after UV renormalization of the
position of the pole in the propagators.

In the present paper, besides from completing our previous calculation
by including the missing fermion contributions, we will discuss in
some detail general features and technical problems which are
specifically related to these contributions.

The paper is organized as follows. In Section~2 we briefly reconsider
the definition of the pole-mass of the massive gauge-bosons within the
SM and remind the reader of some notation given in I. The required
analytical results for the massless fermion two-loop master-integrals
are presented in Section~3.  In Section~4 we discuss the UV
renormalization of the pole mass and the interrelation between our
results and the one's familiar from the standard renormalization group
approach. In particular, we performed several cross-checks of the
singular $1/\ep^2$- and $1/\ep$-terms. General aspects as well as
numerical results for the finite parts are discussed in
Section~5. Some technical details and a number of our analytical results
will be presented in Appendices. In Appendix~A we present a set of
non--standard binomial sums which are needed for the $\ep$--expansion
of some of the hypergeometric functions entering the
master--integrals. The one--loop fermion contribution to the pole
masses of the gauge--bosons are reproduced in Appendix~B. The reducible
two--loop corrections obtained by mass renormalization of the
one--loop fermion contributions may be found in Appendix~C. In
Appendix~D the bare two--loop on--shell self--energy contribution of
the massless fermions are given in exact analytical form.
The corresponding contributions involving the top quark are presented
in terms of the first few coefficients of the expansion in
$\sin^2 \theta_W$ and specific mass ratios in Appendix~E.

\section{Pole mass: definition and calculation}

The position of the pole $s_P$ of the propagator of a massive gauge-boson
in a quantum field theory is a solution for $p^2$ at which the inverse
of the connected full propagator equals zero, i.e.,
\begin{equation}
s_P - m^2 - \Pi(s_P,m^2,\cdots) = 0,
\label{pole}
\end{equation}
where $\Pi(p^2,\cdots)$ is the transversal part of the one-particle
irreducible self-energy. The latter depends on all SM parameters but,
in order to the keep notation simple, we have indicated explicitly
only the dependence on the external momentum $p$ and in some cases
also $m$, where $m$ is the mass of the particle under
consideration. This can be either the bare mass or the
renormalized mass defined in some particular renormalization scheme.

Generally, the pole $s_P$ is located in the complex plane of $p^2$
and has a real and an imaginary part. By writing
\begin{equation}
s_P \equiv M^2 - {\rm i } M \Gamma \EC
\label{def}
\end{equation}
the real part defines $M$ which we call the pole mass
while the imaginary part is related to the width $\Gamma$ of the
particle.  This is the natural generalization of the physical mass of
a stable particle, which is defined by the mass of its asymptotic
scattering state~\cite{Lehmann:1954rq,Veltman}.

For the remainder of the paper we will adopt the following notation:
capital $M$ always denotes the pole mass; lower case $m$ stands for
the renormalized mass in the $\overline{\rm MS}$ scheme, while $m_0$
denotes the bare mass. In addition we use $e$, $g$ and $g_s$ to denote
the $U(1)_{\rm em}$, $SU(2)_{\rm L}$ and $SU(3)_{\rm c}$ couplings of
the SM in the $\overline{\rm MS}$ scheme.

In perturbation theory (\ref{pole}) is to be solved order by order.
To two loops we have the solution\footnote{Similarly,
it is easy to find the on-shell wave-function renormalization constant
$Z_2$. Up to two loops it is given by the following equation~\cite{Broadhurst}:
$$
Z_2^{-1} = 1 - \Pi^{(1)}{}'(m^2,m^2,\cdots)
- \Pi^{(2)}{}'(m^2,m^2,\cdots) - \Pi^{(1)}(m^2,m^2,\cdots) \Pi^{(1)}{}''(m^2,m^2,\cdots)
\;,
$$ where $m^2$ is a bare or renormalized mass.  For a recent
discussion see also~\cite{wave-function}.  }
\begin{eqnarray}
\hspace{-1cm}
s_P &=& m^2 + \Pi^{(1)}(m^2,m^2,\cdots) + \Pi^{(2)}(m^2,m^2,\cdots)
+ \Pi^{(1)}(m^2,m^2,\cdots) \Pi^{(1)}{}'(m^2,m^2,\cdots),
\label{polemass}
\end{eqnarray}
which yields the pole mass $M^2$ and the width $\Gamma$ at this order.
$\Pi^{(L)}$ is the bare ($m=m_0$) or {\MSb}-renormalized ($m$ the
{\MSb}-mass) $L$-loop contribution to $\Pi$, and the prime denotes the
derivative with respect to $p^2$.

In this paper we show by explicit calculation at the two-loop level
that the fermion contribution (including mixed terms) to the
propagator pole $s_P$ of a gauge boson is a gauge invariant and
infrared stable quantity.  For more details concerning the tensor
decomposition of propagators and abbreviations adopted we refer to
Sec.~2 of I.

In order to find the relationship between the poles of the gauge
boson-propagators and the $\overline{\rm MS}$ masses $m_Z^2,\,m_W^2$
we have to compute the one- and two-loop self-energies for the $Z$-
and $W$-bosons at $p^2=m_Z^2$ and $p^2=m_W^2$,
respectively\footnote{There exist a number of programs for analytical
and/or numerical calculations of two--loop self-energies. A selection
may be found in~\cite{numeric}.}.

For the calculation of the two-loop fermion contributions we again use
the strategy described in our previous paper I.  Let us present here
some basic features.  In order to be able to work with manifestly gauge
independent parameter renormalization constants we have to include the
Higgs tadpole diagrams.  To keep control of gauge invariance we work in
the $R_\xi$ gauge with three different gauge parameters $\xi_W,\,\xi_Z$
and $\xi_\gamma$. However, in order to avoid additional mass parameters
like $\sqrt{\xi_W}m_W$ and $\sqrt{\xi_Z}m_Z$ (ghost masses), we expand
the original propagators in an asymptotic series at $\xi_i=1$. For the
purpose of checking the gauge invariance of our results it turned out to
be sufficient to keep the first two terms of the expansion. As an
additional cross--check we actually kept one more term. 

\begin{figure}[th]
\begin{center}
\centerline{\vbox{\epsfysize=120mm \epsfbox{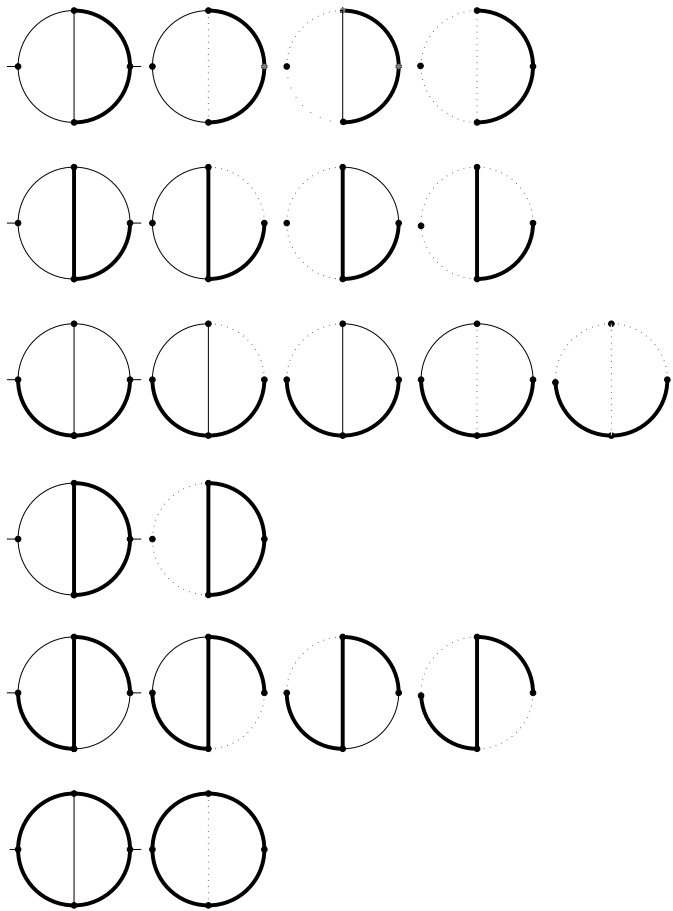}}}
\caption{\label{prototypes} The prototype diagrams and their subgraphs
contributing to the large mass expansion for two-loop diagrams with
heavy propagators. Thick and thin lines correspond to heavy- and
light-mass (massless) particle propagators, respectively.
Dotted lines indicate the lines omitted in the subgraph.}
\end{center}
\end{figure}

For calculating diagrams with massless fermion loops we utilize
Tarasov's recurrence relations~\cite{tarasov-propagator} (a detailed
discussion we postpone to Sec.~3).  For diagrams with top-quark and/or
Higgs-boson propagators we apply asymptotic expansions with respect to
the heavy masses.  In these cases we firstly expand the propagator in
the weak mixing parameter $\sin^2\theta_W = 1 - m_W^2/m_Z^2 \approx
0.25$ and get rid in this way of $m_W$ (or $m_Z$).  All diagrams with
top-quarks and/or Higgs-bosons are divided into several prototypes which
are presented in Fig.~\ref{prototypes}.  The large mass expansion has
been performed with the help of the packages {\bf TLAMM}~\cite{tlamm} or
by the program described in~\cite{top}. In the heavy top limit, the
leading and next--to--leading contributions to the gauge boson
self--energies have been calculated in~\cite{Twolooptop,qcdtadpole} and
~\cite{Twoloopnexttop}, respectively.

As usual, in order to preserve gauge invariance of the regularized
theory, we utilize dimensional regularization~\cite{dimreg} for our
calculation.  In order to avoid spurious anomalies we adopt an
anti-commuting $\gamma_5$ in dimensional
regularization~\cite{dimreg,Bardeen:vi} and impose vector current
conservation by hand in triangle fermion loops which contribute to the
two--loop two--point functions~\cite{gamma5}. The axial vector current
anomalies~\cite{ABJ} cancel for each fermion family by the lepton--quark
duality~\cite{SManomaly}. In contrast to calculations which involve
vertex- and/or box-type diagrams~\cite{freitas}, in our calculation of
self-energies we do not encounter problems with violation of Ward
identities (see also~\cite{2loop-fermion}).

%

\section{Master integrals for the massless fermion contributions}
\label{Mfc}
This part is devoted to the calculation of the two-loop massless
fermion corrections to the pole masses of the gauge bosons. For this
class of corrections it is possible to work out the exact analytical
results without expansion\footnote{In contrast to the bosonic
corrections where all diagrams can be expanded from the very beginning
in the small parameter $\sin^2 \theta_W$ the individual diagrams with
massless fermion-loops develop threshold singularities which behave
like powers of $\ln \sin^2\theta_W$. To control these terms we
need the exact analytical result.}.

\begin{figure}[th]
\centering
{\vbox{\epsfysize=130mm \epsfbox{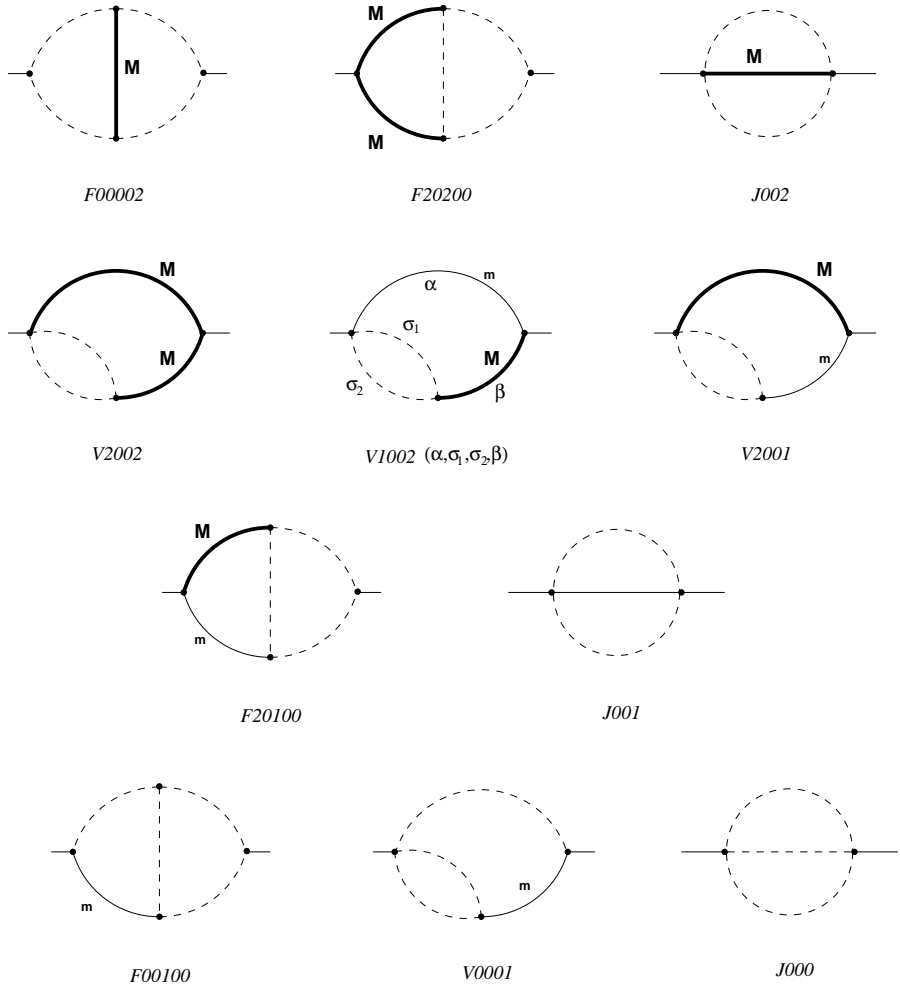}}}
\caption{
Diagrams corresponding to the master--integrals with massless
fermion-loops contributing to the two-loop gauge-boson propagators.
Bold, thin and dashed lines correspond to off-shell massive, on-shell
massive and to massless propagators, respectively.}
\label{master}
\end{figure}

\begin{figure}[bth]
\centering
{\vbox{\epsfysize=25mm \epsfbox{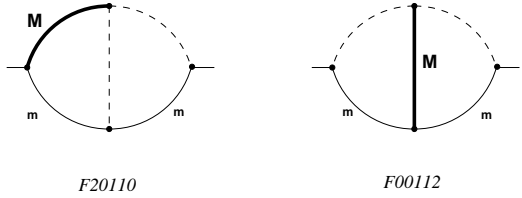}}}
\caption{Diagrams contributing to the two-loop on-shell top-quark
propagator in the limit of massless gauge-bosons.
Bold, thin and dashed lines correspond to off-shell massive,
on-shell massive and to massless propagators, respectively.}
\label{master2}
\end{figure}

In contrast to the previous calculations performed
in~\cite{2loop-fermion}, where results are expressed in terms of the
non--minimal set of scalar two--loop integrals, we use here Tarasov's
recurrence relations~\cite{tarasov-propagator} which allow us to reduce
the number of integrals to a minimal set of master--integrals in our
results. The analytical results for these diagrams are presented in
\cite{2loop-analytic-b}.  Here we present an independent analytical
calculation of the relevant master integrals shown in Fig.~\ref{master}.
Besides the known one's, we consider here two new master-integrals,
shown in Fig.~\ref{master2}, which contribute to the two-loop on-shell
top-quark propagator in the limit of massless gauge-bosons. They may
also be considered as the ``naive'' parts of the asymptotic expansion of
the original diagrams with massive $W$- and $Z$-bosons in the limit,
when the external momentum approaches the heavy
mass-shell~\cite{asymp:onshell}.

For the analytical calculation of these integrals up to the finite parts
in $\varepsilon$
we use the method developed in~\cite{FKV99} (see also~\cite{RS}).  This
approach is based on the possibility to retrieve the analytical results
in terms of sums, as predicted by the differential equation
method~\cite{kotikov}, from several of the first coefficients of the
small momentum expansion \cite{asymptotic}.  Whenever it is possible, we
present the exact analytical results in terms of hypergeometric
functions. Thereby, we apply Mellin--Barnes techniques developed
in~\cite{BD,BFT93}.

Let us introduce now the following notation for the finite sums which
show up in the master integrals: $$ S_a(n) = \sum_{j=1}^n
\frac{1}{j^a}, ~~~~~V_a(n) = \sum_{j=1}^n \frac{1}{\left( 2j \atop
j\right) } \frac{1}{j^a}, ~~~~~W_a(n) = \sum_{j=1}^n \left( 2j \atop
j\right) \frac{1}{j^a}\; .  $$
\noindent
Throughout the rest of this paper, the symbol $S_a$ will always stand
for $S_a(n-1)$ and $\bar{S}_a$ will stand for $S_a(2n-1)$.  The same
notation applies for all the other type of sums, e.g., $V_a=V_a(n-1)$.
We denote all master integrals
as $T_{AB \cdots }$, where the first letter $T=F,V,J$ indicates the
topology in accordance with notation introduced
in~\cite{tarasov-propagator}; indices $A,B,\cdots = 0,1,2$
characterize the relation of the corresponding internal mass to the external
momentum: $0$ indicates a massless line, $1$ corresponds to ``internal mass
equal to external momentum'' and $2$ means that mass and momentum are
different (see~Fig.~\ref{master} for details).
In our normalization each loop is divided by
$(\pi)^{2-\ep}\Gamma(1+\ep).$

\newpage

\Item{$\bf F_{00002}$}
For this integral we use the representation\footnote{Another representation
is given by Eq.~(104) in~\cite{2loop-analytic-b}. } given in~\cite{FKV99}

\begin{eqnarray}
\label{I11int}
p^2 \cdot F_{00002} & = &
      2 \zeta_2 \ln(1+z) + 2\ln(-z){\rm Li}_2(-z) + \ln^2(-z)\ln(1+z)
        + 4\ln(1+z){\rm Li}_2(z)
            \nonumber \\ &&
           - 2{\rm Li}_3(-z)
           - 2{\rm Li}_3(z)
           + 2 S_{1,2}(z^2) - 4S_{1,2}(z) - 4 S_{1,2}(-z)\,,
\end{eqnarray}
where $z = \frac{p^2}{M^2}$ and $S_{n,p}$ are Nielsen
polylogarithms~\cite{Nielsen,Lewin}.  For the on--shell gauge-boson
propagators two points in the variable $z$ are interesting: (i) $z = \cos^2
\theta_W[p^2=m_W^2]$; (ii)$z = 1/\cos^2 \theta_W[p^2=m_Z^2]$.
\Item{$\bf F_{20200}$}
For this integral the following
representation\footnote{The general case with two different internal masses
is given by Eq.~(106) in \cite{2loop-analytic-b} .} is valid

\begin{eqnarray}
\label{I12int}
p^2 \cdot F_{20200} & = &
     \Li{3}{z} - 6 \zeta_3 - \zeta_2 \ln y
     - \frac16\ln^3 y - 4\ln y\, \Li{2}{y}
\nonumber\\ &&
+ 4 \Li{3}{y} - 3 \Li{3}{-y}  + \frac13 \Li{3}{-y^3}\; ,
\end{eqnarray}
where we have introduced the new ``conformal'' variable
\begin{equation}
y = \frac{1-\sqrt{\frac{z}{z-4}}}{1+\sqrt{\frac{z}{z-4}}},
~~~~z = - \frac{(1-y)^2}{y} \equiv \frac{p^2}{M^2}\; .
\label{y:definition}
\end{equation}
Here we are interested in the case $z=1/\cos^2 \theta_W$ only.

\newpage

\Item{$\bf V_{2002}$}
The off-shell integral $V_{X00Y}$ with two different masses and unit
powers of propagators was considered in~\cite{v1001_general}. The
result, Eq.~(46) of~\cite{v1001_general} was presented as a sum of
$F_2$ and $F_4$ hypergeometric functions of two variables. It turns
out, however, that the extraction of the finite part from this result
is difficult.  The analytical result up to finite parts for this
diagram (see Fig.~\ref{master}) is given by\footnote{The general case with
two different internal masses is given by Eq.~(95)
in~\cite{2loop-analytic-b}.}
\begin{eqnarray}
&&
(M^2)^{2\ep}\,V_{2002}
=
  - \frac{1}{2\ep^2} - \frac{1}{2\ep}  - \frac12 - \zeta_2
    + \sum\limits_{n=1}^\infty z^n
        \Biggl\{
         \frac{1}{\ep} \left[ - \frac{1}{n}\frac{1}{{2n \choose n}}
                + 4\frac{1}{n+1}\frac{1}{{2n+2 \choose n+1}} \right]
\nonumber\\ &&
   - \frac{1}{n} - \frac12\frac{1}{n^2}
   - \frac{1}{n}\frac{1}{{2n \choose n}}
     \Biggl[ \frac{3}{2} W_1 + S_1 + 4 \Biggr]
   + \frac{1}{n+1}
   + \frac{1}{n+1}\frac{1}{{2n+2 \choose n+1}}
     \Biggl[ 6 W_1(n) + 4 S_1(n) + 16 \Biggr]
  \Biggr\} ,
\nonumber \\
\label{v2002:series}
\end{eqnarray}
where $z = p^2/M^2$ with $M$ an internal mass.
This result can be written also as
\begin{eqnarray}
&& \hspace{-10mm}
\left(\frac{p^2}{z} \right)^{2\ep}\,V_{2002} =
   -\frac{1}{2\ep^2}
   -\frac{1}{\ep}\Biggl[ \frac52 + \frac{1+y}{1-y}\ln y \Biggr]
   -\frac{19}{2} - \zeta_2
   -\frac12 {\rm Li}_2(z) - \frac{1-z}{z}\ln(1-z)
\nonumber\\ &&
   + \frac{1+y}{1-y} \Biggl[
     \frac72 {\rm Li}_2(-y) - \frac12 {\rm Li}_2(-y^3)
     - \frac54\ln^2 y
       + 2\ln(1+y)\ln y - 4\ln y + \frac32\zeta_2
                   \Biggr] ,
\label{v2002:analytical}
\end{eqnarray}
where $y$ is defined in (\ref{y:definition}).  If we compare this with the
cases considered previously in~\cite{FKV99}, as a novelty, in
Eq.~(\ref{v2002:series}) we encounter series with shifted index of
summation ($n+1$ instead of $n$). The latter give rise to the appearance of
terms of different weight in the finite part. They are connected with
the presence of UV-divergent subgraphs. For this diagram only the point
$z=1/\cos^2 \theta_W$ $[p^2=m_Z^2]$ is of interest here.
\Item{$\bf J_{002}$}
For the off-shell case, we have two master--integrals: one with all
indices equal to one, and a second one with indices 1,1,2.  Their
$\ep$-expansion up to order $O(\ep^2)$ are given by Eqs.~(A1) and (A2)
in Appendix A of~\cite{FJTV}.  For our task only the following finite
parts are needed:
\begin{eqnarray}
\hspace{-5mm}
J_{002}(1,1,1,p^2,m^2) & = &
   (m^2)^{1-2\varepsilon} \Biggl[
   - \frac{1}{2\varepsilon^2}
   + \frac{z-6}{4 \ep}
   - \zeta_2
   -3 + \frac{13}{8} z
   + \frac{1-z^2}{2 z} \ln (1-z)  - \Li{2}{z} \Biggr]\;,
\nonumber \\
\hspace{-5mm}
J_{002}(1,1,2,p^2,m^2) & = &
   (m^2)^{-2\varepsilon}\Biggl[
   - \frac{1}{2\varepsilon^2} - \frac{1}{2\varepsilon}
   + \frac{1}{2} - \zeta_2
   + \frac{1-z}{z} \ln (1-z) -  \Li{2}{z} \Biggr] \;.
\nonumber
\end{eqnarray}
For this diagram again two expansion
points are important: $z=\cos^2 \theta_W$ and $z=1/\cos^2 \theta_W$.

\newpage

\Item{$\bf V_{1002}$}
Let us present some useful details for this integral.
Its Mellin-Barnes representation is
\begin{eqnarray}
&&
V_{1002}(\alpha,\sigma_1,\sigma_2, \beta, p^2,m^2,M^2) \left. \right |_{p^2 = m^2} =
  \nonumber \\ &&
  \int \frac{d^d k_1}{\Gamma(1+\ep)} \frac{d^d k_2}{\Gamma(1+\ep)}
   P^{(\alpha)}(k_2-p,m) P^{(\sigma_1)}(k_1-k_2,0) P^{(\sigma_2)}(k_1,0) P^{(\beta)}(k_2,M) =
  \nonumber \\ &&
 \frac{(\MIN)^2(-m^2)^{d-\alpha-\beta-\sigma_1-\sigma_2}
       \Gamma(\sigma_1 \!+\! \sigma_2 \!-\! \frac{d}{2})\Gamma(\frac{d}{2} \!-\! \sigma_1)
       \Gamma(\frac{d}{2} \!-\! \sigma_2)}
      {\Gamma^2(1+\ep)
       \Gamma(\sigma_1) \Gamma(\sigma_2) \Gamma(\alpha) \Gamma(\beta)
       \Gamma(d \!-\! \sigma_1 \!-\! \sigma_2)}
\frac{1}{2\pi i }\int_{-i \infty}^{i \infty} ds  \left( \frac{M^2}{m^2} \right)^s
  \nonumber \\ &&
\frac{\Gamma(-s) \Gamma(\beta+s)\Gamma(\alpha \!+\! \beta \!+\! \sigma_1 \!+\! \sigma_2 \!+\! s \!-\! d)
\Gamma(2d \!-\! 2\sigma_1  \!-\! 2 \sigma_2  \!-\! 2 \beta \!-\! \alpha \!-\! 2s)}
     {\Gamma(\frac{3d}{2} \!-\! \sigma_1 \!-\! \sigma_2 \!-\! \alpha
\!-\! \beta \!-\! s)}\; ,
\label{v1002:mellin}
\end{eqnarray}
where  $P^{(\sigma)}(p,m) = 1/(p^2-m^2 + {\rm i} \epsilon)^\sigma.$
Closing the integration contour to the left $(s \leq 0)$ we find the
following result for an arbitrary set of indices:
\begin{eqnarray}
&&
V_{1002}(\alpha,\sigma_1,\sigma_2, \beta, p^2,m^2,M^2) \left. \right |_{p^2 = m^2}
=
  \frac{(\MIN)^2
        \Gamma(\sigma_1 \!+\! \sigma_2 \!-\! \frac{d}{2})\Gamma(\frac{d}{2} \!-\! \sigma_1)
        \Gamma(\frac{d}{2} \!-\! \sigma_2)}
       {\Gamma^2(1+\ep)\Gamma(\sigma_1) \Gamma(\sigma_2) \Gamma(\alpha) \Gamma(\beta)
        \Gamma(d \!-\! \sigma_1 \!-\! \sigma_2)} \times
  \nonumber \\ &&
(-M^2)^{d-\alpha-\beta-\sigma_1-\sigma_2}
\Biggl[
\left( \frac{m^2}{M^2} \right)^{d-\alpha-\sigma_1-\sigma_2}
\frac{\Gamma(\beta)\Gamma(\alpha \!+\! \sigma_1 \!+\! \sigma_2 \!-\! d)
      \Gamma(2d \!-\! 2\sigma_1 \!-\! 2\sigma_2 \!-\! \alpha)}
     {\Gamma(\frac{3d}{2} \!-\! \alpha \!-\! \sigma_1 \!-\! \sigma_2)} \times
  \nonumber \\ &&
~{}_{3}F_2 \left(\begin{array}{c|}
      \beta, d \!-\! \sigma_1 \!-\! \sigma_2 \!-\! \frac{\alpha}{2},
       d \!-\! \sigma_1 \!-\! \sigma_2 \!+\! \frac{1-\alpha}{2}
\\ 1 \!+\! d \!-\! \alpha \!-\! \sigma_1 \!-\! \sigma_2,
     \frac{3d}{2} \!-\! \alpha \!-\! \sigma_1 \!-\! \sigma_2
\end{array} ~\frac{4m^2}{M^2} \right)
  \nonumber \\ &&
+ \frac{\Gamma(\alpha)\Gamma(\alpha  \!+\! \beta \!+\! \sigma_1 \!+\! \sigma_2 \!-\!d)
                      \Gamma(d \!-\!\alpha \!-\!\sigma_1 \!-\! \sigma_2)}
     {\Gamma(\frac{d}{2})}
~{}_{3}F_2 \left(\begin{array}{c|}
      \frac{\alpha}{2}, \frac{1+\alpha}{2},
      \alpha \!+\! \beta \!+\! \sigma_1 \!+\! \sigma_2 \!-\! d
\\ 1 \!+\! \alpha \!+\! \sigma_1 \!+\! \sigma_2 \!-\! d, \frac{d}{2}
\end{array} ~\frac{4m^2}{M^2} \right)
\Biggr] \EP
\nonumber \\ &&
\label{v1002:left}
\end{eqnarray}
We introduce the short notation $$ V_{1002}(1,1,1,1,p^2,m^2,M^2)
\left. \right |_{p^2 = m^2} \equiv V_{1002} $$ for the master integral
$\alpha=\beta=\sigma_1=\sigma_2=1$. The result for this particular
integral reads

\begin{eqnarray}
&&
V_{1002} =
-\frac{\left(M^2 \right)^{-2\ep}}{2 (1-\ep) (1-2\ep) \ep^2}
\frac{\Gamma(1-\ep) \Gamma(1+2 \ep)}{\Gamma(1+\ep)}
\times
\nonumber \\ &&
\Biggl[
~{}_{2}F_1 \left(\begin{array}{c|} 1, \frac{1}{2}\\ 2-\ep \end{array} ~4x \right)
-  x^{1-2\ep} \frac{2(1-\ep)(1-4\ep)}{(2-3\ep)(1-3\ep)}
\frac{\Gamma(1-4\ep) \Gamma(1-\ep)}{\Gamma(1-3\ep) \Gamma(1-2\ep)}
~{}_{2}F_1 \left(\begin{array}{c|} 1, \frac{3}{2} - 2 \ep\\ 3 - 3\ep \end{array} ~4x \right)
\Biggr]
\nonumber \\
\end{eqnarray}
where $x=\frac{m^2}{M^2}$.
Using the quadratic transformation for hypergeometric functions
$$
~{}_{2}F_1 \left(\begin{array}{c|}
 1, \frac{1}{2}\\ 2-\ep \end{array} ~4x \right)
= \frac{2}{1+\sqrt{1-4x}}
~{}_{2}F_1 \left(\begin{array}{c|} 1, \ep \\ 2-\ep \end{array}
~\frac{1-\sqrt{1-4x}}{1+\sqrt{1-4x}} \right)
\equiv (1+\chi) \left. _2F_1\left( \begin{array}{c} 1, \; \ep \\
2-\ep \end{array} \right| \chi \right) $$ where $\chi$ is defined in
Eq.~(\ref{chi:definition}), the first term can be reduced to a new
$_{2}F_1$ function, whose all-order $\ep$-expansion is given by
Eq.~(2.14) of~\cite{bastei_ep}:
\begin{eqnarray}
\left. _2F_1\left( \begin{array}{c} 1, \; \ep \\
                 2-\ep \end{array} \right| z  \right) & = &
\frac{1-\ep}{2(1\!-\!2\ep)z}
\Biggl\{ 1+z - (1\!-\!z)^{1 - 2\ep} \crn &&
- 2 (1\!-\!z)^{1- 2\ep} \ep
\sum_{j=1}^\infty \ep^j \sum_{k=1}^{j} (-2)^{j-k} S_{k,j-k+1}(z)
\Biggl\}\;.
\end{eqnarray}
Note, that up to order $\varepsilon^3$ the expansion of the given
hypergeometric function can be extracted
from Eq.~(A.3) of~\cite{FJTV}.
The second term can be reduced to a $_{2}F_1$ function of the type
considered in Appendix A
(see Eq.~(\ref{hypergometric_expansion})). To this end the relation
$$
\frac{\left[1+2(a-b)\ep \right]z}{2(2-b\ep)}
~{}_{2}F_1 \left(\begin{array}{c|} 1, \frac{3}{2}-a\ep\\
3-b\ep \end{array} ~z \right)
=
1 - (1-z)
~{}_{2}F_1 \left(\begin{array}{c|} 1, \frac{3}{2}-a\ep\\
2-b\ep \end{array} ~z \right)
$$
has to be applied.
Combining the relations given above we may represent the integral as a
series:
\begin{eqnarray}
&&
(M^2)^{2 \varepsilon} V_{1002} =
  - \frac{1}{2\varepsilon^2}   - \frac{3}{2\varepsilon}
+  \frac{1}{\varepsilon} \sum\limits_{n=1}^\infty \frac{{2n\choose n}}{n+1}x^n
   \Biggl[- \ln x + 2S_1(n) - 2S_1(2n) + \frac{1}{n+1} \Biggr]
\nonumber\\ &&
+  \sum\limits_{n=1}^\infty \frac{{2n\choose n}}{n+1}x^n
     \Biggl\{
        \ln^2 x
      + \Biggl[ -5S_1(n) + 4S_1(2n) - \frac{3}{n+1} - 2 \Biggr] \ln x
      + \zeta_2
\nonumber\\ &&
      + 3S_2(n) - 4S_2(2n) + 6S_1^2(n) - 10S_1(n)S_1(2n) + 4S_1^2(2n)
\nonumber\\ &&
      + 7\frac{S_1(n)}{n+1} - 6\frac{S_1(2n)}{n+1} + \frac{4}{(n+1)^2}
      + 4S_1(n) - 4S_1(2n)  +  \frac{2}{n+1} \Biggr\}
     - \zeta_2 - \frac72
\label{v1002:series}
\end{eqnarray}
or in analytical form:
\begin{eqnarray}
\label{v1002:analytical}
&&
\left( \frac{p^2}{x} \right)^{2\ep}\,V_{1002} =
   - \frac{1}{2\ep^2}
   - \frac{1}{\ep}\Biggl[ \frac{5}{2} + \chi \ln x  - \frac{(1-\chi^2)}{\chi} \ln (1+\chi)\Biggr]
\nonumber \\ &&
+ \chi \ln^2 x
- \ln x \left(4 \chi -1 + \frac{1-\chi^2}{\chi} \left[ \ln (1-\chi) + 2 \ln (1+\chi) \right] \right)
- \zeta_2
- \frac{19}{2}
+ \chi \zeta_2
\nonumber \\ &&
+ \frac{1-\chi^2}{\chi} \left[
4 \ln (1+\chi)
- 2 \ln (1+\chi) \ln (1-\chi)
- 3\Li{2}{-\chi} - \Li{2}{\chi} - 2 \ln^2 (1+\chi)
                      \right]\; .
\nonumber \\
\end{eqnarray}
The all order $\ep$--expansion for this diagram can be deduced for $x \geq
1/4$.  Closing the contour of the integral~(\ref{v1002:mellin}) on the
right $(s \geq 0)$ we obtain an alternative representation. The latter also
may be obtained from~(\ref{v1002:left}) by analytical continuation of the
hypergeometric function to the inverse argument:
\begin{eqnarray}
&&
-4\ep^2 (1-2\ep) \left(m^2 \right)^{2\ep} V_{1002} =
\left( \frac{1}{x} \right)^{1-2\ep}
\frac{\Gamma(1-\ep) \Gamma(1+2\ep)}{\Gamma(1+\ep)}
~{}_{2}F_1 \left(\begin{array}{c|} 1, \ep \\ \frac{3}{2}
\end{array} ~\frac{1}{4x} \right)
\nonumber \\ &&
+ \frac{2}{ (1-3\ep) }
\frac{\Gamma^2(1-\ep) \Gamma(1+2 \ep)\Gamma(1-4\ep)}
     {\Gamma(1+\ep)\Gamma(1-2\ep)\Gamma(1-3\ep)}
~{}_{2}F_1 \left(\begin{array}{c|} 1, -1+3\ep \\\frac{1}{2}+2\ep
\end{array} ~\frac{1}{4x} \right)
\nonumber \\ &&
-  \frac{4\pi\ep}{(1-2\ep)}
\frac{\Gamma^3(1-\ep) \Gamma(1-4\ep)\Gamma(1+4\ep)}
     {\Gamma^3(1-2\ep)\Gamma(1+\ep)\Gamma(1+2\ep)}
\sqrt{\frac{1}{x}} \left(\frac{2}{x}\right)^{-2\ep}
\left(1-\frac{1}{4x} \right)^{\frac{1}{2}-\ep} \EP
\end{eqnarray}
The $\ep$--expansion of the first hypergeometric function $_{2}F_1$
is given in~\cite{D-ep}.  It only contains log-sine
integrals~\cite{Lewin}.  The second term at first may be transformed
using the Kummer relation to a new function (see Eq.~(4.13)
in~\cite{DK2001}) whose $\ep$--expansion is given by Eq.~(B.13)
in~\cite{DK2001} and includes a $\mbox{Lsc}$
function~(\ref{log-sin-cos}). The analytical continuation of the
$\ep$--expanded result to $x>1/4$ can be constructed in the manner
described in Sec. 2.2. of~\cite{DK2001}.

The point of interest here is $x=\cos^2 \theta_W$ $[p^2=m_W^2]$. In
the rest of this section the expansion parameter $x$ will denote
$x=m^2/M^2$.

\Item{$\bf V_{2001}$}
For this integral we have the series

\begin{eqnarray}
&&
(M^2)^{2\varepsilon} V_{2001} =
 - \frac{1}{2\varepsilon^2}   - \frac{3}{2\varepsilon}
+  \frac{1}{\varepsilon} \sum\limits_{n=1}^\infty \frac{{2n\choose n}}{n+1}x^n
   \Biggl[- \ln x + 2S_1(n) - 2S_1(2n) + \frac{1}{n+1} \Biggr]
\nonumber\\ &&
+ \sum\limits_{n=1}^\infty \frac{{2n\choose n}}{n+1}x^n
     \Biggl\{
        \ln^2 x
      - \Biggl[ 3S_1(n) - 2S_1(2n) + \frac{2}{n+1} + 2 \Biggr] \ln x
      - 3V_2(n) + S_2(n)
\nonumber\\ &&
      + 2S_1^2(n) - 2S_1(n)S_1(2n)
      + 3\frac{S_1(n)}{n+1} - 2\frac{S_1(2n)}{n+1} + \frac{2}{(n+1)^2}
      + 4S_1(n) - 4S_1(2n) +  \frac{2}{n+1} \Biggr\}
\nonumber\\ &&
+  \sum\limits_{n=1}^\infty x^n \Biggr[
      \frac{1}{(n+1)^2} + \frac{1}{n+1} - \frac{1}{n} \Biggr]
- \zeta_2 - \frac{7}{2}
\label{v2001:series}
\end{eqnarray}
and the analytical solution
\begin{eqnarray}
&&
\left( \frac{p^2}{x} \right)^{2\ep}\,V_{2001} =
   - \frac{1}{2\ep^2}
   - \frac{1}{\ep}\Biggl[ \frac{5}{2} + \chi \ln x  - \frac{(1-\chi^2)}{\chi} \ln (1+\chi)\Biggr]
\nonumber\\ &&
+ \chi \ln^2 x - \ln x \left[ 4 \chi + \frac{(1-\chi^2)}{\chi} \ln (1-\chi^2) \right]
- \zeta_2
-\frac{19}{2}
\nonumber\\ &&
- \frac{(1-\chi^2)}{2 \chi}  \left[ 4 \ln(1-\chi) \ln (1+\chi) - 8 \ln (1+\chi) \right]
- \frac{(1-x)}{x} \ln (1-x)
\nonumber\\ &&
- \frac{(1-\chi^2)}{2 \chi}
\left[5 \Li{2}{\chi} + 8 \Li{2}{-\chi}  - \Li{2}{\chi^3}\right]
- \frac{1}{2x} \Li{2}{x} \EP
\label{v2001:analytical}
\end{eqnarray}
The physical point is $x=\cos^2\theta_W$.

\newpage

\Item{$\bf F_{20100}$}
The series and the analytical solution for this
integral are given by

\begin{eqnarray}
M^2 F_{20100} & = &
  \frac{1}{x} \sum\limits_{n=1}^\infty {2n\choose n}\frac{x^n}{n}
     \Biggl\{  - \frac14 \ln^2 x + \frac12 \ln(-x)\ln x
     + \Biggl[ 2 S_1 - 2 \bar{S}_1  + \frac1n \Biggr]\ln x
\nonumber\\ &&
     + \Biggl[ -S_1 + \bar{S}_1 - \frac1n \Biggr] \ln(-x)
     - \zeta_2
     - \frac32 V_2
     - \frac32 S_2
     + 3\bar{S}_2
\nonumber\\ &&
     - 3S_1^2 + 6S_1\bar{S}_1 - 3\bar{S}_1^2
     - 4\frac{S_1}{n} + 4\frac{\bar{S}_1}{n}
     - \frac32\frac{1}{n^2}
     - 2\frac{1}{{2n\choose n}}\frac{1}{n^2}
     \Biggr\}
\label{f20100:series}
\end{eqnarray}
and
\begin{eqnarray}
&&
p^2 F_{20100}  =
\frac{2}{3} \ln^3 (1+\chi)
-2 \zeta_2 \ln(1+\chi)
- \ln^2(1+\chi) \ln(-\chi)
+ \ln \chi \ln (-\chi) \ln(1+\chi)
\nonumber \\ &&
-\frac{1}{2} \ln^2 \chi \ln(1+\chi)
+ 2 \ln \chi \Li{2}{-\chi}
- 2 \Li{3}{-\chi}
-\frac{1}{6} \Li{3}{\chi^3}
+\frac{3}{2} \Li{3}{\chi}
- \frac{1}{2} \Li{3}{x} \EC
\nonumber \\
\label{f20100:analytical}
\end{eqnarray}
respectively. The physical point is again $x=\cos^2\theta_W$.
\Item{\bf ON-SHELL2}
We are interested also in the integrals
$F_{00200}(1,1,1,1,1,p^2,m^2)$, $V_{0002}(1,1,1,1,p^2,m^2)$ and
$J_{000}(1,1,1,p^2)$ when the external momentum is on-shell. Up to
finite parts these diagrams have been calculated
in~\cite{2loop-analytic-b} for arbitrary values of the external
momentum.  For our purpose it is important, that the limit $p^2 \to
m^2$ exists and is smooth, such that we can substitute the on-shell
values of these diagrams\footnote{We note that the sign of the
imaginary parts of the following diagrams: $({\bf F_{00100}}, {\bf
V_{0110}}, {\bf V_{0010}}, {\bf V_{0000}}, {\bf J_{000}})$ collected
in Appendix C of~\cite{onshell2} must be changed in order to get the
correct answer.} without problems:
\begin{eqnarray}
&&
F_{00100}(1,1,1,1,1,m^2,m^2) =   \frac{1}{m^2} \Biggl[ -3 \zeta_3 + 2 i \pi \zeta_2 \Biggr] \;,
\nonumber \\ &&
V_{0001}(1,1,1,1,m^2,m^2)  =    (m^2)^{-2\ep} \Biggl[- \frac{1}{2 \ep^2} - \frac{5}{2 \ep} -  \zeta_2
                                                  - \frac{19}{2} - i \pi \Biggr] \;,
\nonumber \\ &&
J_{001}(1,1,1,m^2)  =  (m^2)^{1-2\ep} \Biggl[
                 - \frac{1}{2\ep^2}  - \frac{5}{4\ep} - \frac{11}{8} - 2 \zeta_2
                                      \Biggr]
\nonumber \\ &&
J_{000}(1,1,1,m^2)  = (m^2)^{1-2\ep} \Biggl[ \frac{1}{4 \ep} + \frac{13}{18} - i \pi \frac{1}{2} \Biggr]
\nonumber \;.
\end{eqnarray}

\newpage

\Item{$\bf F_{20110}$}
This master-integral shows up as a Higgs contribution to the pole of
the top-quark propagator in the approximation of massless $W$- and
$Z$-bosons and a massive Higgs-boson $(m_H^2 = M^2,\: m_t^2 = m^2\gg
M_W^2,\: M_Z^2)$.  It may be obtained also as the leading term of a
Taylor expansion of the original diagram with massive $W$- and
$Z$-bosons in the limit, when the external momentum is on the heavy
mass-shell~\cite{asymp:onshell}.  For this integral we find the
following series representation
\begin{eqnarray}
M^{2} F_{20110}  &=&
  \frac{1}{x} \sum\limits_{n=1}^\infty {2n\choose n} \frac{x^n}{n}
     \Biggl[ \frac14 \ln^2 x -  \frac1n \ln x - \frac12\zeta_2
\nonumber \\ &&
             - \frac12 S_2 + \bar{S}_2 - S_1^2 + 2S_1\bar{S}_1 - \bar{S}_1^2
             + \frac32\frac{1}{n^2}
      \Biggl]\;,\label{f20110:series}
\end{eqnarray}
and the analytical result
\begin{eqnarray}
&&
m^2 F_{20110}  =
- \zeta_2 \ln(1+\chi)
+ \frac{1}{2} \ln^2 \chi \ln(1+\chi)
+ 2 \ln \chi \Li{2}{-\chi}
- 3 \Li{3}{-\chi} .
\end{eqnarray}
The on-shell result coincides with~\cite{FKK99}.
\Item{$\bf F_{00112}$}
This master-integral again is important for the calculation of the
Higgs correction to the pole-mass of the top-quark.
Its representation is
\begin{eqnarray}
M^{2} F_{00112} & = &
  \frac{1}{x} \sum\limits_{n=1}^\infty {2n\choose n} \frac{x^n}{n}
     \Biggl\{ \frac12 \ln^2 x
     + \Biggl[ -2 S_1 + 2\bar{S}_1 - 3\frac1n \Biggr] \ln x
     + \zeta_2
\nonumber \\ &&
      + S_2 - 2\bar{S}_2 + 2 S_1^2 - 4 S_1\bar{S}_1 + 2 \bar{S}_1^2
      + 6\frac{S_1}{n} - 6\frac{\bar{S}_1}{n} + 6\frac{1}{n^2}
      \Biggl\} ,
\end{eqnarray}
or, in analytical form,
\begin{eqnarray}
&&
m^2 F_{00112}  =
2 \zeta_2 \ln(1+\chi)
+ \ln^2 \chi \ln(1+\chi)
+ 2 \ln \chi \Li{2}{-\chi} .
\end{eqnarray}
The on-shell result again coincides with the one given in~\cite{FKK99}.

\section{Renormalization}

The pole mass is a gauge independent and infrared stable quantity. For
the SM this has been shown to all orders of perturbation
theory~\cite{pole:SM}. In order to calculate the pole mass in terms of
renormalized quantities at the two-loop level we need to calculate the
one-loop renormalization constants for all physical parameters (charge
and masses) as well as the two-loop mass renormalization constant. Not
needed are the wave-function renormalizations and the unphysical sector
renormalizations. In order to obtain a gauge invariant result in the SM,
however, we have to add in a proper way the tadpole
contributions~\cite{FJ}.

We first perform the UV-renormalization within the $\overline{\rm MS}$
scheme in order to obtain finite results. In a next step we work out the
relation between the on--shell and $\overline{\rm MS}$ parameters. We
adopt the convention that the $\overline{\rm MS}$ parameters are defined
by multiplying each $L$-loop integral by the factor
$(\exp({\gamma})/4\pi)^{\varepsilon L}$. As usual we denote by $\mu$ the
\MSb renormalization scale.  For the RG functions we use the following
definitions : for all dimensionless coupling constants, like
$g,g',g_s,e,\lambda$, the $\beta$-function is given by $\mu^2
\frac{\partial}{\partial \mu^2} g = \beta_g$ and for all mass
parameters (a mass or the Higgs v.e.v. $v$) the anomalous dimension
$\gamma_{m^2}$ is given by $\mu^2
\frac{\partial}{\partial \mu^2} \ln m^2 = \gamma_{m^2}.$
\subsection{One-loop charge renormalization}
The relationship between the bare charge $e_0$ and the $\overline{\rm
MS}$ charge\footnote{ All $\overline{\rm MS}$--parameters, like $e,g,g'$ are 
$\mu$-dependent quantities.} $e$
reads
\begin{equation}
 e_0= \mu^\ep
e\:\left(1+Z_{\overline{\rm MS}}^{(1)}/\varepsilon +O(e^4) \right)\; ,
\end{equation}
with
\begin{equation}
Z_{\overline{\rm MS}}^{(1)} =
\frac{e^2}{16\pi^2} \left(
- \frac{7}{2} + \frac{2}{3} n_F \Biggl[ \frac{5}{9} N_c + 1  \Biggr]
\right) \;,
\label{e0to0}
\end{equation}
where $n_F$ is the number of fermion families
and $N_c$ the number of colors. In the SM we have $n_F=3$ and $N_c=3$.
The corresponding $\beta$-function reads
\begin{equation}
\mu^2 \frac{\partial}{\partial \mu^2} e =
\beta_e = e^3 \left( \frac{\beta_g}{g^3}
   + \frac{\beta_{g'}}{g'^3} \right)
   = \frac{e^3}{16\pi^2} \left(- \frac{7}{2} +
\frac{2}{3} n_F \Biggl[\frac{5}{9} N_c + 1\Biggr]  \right)+ {\it O} (e^5) \;.
\label{beta_e}
\end{equation}
where $g$ and $g'$ denote the $SU(2)_L$ and $U(1)_Y$ gauge couplings,
respectively.
\subsection{Mass renormalization}
The mass renormalization constants $(Z_{m})_V$ at two loops may be
written in the form
\begin{equation}
m_{V,0}^2 =  m_V^2(\mu)\;(Z_{m})_V = m_V^2(\mu)\:
\left( 1 + \frac{g^2(\mu)}{16\pi^2\ep} Z_V^{(1,1)}
+ \frac{g^4(\mu)}{(16\pi^2)^2\ep} Z_V^{(2,1)} +
\frac{g^4(\mu)}{(16\pi^2)^2\ep^2} Z_V^{(2,2)}
\right),
\label{baremsb}
\end{equation}
where $V$ stands for any of the bosons $Z$, $W$ or $H$. We shall use
the same notation for fermions $V \to f$, where we are interested in
particular in the top quark $f=t$. In addition to the masses, we have
one coupling constant as a free parameter of the SM which we have
chosen above to be the electric charge strength $e=g\sin\theta_W$.
The one-loop mass counter-terms are well known~\cite{FJ}. We divide
all corrections into bosonic and fermionic parts, $Z_V =
Z_{V,\mbox{boson}} + Z_{V,\mbox{fermion}}$. The purely bosonic
contributions have been given in I. The fermion-loop corrections yield
the following \MSb renormalization constants

\begin{eqnarray}
\label{one-loop_H}
\hspace{-5mm}
Z_{H,\mbox{fermion}}^{(1,1)} & = &
\sum_{\mbox{leptons}}       \frac{1}{2} \frac{m_l^2}{m_W^2}
+  N_c \sum_{\mbox{quarks}}    \frac{1}{2} \frac{m_q^2}{m_W^2}
\;,
\\\hspace{-5mm}
\label{one-loop_W}
Z_{W,\mbox{fermion}}^{(1,1)} & = &
       \sum_{\mbox{leptons}} \left( 2 \frac{m_l^4}{m_W^2 m_H^2}
                                -   \frac{1}{2} \frac{m_l^2}{m_W^2} \right)
+  N_c \sum_{\mbox{quarks}} \left( 2 \frac{m_q^4}{m_W^2 m_H^2}
                           - \frac{1}{2} \frac{m_q^2}{m_W^2}
                        \right) \nonumber \\ &&
+  \frac{1}{3} n_F \left( N_c + 1 \right) ,
\\ \hspace{-5mm}
Z_{Z,\mbox{fermion}}^{(1,1)} & = &
Z_{W,\mbox{fermion}}^{(1,1)}
+  \frac{1}{3} n_F \Biggl[
   \frac{m_Z^2}{m_W^2} \left( \frac{11}{9} N_c + 3  \right)
+  \frac{m_W^2}{m_Z^2} \left( \frac{20}{9} N_c + 4   \right)
-  \frac{31}{9} N_c - 7
\Biggr]
\nonumber \\
\label{one-loop}
\end{eqnarray}
\begin{eqnarray}
\hspace{-5mm}
Z_{t-{\rm quark}}^{(1,1)} & = &
\frac{2}{3}
- \frac{2}{3} \frac{m_Z^2}{m_W^2}
- \frac{3}{2} \frac{m_Z^4}{m_W^2 m_H^2}
- \frac{3}{4} \frac{m_H^2}{m_W^2}
- 3 \frac{m_W^2}{m_H^2}
+ \frac{3}{4} \frac{m_t^2}{m_W^2}
- \frac{3}{4} \frac{m_b^2}{m_W^2}
\nonumber \\ \hspace{-5mm} &&
+ 2 \sum_{\rm leptons} \frac{m_l^4}{m_W^2 m_H^2}
+ 2 N_c \sum_{\rm quarks} \frac{m_q^4}{m_W^2 m_H^2} \;,
\label{top-one-loop}
\end{eqnarray}
where sums run over all leptons $(e,\mu,\tau)$ (leptons) or quarks
$(u,d,s,c,b,t)$ (quarks). All masses here are \MSb masses and depend on
the renormalization scale $\mu$: $m^2_V=m^2_V(\mu)$. In general, we will
divide the two-loop fermion contribution to the mass-renormalization
constants into two parts: the first one includes the contribution from
$N_m$ massless fermion-families and the second one the contribution of
$N_h$ lepton-quark families exhibiting one massive u-quark of mass $m_t$
(massive top-quark). The number of fermion families is $n_F = N_m +
N_h$, where in the SM we have $N_m=2 $ and $N_h=1$. The results are
given by the following exact expressions
\begin{eqnarray}
\label{Z_21_w}
Z_{W, \mbox{fermion}}^{(2,1)} & = &
\Biggl[
- \frac{m_Z^4}{m_W^4} \left( \frac{55}{432} N_c  + \frac{5}{16} \right)
+ \frac{m_Z^2}{m_W^2} \left( \frac{29}{108} N_c + \frac{3}{4}  \right)
+ \frac{343}{216} N_c
+ \frac{31}{24}
\nonumber \\ &&
- \frac{m_Z^4}{m_W^2 m_H^2}  \left(\frac{44}{27} N_c  + 4\right)
+ \frac{m_Z^2}{m_H^2}  \left(\frac{40}{27} N_c  + \frac{8}{3} \right)
+ \frac{m_W^2}{m_H^2}  \frac{4}{3} (N_c + 1)
\nonumber \\ &&
+ \frac{m_Z^6}{m_W^4 m_H^2}  \left(\frac{22}{27} N_c + 2  \right)
\Biggr] n_F
+ Z_{W, \mbox{top}}^{(2,1)}
\\
\label{Z_22_w}
Z_{W, \mbox{fermion}}^{(2,2)} & = &
\Biggl[
    - \frac{1}{4} \frac{m_H^2}{m_W^2} (N_c+1)
    + \frac{m_Z^4}{m_W^4} \left( \frac{11}{72} N_c + \frac{3}{8} \right)
    - \frac{m_Z^2}{m_W^2} \left( \frac{1}{18} N_c +  \frac{1}{2} \right)
\nonumber \\ &&
    + \frac{1}{9} (N_c+1)^2 n_F
    - \frac{15}{4} N_c
    - \frac{127}{36}
    + \frac{m_Z^4}{m_W^2 m_H^2} \left( \frac{13}{18} N_c + \frac{5}{2} \right)
\nonumber \\ &&
    - \frac{m_Z^2}{m_H^2} \left(\frac{10}{9} N_c + 2  \right)
    - \frac{m_W^2}{m_H^2} 2 (N_c + 1)
     -\frac{m_Z^6}{m_H^2 m_W^2} \left(\frac{11}{18} N_c + \frac{3}{2}  \right)
\Biggr] n_F
\nonumber \\ &&
+ Z_{W, \mbox{top}}^{(2,2)}
\nonumber \\
\\
Z_{Z,\mbox{fermion}}^{(2,1)} & = &
\label{Z_21_z}
\Biggl[
\frac{1}{8} N_c + \frac{17}{8}
- \frac{m_Z^2}{m_W^2} \left(\frac{73}{216} N_c + \frac{19}{8} \right)
+ \frac{m_W^2}{m_Z^2} \left(\frac{301}{162} N_c + \frac{7}{6} \right)
\nonumber \\ &&
+ \frac{m_W^2}{m_H^2} \frac{4}{3} ( N_c + 1)
+ \frac{m_Z^4}{m_W^4} \left(\frac{109}{1296} N_c + \frac{13}{16} \right)
+ \frac{m_Z^2}{m_H^2} \left(\frac{40}{27} N_c + \frac{8}{3} \right)
\nonumber \\ &&
- \frac{m_Z^4}{m_W^2 m_H^2} \left(\frac{44}{27} N_c + 4 \right)
+ \frac{m_Z^6}{m_W^4 m_H^2} \left(\frac{22}{27} N_c + 2 \right)
\Biggr] n_F
+ Z_{Z, \mbox{top}}^{(2,1)}
\\
\label{Z_22_z}
Z_{Z,\mbox{fermion}}^{(2,2)} & = &
\Biggl[
  \frac{m_H^2}{m_W^2} \left(\frac{11}{18} N_c + \frac{3}{2} \right)
- \frac{m_H^2 m_Z^2}{m_W^4} \left(\frac{11}{36} N_c + \frac{3}{4} \right)
- \frac{m_H^2}{m_Z^2} \left( \frac{5}{9} N_c + 1 \right)
+ \frac{29}{108} N_c
\nonumber \\ &&
+ \frac{1}{243} (27 + 11 N_c)^2 n_F
- \frac{m_Z^2}{m_W^2}  \left(\frac{77}{108} N_c + \frac{7}{4} + \frac{1}{243} (27 + 11 N_c)^2 n_F  \right)
\nonumber \\ &&
+ \frac{m_Z^4}{m_W^4}  \left(\frac{385}{648} N_c + \frac{35}{24} + \frac{1}{729} (27 + 11 N_c)^2 n_F  \right)
- \frac{m_Z^2}{m_H^2}  \left(\frac{31}{9} N_c + 7 \right)
- \frac{1}{4}
\nonumber \\ &&
- \frac{m_W^2}{m_Z^2}  \left(\frac{308}{81} N_c + \frac{28}{9} + \frac{8}{729} (N_c+9) (5 N_c+9) \right)
+ \frac{m_W^2}{m_H^2}  \left(\frac{13}{9} N_c + 5 \right)
\nonumber \\ &&
- \frac{m_W^4}{m_Z^2 m_H^2} \left(\frac{20}{9} N_c + 4 \right)
+ \frac{m_Z^4}{m_W^2 m_H^2} \left(\frac{22}{9} N_c + 6 \right)
- \frac{m_Z^6}{m_W^4 m_H^2} \left(\frac{11}{9} N_c + 3 \right)
\Biggr] n_F
\nonumber \\ &&
+ Z_{Z, \mbox{top}}^{(2,2)}
\; \; ,
\nonumber \\ &&
\end{eqnarray}
where
\begin{eqnarray}
Z_{W, \mbox{top}}^{(2,1)} & = &
\Biggl[
- \frac{5}{2} \frac{m_t^6}{m_W^4 m_H^2}
- \frac{7}{32} \frac{m_t^4}{m_W^4}
- \frac{4}{9} \frac{m_t^4}{m_W^2 m_H^2}
+ \frac{4}{9} \frac{m_t^4 m_Z^2}{m_W^4 m_H^2}
- \frac{85}{288} \frac{m_t^2 m_Z^2}{m_W^4}
\nonumber \\ &&
+ \frac{19}{12} \frac{m_t^2 m_Z^4}{m_W^4 m_H^2}
+ \frac{3}{8} \frac{m_t^2 m_H^2}{m_W^4}
- \frac{20}{3} \frac{m_t^2 m_Z^2}{m_W^2 m_H^2}
- \frac{43}{144} \frac{m_t^2}{m_W^2}
+ \frac{35}{6} \frac{m_t^2}{m_H^2}
\Biggr] N_c
\\
Z_{W, \mbox{top}}^{(2,2)} & = &
\Biggl[
4 N_c \frac{m_t^8 }{m_W^4 m_H^2}
+ \left( \frac{3}{2} - N_c \right) \frac{m_t^6 }{m_W^4 m_H^2}
- \frac{39}{16}  \frac{m_t^4 }{m_W^4 }
+ \frac{1}{6} \frac{m_t^4 m_Z^2 }{m_W^4 m_H^2}
- 6 \frac{m_t^4 m_Z^4 }{m_W^4 m_H^4}
\nonumber \\ &&
+ \left(\frac{2}{3} (N_c +1 ) n_F - 10 \right)
  \frac{m_t^4}{m_W^2 m_H^2}
- \frac{1}{48}  \frac{m_t^2 m_Z^2 }{m_W^4 }
+ \frac{3}{2}  \frac{m_t^2 m_Z^4 }{m_W^4 m_H^2}
- 12  \frac{m_t^4 }{m_H^4 }
\nonumber \\ &&
+ 3 \frac{m_t^2 }{m_H^2 }
- \left( \frac{1}{6} n_F (N_c + 1 )- \frac{73}{24} \right)
  \frac{m_t^2}{m_W^2}
\Biggr] N_c
\\
Z_{Z, \mbox{top}}^{(2,1)} & = &
\Biggl[
- \frac{5}{2} \frac{m_t^6}{m_W^4 m_H^2}
- \frac{7}{32} \frac{m_t^4}{m_W^4}
- \frac{4}{9} \frac{m_t^4}{m_W^2 m_H^2}
+ \frac{4}{9} \frac{m_t^4 m_Z^2}{m_W^4 m_H^2}
- \frac{17}{32} \frac{m_t^2 m_Z^2}{m_W^4}
- \frac{13}{36} \frac{m_t^2}{m_Z^2}
\nonumber \\ &&
+ \frac{19}{12} \frac{m_t^2 m_Z^4}{m_W^4 m_H^2}
+ \frac{3}{8} \frac{m_t^2 m_H^2}{m_W^4}
- \frac{20}{3} \frac{m_t^2 m_Z^2}{m_W^2 m_H^2}
+ \frac{43}{144} \frac{m_t^2}{m_W^2}
+ \frac{35}{6} \frac{m_t^2}{m_H^2}
\Biggr] N_c
\\
Z_{Z, \mbox{top}}^{(2,2)} & = &
\Biggl[
  4 N_c                     \frac{m_t^8}{m_W^4 m_H^4}
+ \left(\frac{3}{2} - N_c  \right) \frac{m_t^6}{m_W^4 m_H^2}
- \frac{39}{16}               \frac{m_t^4}{m_W^4}
- 6                         \frac{m_t^4 m_Z^4}{m_W^4 m_H^4}
- 12                        \frac{m_t^4}{m_H^4}
\nonumber \\ &&
- \left(4 n_F + \frac{44}{27} n_F N_c - \frac{11}{3}  \right)
                            \frac{m_t^4}{m_W^2 m_H^2}
+ \left( 2 n_F + \frac{22}{27} n_F N_C + \frac{1}{2} \right)
                            \frac{m_t^4 m_Z^2}{m_W^4 m_H^2}
\nonumber \\ &&
+ \left( \frac{8}{3}n_F + \frac{40}{27} n_F N_c - 14 \right)
                            \frac{m_t^4}{m_Z^2 m_H^2}
- \left(\frac{1}{2} n_F + \frac{11}{54} n_F N_c + \frac{5}{48} \right)
                             \frac{m_t^2 m_Z^2}{m_W^4}
\nonumber \\ &&
+ \frac{3}{2}                  \frac{m_t^2 m_Z^4}{m_W^4 m_H^2}
+ 3                          \frac{m_t^2}{m_H^2}
+ \left( n_F + \frac{11}{27} n_F N_c - \frac{3}{8} \right)
                             \frac{m_t^2}{m_W^2}
\nonumber \\ &&
- \left( \frac{2}{3} n_F + \frac{10}{27} n_F N_c - \frac{7}{2} \right)
                               \frac{m_t^2}{m_Z^2}
\Biggr] N_c \; \;.
\label{Z_22_z_t}
\end{eqnarray}
The renormalization group equation may be utilized to verify the
higher order pole terms $1/\ep^2$.  It is given by Eq.~(4.16) of I
$$
\left( Z_V^{(1,1)} \right)^2 +
\frac{16 \pi^2 }{g^2}
2 \frac{\beta_g^{(1)} Z_V^{(1,1)}}{g}
+ \sum_i Z_{m_i}^{(1,1)}  m_i^2 \frac{\partial}{\partial m_i^2}
          Z_V^{(1,1)} = 2  Z_V^{(2,2)}\;.
$$
The value of
$\beta_g^{(1)} =\frac{g^3}{16 \pi^2}\;
\left( - \frac{43}{12} + \frac{1}{6} ( N_c+1) n_F \right)$
may be calculated from the relation
$$
\beta_g^{(1)} \sin\theta_W =
   \frac{1}{2} g \frac{ \cos^2 \theta_W}{\sin \theta_W}
    \left( \frac{g^2}{16 \pi^2 } \right)
    \left(Z_W^{(1,1)} - Z_Z^{(1,1)} \right) + \beta_e^{(1)} \;,
$$
where $\beta_e^{(1)}$ is given in (\ref{beta_e}), and $\cos^2 \theta_W
= m_W^2/m_Z^2$.  An additional relation which holds for the
$1/\ep$--terms is the following~(see Eq.~(4.19) in I):
$$
g^2 \frac{\beta_{g'}^{(2)}}{g'^3}
\sin^4\theta_W - \frac{\beta_g^{(2)}}{g} \sin^2\theta_W \cos^2\theta_W
= - \left( \frac{g^2}{16 \pi^2} \right)^2 \left( Z_W^{(2,1)} -
Z_Z^{(2,1)}\right) \cos^2\theta_W \;.
$$
The two-loop $\beta$-functions for $g$ and $g'$ are given in
\cite{RG_2loop} and read
\begin{eqnarray}
&& \hspace{-10mm}
\beta_{g'} \left. \right|_{N_c=3}  =
  \left(  \frac{1}{12} + \frac{10}{9} n_F \right) \frac{g'^3}{16 \pi^2}
+ \left( \frac{1}{4} + \frac{95}{54} n_F \right) \frac{g'^5}{(16 \pi^2)^2}
+ \left( \frac{3}{4} + \frac{1}{2} n_F \right) \frac{g'^3 g^2}{(16 \pi^2)^2}
\nonumber \\&& \hspace{10mm}
- \frac{17}{24}\frac{m_t^2}{m_W^2}\frac{g'^3 g^2}{(16 \pi^2)^2}
,
\nonumber \\&& \hspace{-10mm}
\beta_g \left. \right|_{N_c=3}  =
  \left( - \frac{43}{12} + \frac{2}{3} n_F \right) \frac{g^3}{16 \pi^2}
+ \left( - \frac{259}{12} + \frac{49}{6} n_F \right) \frac{g^5}{(16 \pi^2)^2}
+ \left( \frac{1}{4} + \frac{n_F}{6} \right) \frac{g^3 g'^2}{(16 \pi^2)^2}
\nonumber \\&& \hspace{10mm}
- \frac{3}{8}\frac{m_t^2}{m_W^2}\frac{g^5}{(16 \pi^2)^2} \;.
\nonumber
\end{eqnarray}

In all cases we could verify our results to satisfy the above RG
equations. We also have established that our two-loop RG equations
(\ref{Z_21_w})-(\ref{Z_22_z_t}), calculated in the broken phase, are
related to the ones found in the unbroken theory a long time
ago~\cite{Jones}\footnote{Only recently, these results
have been confirmed in~\cite{2-loop:potential} by an independent
calculation.}. On the one hand the relationship between the UV
counterterms and the RG equations in the broken and the unbroken phases
for massless coupling constants is a trivial consequence of the Higgs
mechanism (spontaneous symmetry breaking), which by definition must
preserve the Ward--Takhhashi-- and the Slavonv--Taylor--identities.  On
the other hand the Higgs field vacuum expectation value $v$ appears only
in the broken phase and its renormalization is looking very different
from the ones one has to perform in the symmetric phase.  In the broken
phase the \MSb parameter $v(\mu^2)$ may be defined by the ratio
$v^2(\mu^2)= 4 m_W^2(\mu^2)
\sin^2 \theta_W (\mu^2)/e^2(\mu^2)$ and the \MSb Fermi
constant\footnote{Note that our definition of the \MSb Fermi constant is
different from the one used in~\cite{deltar_MS}.} by $
\sqrt{2}G_F(\mu^2) = 1/v^2(\mu^2)
$.
The relationship with the unbroken phase follows if we define,
alternatively, $v^2(\mu^2)=m^2(\mu^2)/\lambda(\mu^2)$, where
$m^2$ and $\lambda$  are the \MSb parameters of the scalar potential in the
symmetric phase.
As a consequence the following relations hold between the RG functions of the
broken and the unbroken phase:
\begin{equation}
\gamma_{v^2}= \gamma_W - \frac{\cos^2 \theta}{\sin^2 \theta} \left(\gamma_W - \gamma_Z \right)
- 2 \frac{\beta_e}{e}
=\gamma_{m^2} - \frac{\beta_\lambda}{\lambda} \; .
\label{v2beta}
\end{equation}
Our results allow us to confirm the $\beta$--function for the
combination $\gamma_{m^2} - \beta_\lambda/\lambda$, not however, the
two-loop RG equations for $\lambda$ and $m^2$ independently. Given the
RG equation for $v^2$, we are able to write the two--loop RG equation
for the top--quark and Higgs masses using the two--loop
result~\cite{Jones} for the corresponding coupling,
i.e. $\gamma_{m_H^2} = \gamma_{m^2}$ (for details we refer
to~\cite{RG:2loop}).

These considerations also shed light on the role of the tadpole
contributions in parameter renormalizations. The zero momentum
transfer Higgs propagator which multiplies the tadpole loops implies
(in the perturbation expansion) non--analytic terms proportional to
$1/m_H^2 \sim 1/\lambda $.
In view of (\ref{v2beta}) and the mass coupling
relations, like $m_W=gv/2$ etc., it is obvious that such terms must
appear in mass counter--terms (see above and Sec.~4.2 of I) as well as
in the coefficients of the corresponding $\beta$--functions.

\subsection{QCD corrections}
At the two-loop level there are also diagrams exhibiting gluon exchange
which are contributing to the gauge-boson propagators.  The exact
analytical result for the $O(\alpha \alpha_s)$ contribution of
quark-pairs to the vacuum polarization function of the gauge-bosons have
been calculated in~\cite{qcd:1} (see also~\cite{qcd:2}).  The
corresponding set of 1PI diagrams yields a gauge independent and
infrared stable (with respect to taking the limit of vanishing fermion's
masses) contribution\footnote{It is interesting to note, that in
contrast to the two--loop electroweak massless fermion contributions,
where it suffices to expand all master-integrals up to the finite parts,
the master-integrals $J_{002}$ and $J_{022}$, showing up as additional
basic integrals when considering the QCD corrections to the $W$- and
$Z$-propagators, in this case must be expanded up to terms proportional
to $\ep^2$ and $\ep$, respectively. The situation here is similar the
one discussed in~\cite{FJTV}.}. For the sake of RG invariance, in
contrast to the usual conventions, we include tadpole contributions
which cancel in physical observables like $\Delta r$, $\Delta
\rho$ etc.~\cite{Veltman:screening} (see also~\cite{qcdtadpole}). For
each quark species, there is one $O(\alpha \alpha_s)$ tadpole diagram
which is gauge invariant. For a quark with mass $m_q$ its bare
contribution to the location of the pole of the gauge boson propagators
reads 
$$
\Delta_t = - \frac{g}{16 \pi^2 } \frac{g_s^2}{16 \pi^2}
16 N_c C_f \frac{1}{(d-4)^2} \frac{(d-1)}{(d-3)}
\frac{m_q^4}{m_W} (m_q^2)^{-2\ep} \;, $$
with $C_f = 4/3$ in the SM.  Taking into account the $O(\alpha_s)$ term
of the \MSb one--loop top--quark mass renormalization constant
$$ Z_{\rm top}^{\alpha_s} = 1 - 2 \frac{\alpha_s}{4 \pi} 3 C_f
\frac{1}{\ep} $$ we find
\begin{eqnarray}
Z_W^{\alpha_s} & = &
1 + \frac{g^2}{(16 \pi^2)} \frac{\alpha_s}{4 \pi} N_c C_f
\Biggl[
  \frac{1}{\ep} \left(
4 \frac{m_t^4}{m_H^2 m_W^2}  - \frac{5}{4} \frac{m_t^2}{m_H^2}
+ \frac{1}{2}n_F
                \right)
+ \frac{1}{\ep^2}
 \left( -12 \frac{m_t^4}{m_H^2 m_W^2}  + \frac{3}{2} \frac{m_t^2}{m_H^2 } \right)
\Biggr]
\nonumber \\
Z_Z^{\alpha_s} & = & 1 + \frac{g^2}{(16 \pi^2)} \frac{\alpha_s}{4 \pi} N_c C_f
\Biggl[
 \frac{1}{\ep} \left(
4 \frac{m_t^4}{m_H^2 m_W^2}  - \frac{5}{4} \frac{m_t^2}{m_H^2}
+ \frac{10}{9} n_F    \frac{m_W^2}{m_Z^2}
+ \frac{11}{18} n_F \frac{m_Z^2}{m_W^2}
- \frac{11}{9} n_F
                \right)
\nonumber \\ && \hspace{35mm}
+ \frac{1}{\ep^2}  \left( -12 \frac{m_t^4}{m_H^2 m_W^2}
+ \frac{3}{2} \frac{m_t^2}{m_H^2 } \right)
\Biggr] \; ,
\label{z:qcd}
\end{eqnarray}
where $\alpha_s = g_s^2/4
\pi $ and the first five quarks are treated as massless.  
The terms proportional to $m_t^4$ come from the tadpole contribution and
will cancel in observable quantities. Again, the renormalization group
equations can be used for cross-checking the $1/\ep^2$-- and
$1/\ep$--terms.  The $1/\ep$--part satisfies the following relation (see
Eq.~(4.19) of I): $$
\sin^2\theta_W  \Biggl[ \frac{\beta_{g'}^{\alpha_s}}{g'^3} \sin^2\theta_W
- \frac{\beta_g^{\alpha_s}}{g^3} \cos^2\theta_W
\Biggr]
= - \frac{1}{g^2}
     \left( Z_{W,\ep}^{\alpha_s} -  Z_{Z,\ep}^{\alpha_s}\right) \cos^2\theta_W \;,
$$
where \cite{RG_2loop}
\begin{eqnarray}
&& \hspace{-10mm}
\beta_{g'}^{\alpha_s} \left. \right|_{N_c=3}  =
\frac{22}{9} n_F \frac{g'^3 g_s^2}{(16 \pi^2)^2} \;,
\nonumber \\&& \hspace{-10mm}
\beta_g^{\alpha_s} \left. \right|_{N_c=3}  =
 2 n_F \frac{g^3 g_s^2}{(16 \pi^2)^2}
\nonumber
\end{eqnarray}
and the $Z_{V,\ep^j}^{\alpha_s}$ denote the $1/\ep^j$--parts of the
$Z_V^{\alpha_s}$ renormalization constants~(\ref{z:qcd}).  The terms
proportional to $1/\ep^2$ may be calculated from the relation
\begin{equation}
\frac{g^2}{16 \pi^2} \left( Z_{\rm top}^{\alpha_s} -1 \right )
m_t^2 \frac{\partial}{\partial m_t^2} Z_V^{(1,1)}
= 2  Z_{V,\ep^2}^{\alpha_s} \;,
\label{QCD}
\end{equation}
where the definition of $Z_V^{(i,j)}$ has been given
in~(\ref{baremsb}). Here, we have to take into account that, in the
presence of several dimensionless coupling constants, Eqs.~(4.14) and
(4.15) in I have to be modified accordingly:
\begin{eqnarray}
\gamma_V & = &  \frac{1}{2} \sum_j g_j \frac{\partial}{\partial g_j } Z_V^{(1)},
\\
\left(
\gamma_V +\sum_j \beta_{g_j} \frac{\partial}{\partial g_j }
+ \sum_i \gamma_i m_i^2 \frac{\partial}{\partial m_i^2} \right) Z_V^{(n)}
& = &  \frac{1}{2} \sum_j g_j \frac{\partial}{\partial g_j } Z_V^{(n+1)}.
\label{RG:SM+QCD}
\end{eqnarray}
Note that our Eq.~(\ref{QCD}) explicitly reveals, that
the systematic inclusion of the tadpole contributions is important
for the self--consistency of the renormalization group equations.
\subsection{\MSb renormalization of the propagator pole}

The calculation of the one-loop \MSb renormalized on--shell amplitude
$\hat{\Pi}_V^{(1)}$
is well known (see e.g. \cite{FJ}).
We get it by rewriting the bare expression in terms of \MSb parameters
\begin{eqnarray}
\hat{\Pi}_V^{(1)}
& \equiv &
\lim_{\ep \to 0} \left( m_{0,V}^2 -m^2_V
+  m_{0,V}^2 \frac{g_0^2}{16\pi^2} X_{0,V}^{(1)}\right)
\nonumber \\
& = &
m_V^2(\mu) \frac{e^2}{16\pi^2 \sin^2 \theta_W} \lim_{\ep \to 0}
 \left( \frac{1}{\ep} Z_V^{(1,1)} + X_{0,V}^{(1)}\right)
=  m_V^2(\mu) \frac{e^2}{16\pi^2 \sin^2 \theta_W} X_{V}^{(1)} \;.
\nonumber \\
\label{MS1:subtracted}
\end{eqnarray}
As explained in Sec.~5 of I, we may avoid the consideration of
wave-function renormalization as well as the renormalization of the
ghost sector and of the gauge parameters if we look directly at the full
two-loop \MSb renormalized on--shell amplitude\footnote{In contrast,
one could attempt to renormalize the individual amplitudes
$\Pi_{0,V}^{(2)}$ and $\Pi_{0,V}^{(1)}{}'$ in addition to the known
$\Pi_{0,V}^{(1)}$.}. The latter can be written in the form
\begin{eqnarray}
&&
\Biggl \{ \Pi_{0,V}^{(2)} + \Pi_{0,V}^{(1)} \Pi_{0,V}^{(1)}{}' \Biggr\}_{\overline{MS}}
=  \lim_{\ep \to 0}  \Biggl(
\Pi_{0,V}^{(2)} + \Pi_{0,V}^{(1)} \Pi_{0,V}^{(1)}{}'
\nonumber \\ &&
+ m_V^2(\mu) \frac{1}{\ep} \left( \frac{e^2}{16\pi^2 \sin^2 \theta_W} \right)^2
\Biggl[Z_V^{(1,1)}
+ \Biggl[ \frac{\Delta g^2}{g^2} \Biggr]
+ \sum_j Z_{m^2_j}^{(1,1)} \frac{\partial}{\partial m_j^2} \Biggr] X_{0,V}^{(1)}
\nonumber \\  &&
+ m_V^2(\mu) \left( \frac{e^2}{16\pi^2 \sin^2 \theta_W} \right)^2
\Biggl[ \frac{1}{\ep} Z_V^{(2,1)}  + \frac{1}{\ep^2} Z_V^{(2,2)}
\Biggr]
\Biggr) \; ,
\label{MS2:subtracted}
\end{eqnarray}
where the sum runs over all species of particles $j=Z,\,W,\,H, \ t$
and
$$
\Biggl[ \frac{\Delta g^2}{g^2} \Biggr] =
\frac{\cos^2 \theta_W}{\sin^2 \theta_W} \left(Z_W^{(1,1)} - Z_Z^{(1,1)}
\right) - \left(7-\frac43 n_F\: \left[\frac59 N_c+1 \right] \right) \sin^2 \theta_W \;.  $$
The functions $Z_V^{(i,j)}, X_{V}^{(1)}$ and $ X_{0,V}^{(1)}$ are
defined in (\ref{one-loop_W}-\ref{one-loop}),
(\ref{Z_21_w}-\ref{Z_22_z}), (\ref{X1W}-\ref{X1Z})
and Appendix C of I, respectively. The derivatives of $X_{0,V}^{(1)}$
with respect to the masses may be found in Appendix~\ref{Mrtwo}.

It is interesting to observe that (\ref{MS1:subtracted}) and
(\ref{MS2:subtracted}) account for the renormalization of the real and
the imaginary parts, i.e., for the mass and the width, simultaneously
in a unified form.

\section{Results and discussions}
\setcounter{equation}{0}

We have calculated the location of the pole in terms of the \MSb--mass for the
massive gauge-bosons $Z$ and $W$. As in I we will write the result in
the form
\begin{equation}
\frac{s_P}{m_V^2} =  1
  + \left( \frac{e^2}{16\pi^2\sin^2 \theta_W} \right)\: X^{(1)}_V
  + \left( \frac{e^2}{16\pi^2\sin^2\theta_W} \right)^2 \: X^{(2)}_V \,,
\label{result}
\end{equation}
where both $e$ and $\sinw$ are to be taken in the
$\overline{\rm MS}$ scheme.
%
The one-loop coefficients $X^{(1)}_V $ for $Z,\,W$ and $H$ are known
of course as exact results. We wrote them down for completeness in
Appendix~B. The light (massless) fermion contributions to the
coefficients $X^{(2)}_V$ are given in Appendix~\ref{Fctwo} in exact
analytical from in terms of the master--integrals discussed in
Sec.~\ref{Mfc}. Alternatively, these contributions may be represented
in a $\sinW$--expanded form\footnote{Details concerning this expansion
can be found in~\cite{RG:2loop}.} in the same way as the other
contributions.  In the latter case, after summing all diagrams with
massless fermion loops, all singularities of type $\ln^j \sinW$ cancel
which infers the infrared finiteness of the massless fermion
contribution to the pole--mass. When the top quark is involved we
perform an expansion and present the coefficients as series with
respect to three small parameters. One ``small'' parameter is the weak
mixing parameter $\sinW$, the others are the mass ratios $m_V^2/m_H^2$
and $m_V^2/m_t^2$ where $V=W {\rm \ or \ } Z$. The analytical values
of the first three coefficients are presented in Appendix~E.

Again we have checked, that the relations between the pole-- and the
\MSb--masses, when calculated in the so-called modified $\overline{\rm
MS}$ scheme ($\overline{\rm MMS}$), coincide with the ones obtained in
the ordinary \MSb scheme.

We have developed and implemented into a computer program an efficient
algorithm for the calculation of the massive Feynman diagrams with
external momentum on the mass shell of one of the internal particles.
The necessary analytical continuations which allow us to cover the
hole region of parameter space has also been constructed and
implemented numerically.

After UV renormalization the pole--mass is a finite expression in the
limit $\ep \to 0$. Since also the IR singularities have been
regularized by dimensional regularization, this result implies the
infrared finiteness of the SM contributions to the pole--mass.  As a
consequence, the pole--masses of the gauge--bosons are infrared finite
quantities. Note that, in order to establish the gauge invariance of
the location of the pole $s_p$, the tadpole contributions have to be
taken into account.  Also RG-invariance requires the inclusion of the
tadpoles.

Our calculation proves that the \MSb renormalization scheme, comprised
in~(\ref{MS2:subtracted}), is self consistent and works properly in
case of unstable particles.

In contrast to the mass of a stable particle, the definition of a mass
of an unstable particle is not unambiguous. The natural definition by
$\sqrt{{\rm Re} s_P}$ (see~\ref{pole},\ref{def}), which we
adopted in this paper, is not always used. In particular LEP/SLC
experiments have adopted another definition, which has been used by
most of the theoretical papers on the $Z$ line--shape before LEP/SLC
started to take data (see~\cite{LEPrev} for a pre--LEP status report).
As a consequence the $Z$-- and $W$--masses determined by LEP/SLC
experiments and listed in the particle data tables~\cite{Hagiwara:pw}
correspond to $M^{'2}_V=M_V^2+\Gamma_V^2$ rather than to the
pole--mass $M_V$\footnote{Note that $M'_Z-M_Z \simeq 35 \mv$ and thus
in order to obtain the pole--mass from the ``experimental'' $Z$--mass
one has to subtract about 35 MeV.}. In this case the width $\Gamma'_Z$
entering the $Z$ line--shape must be defined by ${\rm Im} \Pi_Z(s)
\simeq s
\Gamma_Z/M_Z$ and hence is a function of the c.m. energy square
$s$. While at one--loop order the width is determined solely by the
gauge invariant fermion contributions at higher orders a
$s$--dependent widths, which is not an on--shell quantity, will cause
troubles with gauge invariance. However, one still may understand the
LEP measurements to be defined from the quantities defined in
terms of the gauge invariant position of the pole $s_P$ by the
gauge invariant relationship $M^{'2}_V=M_V^2+\Gamma_V^2$. In the
latter case the mass may be considered to be defined via
$M'_V/M_V = \left| s_P \right|/{\rm Re} s_P$.

Our results for the two--loop mass renormalization constants in the
on-shell and the \MSb scheme can be applied for the calculation of
physical quantities in both of these schemes at the two-loop
level. Examples of such calculations, where results of our paper I
have been used, are the computations of the bosonic two--loop
contributions to the muon life--time (usually encoded in the
correction $\Delta r$~\cite{Sirlin}) presented in~\cite{r-boson}.


The numerical evaluation of our results is of interest in several
respects.  First of all, even though the calculated quantities are not
observables, but only a subset of contributions to those, their
numerical magnitude is interesting. It certainly sheds light on the
size of possible corrections as well as on the convergence of the
perturbation expansion. Last but not least, the numerical evaluation
by itself is a challenge: the amplitudes are of remarkable complexity
and the numerical stability and the efficiency of the evaluations are
far from trivial. Although we are largely working with series
expansions already, and not with numerical integration of the master
integrals, we have to apply numerical multiple precision techniques
to be sure that we get the correct answers. To this end numerical
evaluations at the present stage have been performed in MAPLE were we
can work at the desired precision only on the expense of CPU time.

Since we are calculating the on--shell counter terms for the gauge
bosons, it is natural to consider the on--shell scheme with the fine
structure constant $\alpha$ and the masses as input parameters. For
our numerical calculation we have chosen the following input
parameters: $\alpha = 1/137.036$, $M_W=80.419$ GeV, $M_Z=91.188$ GeV.
Neutrinos, leptons and quarks are taken in the massless approximation
and the top--quark mass is taken to be $m_t=174.3 \gv$. For the strong
coupling constant we are using a fixed value
$\alpha_s=\alpha_s(M_Z)=0.1185$. Often the two--loop corrections are
of comparable size to the mass effects of the ``light'' fermions at
one-loop. We then take the following fermion masses for the numerical
evaluation:\\[4mm]
\centerline{%
\begin{tabular}{lcl|lcl|lcl}
\hline
\hline
\multicolumn{9}{c}{Fermion masses in GeV}\\
\hline
$m_e$    &=&0.00051   &     $m_u$&=& 0.003   & $m_d$&=&0.006  \\
$m_\mu$  &=&0.1056583 &     $m_c$&=& 1.2     & $m_s$&=&0.120  \\
$m_\tau$ &=&1.77703   &	    $m_t$&=& 174.3   & $m_b$&=&4.4   \\
\hline
\end{tabular}
}

\vspace*{4mm}

The general features of our expansion have been discussed in I.  While
the series expansion in $\sinW$ for the known fixed value converges
very well, the expansions in $M_V^2/m_H^2$ for large $m_H$ leads to a
strong coupling problem and perturbation theory becomes useless above
about 800 GeV. In general, as is illustrated by Figs.~\ref{f:sex}
and \ref{f:zex}, the first three coefficients yield a good
approximation as long as the asymptotic series behaves well. Of
course, the expansion in $M_V^2/m_H^2$ for smaller values of $m_H$
starts to diverge below about 120 GeV. We have calculated
six terms in each of the different expansion parameters, but only
write down three of them in Appendix~E~\footnote{The full set
of coefficients may be found at {\tt
http://www-zeuthen.desy.de/$\;\widetilde{}\;$kalmykov/pole/pole.html}}.
Figs.~\ref{f:wl}--\ref{f:zh} show the corrections $\Delta_W \equiv
M^2_W/m^2_W(M_W)-1$ and $\Delta_Z \equiv M^2_Z/m^2_Z(M_Z)-1$,
respectively, as a function of the Higgs mass $M_H$ for intermediate
and for heavy Higgs masses. The light fermion contributions are small
relative to the other corrections. The same is true for the QCD
corrections for Higgs masses above about 300 GeV. In Fig.~\ref{f:im} the
imaginary part  and the absolute value of the pole position $s_P$ are
shown.

Very often the inverse of (\ref{result}) is required.  To that end we
have to solve the real part of (\ref{polemass}) iteratively for $m_V^2$
and to express all $\overline{\rm MS}$ parameters in terms of on-shell
ones.

The solution to two loops reads
\begin{eqnarray}
\label{inverse}
m^2_V &=& M_V^2
- {\rm Re} \hat{\Pi}_V^{(1)}
- {\rm Re} \Biggl \{ \Pi_V^{(2)} + \Pi_V^{(1)} \Pi_V^{(1)}{}' \Biggr\}_{\msb}
\nonumber\\
&&
- \sum\limits_j (\Delta m^2_j)^{(1)} \frac{\partial}{\partial m_j^2}  {\rm Re} \hat{\Pi}_V^{(1)}
       - (\Delta e)^{(1)} \frac{\partial}{\partial e} {\rm Re} \hat{\Pi}_V^{(1)}
      \Biggr|_{m_j^2=M_j^2,\, e=e_{\rm OS}},
\label{reverse}
\end{eqnarray}
where the sum runs over all species of particles $j=Z,\,W,\,H, \, t$
and
$$
(\Delta m^2_j)^{(1)}=
-{\rm Re} \hat{\Pi}^{(1)}_j
\Biggr|_{m_j^2=M_j^2,\, e=e_{\rm OS}}
\equiv
- M_V^2 \frac{e^2_{\rm OS}}{16 \pi^2 \sin^2 \theta_W} X_V^{(1)} \Biggr|_{m_j^2=M_j^2}
$$ stands for the
self-energy of the $j$th particle at $p^2=m_j^2$ in the \MSb scheme
and parameters replaced by the on-shell ones.
The analytical results for the derivative term
$\sum\limits_j \delta_j \frac{\partial}{\partial m_j^2}  \hat{\Pi}_V^{(1)}$
can be extracted from the results of Appendix~C.
In (\ref{reverse}) $m^2_V- M_V^2=\left(\delta M_V^2\right)_{\rm finite}$ represents the
finite part of the on--shell mass counter--term. The corresponding
bare one is obtained by adding the \MSb mass counter--term (see section 4.2)
$\delta m_V^2= \left( Z_{\overline{MS}} - 1 \right) m_V^2
= m_{0V}^2-m_V^2$.

The relation (\ref{reverse}) involves a change from the $\overline{\rm
MS}$ to the on-shell (OS) scheme also for the electric charge.
Let us consider therefore, in details,
the relationship between the fine structure constants $\alms$
and $\alos=\alpha$.

For the electroweak couplings we have to calculate the \MSb versions
from their commonly used on-shell values\footnote{The QCD coupling is
parameterized almost always in the \MSs. Therefore $\al_s(\mu^2)$
usually is directly determined experimentally by fitting data to a
suitable perturbative QCD prediction in terms of $\al_s(\mu^2)$.}.
The \MSb version of the fine structure constant is defined as a
solution of the renormalization group equation (see \ref{beta_e}) and
can be calculated from the UV-counterterms of the electrical charge
$e$ (see (\ref{e0to0})).  In perturbation theory the naive relation
between \MSb and on-shell values of $\alpha$ is determined by the
Thomson limit of Compton scattering and can be written as
\begin{eqnarray}
\alpha(\mu^2) & = & \alpha \Biggl\{ 1 +  \frac{\alpha}{4 \pi}
\Biggl[ 7  \ln \left( \frac{m_W^2}{\mu^2} \right)-\frac{2}{3}
+ \frac{4}{3} \sum Q_f^2 N_{cf} \ln \frac{\mu^2}{m_f^2}
\Biggr]
\Biggr\} \;,
\label{alpha_ms}
\end{eqnarray}
where the sum goes over all fermions $f$ with $N_{cf}=1$ for leptons
and $N_{cf}=3$ for quarks and
$Q_f$ is the fermion charge equal to 1 for leptons, -1/3 and 2/3 for
d- and u-quarks, correspondingly.

The terms proportional to logarithms of the fermion masses
originate from evaluating the finite part of the derivative of
the photon propagator at zero momentum transfer, which we denote
as $\Pi^f_{\gamma \gamma}(0)$\footnote{At zero momentum fermions contribute
to the charge only via $\Pi_{\gamma \gamma}(0)$.}.
However, the low energy contribution of the five light quarks
$(u,d,s,c,b)$ cannot be calculated in perturbation theory.  The free
quark loops are strongly (non-perturbatively) modified by the strong
interaction at low energies. In order to evaluate
$\Pi^f_{\gamma \gamma}(0)$ we may write it as
\begin{equation}
\Pi^f_{\gamma \gamma}(0) = {\rm Re} \Pi^f_{\gamma \gamma}(q^2) -
\Biggl[{\rm Re} \Pi^f_{\gamma \gamma}(q^2)   - \Pi^f_{\gamma \gamma}(0)
\Biggr] \;,
\label{photon}
\end{equation}
where $q^2$ is chosen sufficiently large (typically $M_Z^2$)
such that the first term on the r.h.s. can be calculated in
perturbative QCD.
In the limit of large momentum transfer $s \gg m_f^2$
the result is
\begin{equation}
{\rm Re}\Pi^f_{\gamma \gamma}(s) =
\frac{4}{3}\sum_f Q_{f}^2 N_{cf} \left( \ln \frac{s} {\mu^2}-\frac{5}{3} \right) \;,
\label{photonpert}
\end{equation}
which we use in the relation (\ref{alpha_ms}) for the perturbative
light quarks contributions $(s=M_Z^2)$.

On the grounds of analyticity and unitarity, the
second non--perturbative term in (\ref{photon}) can be determined by
evaluating a dispersion integral over the known experimental
$e^+ e^- \to {\rm hadrons}$ cross--section (for details see
\cite{hadrons:1}).  Usually, the cross--sections are represented in
terms of the cross--section ratio
$$
R(s)=\frac{\sigma_{tot} (\epm \to \gamma^* \to {\rm hadrons})}
     {\sigma (\epm \to \gamma^* \to \mu^+ \mu^-)}\;,
$$
where $\sigma (\epm \to \gamma^* \to \mu^+ \mu^-)=\frac{4\pi
\alpha ^2}{3s}$ at tree level. In terms of $R(s)$
we obtain
$$ \Delta \Pi_{\gamma \gamma}(q^2)\equiv
\Biggl[{\rm Re} \Pi_{\gamma \gamma}(q^2)   - \Pi_{\gamma \gamma}(0)
\Biggr] = -\frac{4q^2}{3}
\int_{4 m_\pi^2}^\infty ds \frac{R(s)}{s\:(s - q^2 - i \epsilon)}.
$$
The shift in the fine structure constant in then
$\Delta \alpha(s)= 4\pi \alpha \times \frac{1}{16\pi^2} \Delta
\Pi_{\gamma \gamma}(s)$. Using the experimental
data for $R(s)$ up to $\sqrt{s}=E_{cut}=5$ GeV and for the $\Upsilon$
resonances region between 9.6 and 13 GeV and perturbative QCD from 5.0
to 9.6 GeV and for the high energy tail above 13 GeV one gets
\ba
\Delta \alpha _{\rm hadrons}^{(5)}(M^2_Z) &=& 0.027572 \pm 0.000359\;\;;\;\;\;
\alpha^{-1}(M_Z^2)=128.952 \pm 0.049
\ea
at $M_Z=$ 91.19 GeV. For numerical estimations we use the results of
\cite{hadrons:2}.

The shift in $\alpha$ calculated by the
dispersion relation corresponds to the on-shell scheme.
Accordingly, since
$\hat{\Pi}^{(1)}$ depends on $e$ by an overall factor $e^2$ only,
we have

$$
(\Delta e)^{(1)} \frac{\partial}{\partial e}\hat{\Pi}_V^{(1)}
=
\left\{
  \delta \alpha_{\rm bos}
+ \delta \alpha_{\rm lep}
+ \delta \alpha_{\rm top}
+ \Delta \alpha _{\rm hadrons}^{(5)}(M^2_Z) -\delta \Delta \alpha _{\rm udscb}(M^2_Z)\right\} \hat{\Pi}_V^{(1)}\;,
$$
where
$$ \delta \alpha_{\rm bos} = \frac{\alpha}{4\pi} \left( 7\ln\frac{M_W^2}{\mu^2}-\frac{2}{3} \right),\;\;
 \delta \alpha_{\rm lep} = - \frac{\alpha}{3\pi}
\sum\limits_{\ell=e,\mu,\tau}\ln \frac{m^2_\ell}{\mu^2},\;\;
 \delta \alpha_{\rm top} =  - \frac{4 \alpha}{9 \pi} \ln \frac{m^2_t}{\mu^2}$$
and  $$\delta \Delta \alpha _{\rm udscb}(M^2_Z)=
\frac{11\alpha}{9\pi} \left(\ln\frac{M_Z^2}{\mu^2}-\frac{5}{3} \right)$$
the perturbative subtraction-term (\ref{photonpert}) form the 5 light quarks.
The quark loops contributing to the on--shell gauge boson
self--energies are evaluated here at a high energy scale and hence it
should be save to calculate them in perturbation
theory\footnote{Possibilities to evaluate them by non--perturbative
methods via dispersion relations have been discussed
in~\cite{hadrons:1}.}.

Finally, we analyze the Higgs mass dependence of the \MSb parameter
$\sin^2\theta$ defined as
\begin{equation}
\label{sinus}
\sin^2 \theta_W =  1 - \frac{m_W^2}{m_Z^2} = \frac{g'^2}{g^2+g'^2}.
\end{equation}
The Fig.~\ref{sinofmh} shows the correction to
$\delta_{\sin^2\Theta}=
\sin^2\theta^{\msb}_W/\sin^2\theta^{\rm OS}_W-1$ as a function of the
Higgs mass. Since the unphysical $m_H^4$ terms drop out by virtue of
Veltman's screening theorem, the corrections do not blow up so
dramatically in the strong coupling regime of large Higgs masses. In the
region displayed between 150 and 800 GeV the two--loop correction stays
below about 0.6\% and has a sign opposite to the one--loop
correction. The latter is one order of magnitude larger.

\section{Conclusion}
We have calculated the full SM two--loop radiative corrections to the
pole masses of the gauge bosons $W$ and $Z$.
A number of conceptual problems dealing with the renormalization of
unstable particles could be studied by explicite calculations: {\bf
i)} The position of the complex pole $s_P$ of the gauge boson
propagators is manifestly gauge invariant after taking into account
the Higgs tadpole contributions.  {\bf ii)} The renormalized on-shell
self--energies are infrared finite.  {\bf iii)} Our calculation proves
that the \MSb renormalization scheme, comprised
in~(\ref{MS2:subtracted}), is self consistent and works properly in
case of unstable particles.  {\bf iv)} Up to two--loops we explicitely
confirm that the UV singularities and the related RG equations
of the broken phase are completely determined by the unbroken
phase~\cite{RG:2loop}.  {\bf v)} The inclusion of the tadpoles is also
required from the point of view of the renormalization group
invariance.  {\bf vi)} Our results for the 2-loop mass renormalization
constants in the on-shell and the \MSb scheme can be applied in
calculations of physical quantities in both of these schemes at the
2-loop level.  {\bf vi)} A compact expression for the massless fermion
contributions in terms of an irreducible set of master--integrals in
given. {\bf vii)} A new technique for calculating the $\ep$--expansion
of some types of hypergeometric functions has been developed.

Note that all results concerning general properties (like gauge
invariance etc.) have been checked analytically. Since exact
analytical results for many of the multi--scale master--integrals are
not yet available, in this work we had to resort to asymptotic
expansion techniques. The exact results in terms of a basis of
master--integrals will by given elsewhere.  For the latter one
dimensional integral representations are available which are suitable
for the numerical evaluations \cite{numeric}.  However, for numerical
evaluations in the heavy Higgs region above about 200 GeV, the
computation in terms of our series expansions is numerically much
more stable and much more efficient.

\vspace{1.0cm}
\noindent
{\bf Acknowledgments.}  We are grateful to D.~Bardin, A.~Davydychev,
O.~V.~Tarasov and B.~Tausk for useful discussions.  M.~K.'s research
was supported in part by INTAS-CERN grant No.~99-0377 and by the
Australian Research Council grant No.~A00000780. We thank Oleg Tarasov
for carefully reading the manuscript.

\section*{Appendix}
\appendix
\section{Sums with binomial coefficients in the numerator}
\setcounter{equation}{0}
Let us present in the following some useful formulae for the binomial
sums which occur in the master-integrals~(\ref{f20100:series})
and~(\ref{f20110:series}).  The sums which involve only harmonic
coefficients, were investigated in~\cite{kk}, while the sums with
binomial coefficients in the denominator\footnote{In~\cite{KV00} sums
of this type were called {\it multiple binomial sums}. Here we will
use the more adequate name {\it inverse multiple binomial sums}.} were
considered in~\cite{FKV99,KV00}. In this Appendix we present some
results for sums with binomial coefficients in the numerator. A sum
$F_a$ of weight $a$ has form

\begin{eqnarray}
F_a(x) \equiv \sum_{k=1}^\infty \left( 2k \atop k\right) f(k) \frac{x^k}{k^a}
\label{binomial}
\end{eqnarray}
where $f(k)$ is a product of some finite sums (e.g. $S_a, \bar{S}_b$,
etc.).  It is convenient to introduce a new variable $\chi$ related to
$x$ via
\begin{eqnarray}
\chi & = & \frac{1-\sqrt{1-4x}}{1+\sqrt{1-4x}},
~~~~x = \frac{\chi}{(1+\chi)^2} ,
~~~~\sqrt{1-4x} = \frac{1-\chi}{1+\chi} .
\label{chi:definition}
\end{eqnarray}
A general relation for sums of type~(\ref{binomial}) is then given by
\begin{eqnarray}
F_{a-1}(x) =
x \frac{d}{dx}  F_a(x) =
\frac{1+\chi}{1-\chi} \chi \frac{d}{d \chi}  \tilde{F}_a(\chi) .
\label{diff}
\end{eqnarray}
Relevant analytical results for sums of the type considered are:
\begin{eqnarray}
&&
\label{sum000}
\sum\limits_{k=1}^\infty \left( 2k \atop k\right)  \frac{x^k}{k^4}  =
    - 8 \Snp{1,2}{-\chi} \ln (1+\chi)
    + 4 \Li{3}{-\chi} \ln (1+\chi)
    - 4 \Li{2}{-\chi} \ln^2 (1+\chi)
\nonumber \\ && \hspace{3.0cm}
    - \frac{2}{3} \ln^4 (1+\chi)
    - \Li{4}{-\chi}
    + 4 \Snp{2,2}{-\chi}
    - 8 \Snp{1,3}{-\chi} ,
\\ &&
\sum_{k=1}^\infty \left( 2k \atop k\right) \frac{x^k}{k^2} S_1 =
- \frac{4}{3} \ln^3 (1+\chi) +  \frac{1}{2} \Li{3}{\chi^2}
- 2 \ln (1+\chi) \Li{2}{\chi^2}
\nonumber \\ && \hspace{3.0cm}
+ 4 \Snp{1,2}{\chi} -2 \Snp{1,2}{\chi^2} ,
\\&&
\sum_{k=1}^\infty \left( 2k \atop k\right) \frac{x^k}{k^2} \bar{S}_1 =
- \frac{2}{3} \ln^3 (1+\chi)
+ 2 \Li{3}{\chi}
- 4 \ln (1+\chi) \Li{2}{\chi}
\nonumber \\&& \hspace{3cm}
+ 4 \Snp{1,2}{\chi} + 6 \Snp{1,2}{-\chi} - 2 \Snp{1,2}{\chi^2} ,
\\ &&
\sum_{k=1}^\infty \left( 2k \atop k\right) \frac{x^k}{k} S_2
=
-\frac{4}{3} \ln^3 (1+\chi)
- 4 \ln (1+\chi) \Li{2}{-\chi}
- 4 \Snp{1,2}{-\chi} ,
\\ &&
\label{sum12000}
\sum_{k=1}^\infty \left( 2k \atop k\right) \frac{x^k}{k} S_1^2
=
\frac{4}{3} \ln^3 (1+\chi)
+ 8 \ln (1+\chi) \Li{2}{\chi}
+ 4 \ln (1+\chi) \Li{2}{-\chi}
\nonumber \\&& \hspace{3cm}
- 4 \Snp{1,2}{-\chi} + 4 \Snp{1,2}{\chi^2} ,
\\ &&
\label{sum11100}
\sum_{k=1}^\infty \left( 2k \atop k\right) \frac{x^k}{k} S_1  \bar{S}_1
=
\frac{2}{3} \ln^3 (1+\chi)
+ 6 \ln (1+\chi) \Li{2}{\chi}
- 6 \Snp{1,2}{-\chi}
\nonumber \\ && \hspace{3.0cm}
+ 4 \Snp{1,2}{ \chi}
+ 2 \Snp{1,2}{\chi^2} ,
\\ &&
\label{sum-combination}
\sum_{k=1}^\infty \left( 2k \atop k\right)  \frac{x^k}{k}
\Biggl[
\bar{S}_2 - \bar{S}_1^2
\Biggr]
=
- \frac{2}{3} \ln^3 (1+\chi)
-4 \ln (1+\chi) \Li{2}{\chi}
- 8 \Snp{1,2}{ \chi} \; ,
\\ &&
\label{V21}
\sum_{k=1}^\infty \left( 2k \atop k\right) \frac{x^k}{k} V_2  =
- \Li{3}{\chi}
- 2 \Li{3}{-\chi}
+ \frac{1}{9} \Li{3}{\chi^3} - \Li{3}{x} .
\end{eqnarray}
All these low-weights sums may be obtained by applying~(\ref{diff}).

\noindent
The following  representation is often useful for sums of the type
just considered:

\begin{eqnarray}
\sum_{n=1}^\infty \left( 2n \atop n \right)  \frac{x^n}{n^a} f(n) =
\frac{1}{(a-2)!} \int_0^x
\sum_{n=1}^\infty \left( 2n \atop n\right) \frac{z^n}{n} f(n)
\Biggl[\ln x - \ln z \Biggr]^{a-2} \frac{dz}{z} \;,
\label{aux-representation}
\end{eqnarray}
where $f(n)$ is an arbitrary combination of finite sums.
For particular cases, $f(n) = 1, S_1, \bar{S}_1,$ $S_2, S_1 \bar{S}_1,
\bar{S}_2 - \bar{S}_1^2$ or $V_2$,
the corresponding expressions extracted from~(\ref{sum000})-(\ref{V21})
allow us to perform the sum under the integral
in~(\ref{aux-representation}) exactly and to obtain one-fold integral
representations for a large class of binomial sums.  For $0 \le x \le
1/4$ , a further simplification is possible by substituting, $z=
\frac{1}{4} \sin^2 \phi$. It leads to the following representation:

\begin{eqnarray}
\sum_{n=1}^\infty \left( 2n \atop n \right)  \frac{x^n}{n^a} f(n) & = &
\frac{2}{(a-2)!} \int_0^{\arcsin \sqrt{4x}}
d \phi \frac{\cos \phi}{\sin \phi}
\Biggl[\ln (4x)  -\ln ( \sin^2 \phi )\Biggr]^{a-2}
\times
\nonumber \\ &&
\sum_{n=1}^\infty \left( 2n \atop n\right) \frac{f(n)}{n}
\left( \frac{\sin^2 \phi}{4}\right)^n
\nonumber \\
& \equiv & \Sigma (f,\arcsin \sqrt{4x}) ~~,
\label{small}
\end{eqnarray}
where $\Sigma (f,\theta)$ is the short notation for this sum.
Unfortunately, we could not find corresponding analytical results for
these sums for arbitrary $x\; (x\le 1/4)$ in terms of known functions,
like generalized Nielsen polylogarithms. Not even for the simplest
case $f(n) = 1$, which yields

\begin{eqnarray}
\sum_{n=1}^\infty \left( 2n \atop n \right)  \frac{x^n}{n^a}  & = &
- \frac{2^{a+1}}{(a-2)!} \int_0^{\arcsin \sqrt{4x}}
\ln \left( \cos \frac{\theta}{2} \right)
\Biggl[\ln (2 \sqrt{x})
- \ln \left( \sin \theta \right)
\Biggr]^{a-2}
d \Biggl(  \ln \sin \theta \Biggr)
\; ,
\nonumber
\end{eqnarray}
a solution could be found. The problem here may be
compared with the one encountered in the context of the inverse
binomial sums, considered in~\cite{KV00,BBK}.

For $x > 1/4$ the representation
\begin{eqnarray}
\sum_{n=1}^\infty \left( 2n \atop n \right)  \frac{x^n}{n^a} f(n)
& = &
\Sigma \left ( f, \frac{\pi}{2} \right)
+
\frac{1}{(a-2)!} \int_{\frac{1}{4}}^x
\sum_{n=1}^\infty \left( 2n \atop n\right) \frac{z^k}{n} f(n)
\Biggl[\ln x - \ln z \Biggr]^{a-2} \frac{dz}{z}
\nonumber \\
& = &
\Sigma \left ( f, \frac{\pi}{2} \right)
+
\frac{2^{a-2}}{(a-2)!} \int_0^{2 \arccos \frac{1}{\sqrt{4x}}}
d \theta \frac{\sin \frac{\theta}{2}}{\cos \frac{\theta}{2}}
\Biggl[\ln \sqrt{x} + \ln \left( 2 \cos \frac{\theta}{2} \right) \Biggr]^{a-2}
\times
\nonumber \\ && \hspace{3cm}
\sum_{n=1}^\infty \left( 2n \atop n\right) \frac{f(n)}{n}
\left( \frac{1}{4 \cos^2 \frac{\theta}{2}} \right)^n
\label{large}
\end{eqnarray}
is available, where we have introduced a new variable $z=1/(4 \cos^2
\frac{\theta}{2})$ and $\Sigma \left ( f, \frac{\pi}{2}
\right)$ is a constant, related to some (combination of) hypergeometric
functions ${}_P F_Q$ of argument unity.  For each given choice of
$f(n)$ these constants are different, but in all cases they are
expressible in terms of ``even basis'' elements
(see~\cite{euler-basis} and Appendix~B.1 in~\cite{DK2001}).  We have
checked this statement by high-precision calculations, which allow for
``numerical proofs'' in the manner explained in~\cite{KV00}.  The second
term of (\ref{large}) can be calculated analytically for some
particular cases. For example, for $f(n) = 1$ we have
\begin{eqnarray}
\sum_{n=1}^\infty \left( 2n \atop n \right)  \frac{x^n}{n^a}
& = &
\frac{1}{2}~ {}_{a+2}F_{a+1}\left(\begin{array}{c|}
\tfrac{3}{2},  \{ 1 \}_{a+1} \\
\{ 2 \}_{a+1}
\end{array} ~1 \right)
-
\frac{2^a}{(a-2)!} \sum_{j=0}^{a-2} \left( a-2 \atop j\right)
\left(\ln \sqrt{x} \right)^{a-2-j} \times
\nonumber \\ &&
\Biggl\{
\frac{1}{j+2} \Biggl[\ln^{j+2} \left( 2 \cos \frac{\theta}{2} \right)  - \ln^{j+2} 2 \Biggr]
\nonumber \\ &&
+ \frac{\mbox{i} \sigma}{2(j+1)}
\Biggl[\Ls{j+2}{\pi-\theta} -\Ls{j+2}{\pi} - \theta \ln^{j+1} \left( 2 \cos \frac{\theta}{2} \right)
\Biggr]
\Biggr\}\; ,
\nonumber \\ &&
\label{f=1}
\end{eqnarray}
where we have used $\sum\limits_{k=1}^\infty \left( 2k \atop k\right)
\frac{x^k}{k} = 2\ln(1+\chi) $, $a \geq 2$, $\theta = 2 \arccos
\frac{1}{\sqrt{4x}}\, ,$ $\sigma = \pm 1$ and $\Ls{n}{\theta}$ is
so-called log-sine integral \cite{Lewin} defined by
\begin{equation}
\LS{j}{k}{\theta} =   - \int\limits_0^\theta {\rm d}\phi \;
   \phi^k \ln^{j-k-1} \left| 2\sin\frac{\phi}{2}\right| \, ,
~~~
\Ls{j}{\theta} \equiv \LS{j}{0}{\theta} \;.
\label{log-sin}
\end{equation}
The values of $\Ls{j}{\pi}$ can be expressed in terms of
$\zeta$--function for any $j$~\cite{Lewin}.  The sign $\sigma $ is
defined via the relation $\chi \equiv e^{- {\rm i } \sigma
\theta}$. The choice of the sign derives from the causal ``+{\rm
i}0''--prescription for the propagator or as the sign of the square
root of $\cos^2 \frac{\theta}{2}$. In the physical region of interest
here we have $\sigma = -1$.

For $x > 1/4$, an non--zero imaginary part develops for the sum
(\ref{f=1}).  It can be calculated in closed form:
\begin{eqnarray}
{\it Im} \sum_{n=1}^\infty \left( 2n \atop n \right)  \frac{x^n}{n^a}
& = &
- \sigma \frac{2^{a-1}}{(a-2)!} \sum_{j=0}^{a-2} \left( a-2 \atop j\right)
\frac{\left(\ln \sqrt{x} \right)^{a-2-j} }{j+1} \times
\nonumber \\ &&
\Biggl[\Ls{j+2}{\pi-\theta} -\Ls{j+2}{\pi} - \theta \ln^{j+1} \left( 2 \cos \frac{\theta}{2} \right)
\Biggr]_{\theta = 2 \arccos \frac{1}{\sqrt{4x}} }.
\nonumber \\ &&
\end{eqnarray}
On the mass--shell $(x=1, \theta = \frac{2 \pi}{3}, \sigma = -1)$ we find
the result
\begin{eqnarray}
\sum_{n=1}^\infty \left( 2n \atop n \right)  \frac{1}{n^a}
& = &
\frac{1}{2}~ {}_{a+2}F_{a+1}\left(\begin{array}{c|}
\tfrac{3}{2},  \{ 1 \}_{a+1} \\
\{ 2 \}_{a+1}
\end{array} ~1 \right)
+ \frac{2^a}{(a-2)!} \frac{\ln^a 2}{a}
\nonumber \\ &&
+ \mbox{i} \frac{2^{a-1}}{(a-1)!}
\Biggl[\Ls{a}{\frac{\pi}{3}} -\Ls{a}{\pi}  \Biggr] .
\end{eqnarray}

In general, however, analytical results for
these types of sums are not expressible in terms of generalized
Nielsen polylogarithms. For example, we have

\begin{eqnarray}
&&
\sum_{n=1}^\infty \left( 2n \atop n \right)  \frac{x^n}{n^a} \bar{S}_1
=
\Sigma \left ( \bar{S}_1, \frac{\pi}{2} \right)
+
\frac{2^{a-2}}{(a-2)!} \int_0^{2 \arccos \frac{1}{\sqrt{4x}}}
d \theta \frac{\sin \frac{\theta}{2}}{\cos \frac{\theta}{2}}
\Biggl[\ln \sqrt{x} + \ln \left( 2 \cos \frac{\theta}{2} \right) \Biggr]^{a-2}
\times
\nonumber \\ && \hspace{3.0cm}
\Biggl\{
\Biggl[ 2 \zeta_2 - \pi \theta + \frac{\theta^2}{4} + \ln^2 \left( 2 \cos \frac{\theta}{2} \right)
\Biggr]
- \mbox{i}
\Biggl[
2 \Cl{2}{\theta} + \theta \ln \left( 2 \cos \frac{\theta}{2} \right)
\Biggr]
\Biggr\}
\nonumber \\ && =
\Sigma \left ( \bar{S}_1, \frac{\pi}{2} \right)
- \frac{2^{a-1}}{(a-2)!} \sum_{j=0}^{a-2} \left( a-2 \atop j\right)
\frac{\left(\ln \sqrt{x} \right)^{a-2-j} }{j+1} \times
\nonumber \\ &&
\Biggl\{
\Biggl[ 2 \zeta_2 - \pi \theta + \frac{\theta^2}{4} \Biggr]
\ln^{j+1} \left( 2 \cos \frac{\theta}{2} \right)
+   \frac{\pi}{2} \Biggl[ \Ls{j+2}{\pi-\theta} -\Ls{j+2}{\pi} \Biggr]
- 2 \zeta_2 \ln^{j+1} 2
\nonumber \\ &&
+ \frac{1}{2} \Biggl[ \LS{j+3}{1}{\pi-\theta} -\LS{j+3}{1}{\pi} \Biggr]
+ \frac{j+1}{j+3}
\Biggl[
\ln^{j+3} \left( 2 \cos \frac{\theta}{2} \right) - \ln^{j+3} 2
\Biggr]
\nonumber \\ &&
- 2 \mbox{i} \sigma
\Biggr[
\Cl{2}{\theta} \ln^{j+1} \left( 2 \cos \frac{\theta}{2} \right)
- \Lsc{2,j+2}{\theta}
\Biggl]
\nonumber \\ &&
- \mbox{i} \sigma \frac{j+1}{j+2}
\Biggl[ \Ls{j+3}{\pi} -\Ls{j+3}{\pi-\theta}
+ \theta \ln^{j+2} \left( 2 \cos \frac{\theta}{2} \right) \Biggr]
\Biggr\}_{\theta = 2 \arccos \frac{1}{\sqrt{4x}} } \; ,
\end{eqnarray}
where $\sum\limits_{k=1}^\infty \left( 2k \atop k\right) \frac{x^k}{k}
\bar{S}_1 = 2 \Li{2}{\chi} + \ln^2 (1+\chi) $ and $\Lsc{i,j}{\theta}$
are new type of log-sine integrals introduced in~\cite{DK2001} (where the
properties of these functions are given in Appendix A.2) and defined by
\begin{equation}
\Lsc{i,j}{\theta} = -\int\limits_0^\theta
{\mbox d} \phi
\ln^{i-1} \left| 2\sin\frac{\phi}{2} \right|
\ln^{j-1} \left| 2\cos\frac{\phi}{2} \right|  \; .
\label{log-sin-cos}
\end{equation}
For $i,j>2$ these functions cannot be written in terms of Nielsen
polylogarithms, but are related to {\em harmonic polylogarithms}
which have been considered in~\cite{RV00}.

Using the results for the binomial series given above, we are able to
calculate analytically the required first few coefficients of the
$\ep$--expansion of hypergeometric functions of the following type

\begin{equation}
_{Q}F_P\left(\begin{array}{c|}
\tfrac{3}{2} + b_1 \ep,\ldots, \tfrac{3}{2} + b_I \ep, \;
1+a_1\ep, \ldots, 1+a_K \ep, \; 2+d_1 \ep \ldots, 2+d_L \ep \\
\tfrac{3}{2} + f_1 \ep, \ldots, \tfrac{3}{2} + f_J \ep, \;
1+e_1\ep, \ldots, 1+e_{R}\ep, \;
2+c_1\ep, \ldots, 2+c_{Q-J-R} \ep
\end{array} ~z\right) \EP
\label{PFQ}
\end{equation}
Rewriting this function as an infinite series and using the well-know
representation $$
\frac{\Gamma(j+a\ep)}{\Gamma(1+a\ep)}
= (j-1)! \; \exp\left[ -\sum_{k=1}^{\infty} \frac{(-a\ep)^k}{k} S_k(j-1) \right] \; ,
$$
we obtain (for details we refer to Appendix~B of~\cite{DK2001})

\begin{eqnarray}
&& \hspace*{-7mm}
_{P+1}F_P\left(\begin{array}{c|}
\{ \tfrac{3}{2} +b_i\ep\}_0^{J+1}, \;
\{ 1+a_i\ep\}_0^K, \; \{ 2+d_i\ep\}_0^L  \\
\{ \tfrac{3}{2} + f_i\ep\}_0^J, \;
\{ 1+e_q\ep \}_0^R,
\{ 2+c_i\ep \}_0^{K+L-R}
\end{array} ~  4z \right)
= \frac{1}{2 z}
\times
\nonumber \\ && \hspace*{-5mm}
\Pi_{s=1}^{K+L-R} (1+c_s\ep)
\Pi_{i=1}^{L} \frac{1}{(1+d_i\ep)}
\Pi_{r=1}^{J+1} \frac{1}{(1 + 2 b_r\ep)}
\Pi_{k=1}^{J} (1 + 2 f_k\ep)
\sum_{j=1}^\infty \left( 2j \atop j \right)   \frac{z^j}{j^{K-R-1}}
\times
\Delta \;,
\nonumber \\ &&
\label{hypergometric_expansion}
\end{eqnarray}
where $P=K+L+J$
and
\begin{eqnarray}
&& \hspace*{-7mm}
\Delta =
\exp \left( \sum_{k=1}^{\infty} \frac{(-\ep)^k}{k}
\left\{   S_k T_k + 2^k U_k \bar{S}_k + \frac{Y_k}{j^k}     \right\}     \right)
=
\Biggl(
1
- \ep \left\{
   S_1 T_1
+ \frac{Y_1}{j}
+ 2 U_1 \bar{S}_1
\right\}
\nonumber \\ &&
+  \ep^2 \Biggl\{
\frac{1}{2 j^2} \left[ Y_2 + Y_1^2 \right]
+ \frac{Y_1}{j}
 \left[ S_1 T_1
+ 2 U_1 \bar{S}_1  \right]
+  2 U_1 S_1 \bar{S}_1 T_1
+ 2 U_2 \bar{S}_2
+ 2 U_1^2 \bar{S}_1^2 \nonumber \\ &&~~~~~~
+ \frac{1}{2} S_2  T_2
+ \frac{1}{2} S_1^2 T_1^2
\Biggr\}
+ {\rm O} (\ep^3)
\Biggr) \;.
\end{eqnarray}
Here, we introduced new constants $A_k,B_k,C_k,D_k,E_k,F_k,T_k,U_k$:
$A_k \equiv \sum_{i=1}^{K} a_i^k$, $B_k \equiv \sum_{i=1}^{J-1}
b_i^k$, $C_k \equiv \sum_{i=1}^{K+L-R-2} c_i^k$, $D_k \equiv
\sum_{i=1}^{L} d_i^k$, $E_k \equiv \sum_{i=1}^{R} e_i^k$, $F_k \equiv
\sum_{i=1}^{J} f_i^k$, $T_k \equiv B_k + C_k + E_k - A_k - D_k - F_k$,
$U_k \equiv F_k - B_k$, $Y_k \equiv C_k - D_k$. With the expansion
just worked out, analytical results are available for the first three
coefficients of the $\ep$--expansion of the hypergeometric function
(\ref{PFQ}).  In particular, we find

\begin{eqnarray}
&& \hspace*{-7mm}
_{2}F_1\left(\begin{array}{c|}
\tfrac{3}{2} -  2 \ep \; , 1  \\ 2-3 \ep
\end{array} ~  4z \right)
= \frac{1}{2z} \frac{1 -3  \ep}{1 -4 \ep} \frac{1+\chi}{1-\chi}
\Biggl[ \frac{2 \chi}{1+\chi}
- 2 \Biggl \{ \ln(1-\chi) + 2 \ln(1+\chi) \Biggr\} \ep
\nonumber \\ &&
+ 2 \ep^2 \Biggl \{
         3 \Biggl( \Li{2}{\chi} + 2 \Li{2}{-\chi} \Biggr)
      + \Biggr( \ln (1-\chi)  + 2 \ln (1+\chi) \Biggl)^2
       \Biggr\}
+ {\rm O} (\ep^3)
\Biggr] \EP
\end{eqnarray}

For the analytical representation of the $\ep$--expansion of V-type
integrals we need also sums with shifted arguments (see
Eqs.~(\ref{v1002:series}) and (\ref{v2001:series}) ). The following
two types of sums have to be considered:
\begin{eqnarray}
I_{a_1,\ldots,a_p; \; b_1,\ldots,b_q;c}^{\; i_1,\ldots,i_p; \;j_1,\ldots,j_q}(x)
& \equiv &
\sum_{n=1}^\infty \left( 2n \atop n\right) \frac{x^n}{n^c}
[S_{a_1}(n\!-\!1)]^{i_1}\ldots [S_{a_p}(n\!-\!1)]^{i_p}\;
[S_{b_1}(2n\!-\!1)]^{j_1}\ldots [S_{b_q}(2n\!-\!1)]^{j_q},
\nonumber \\
J_{a_1,\ldots,a_p; \; b_1,\ldots,b_q;c}^{\; i_1,\ldots,i_p; \;j_1,\ldots,j_q}(x)
& \equiv &
\sum_{n=0}^\infty \left( 2n \atop n\right) \frac{x^n}{(n+1)^c}
[S_{a_1}(n)]^{i_1}\ldots [S_{a_p}(n)]^{i_p}\;
[S_{b_1}(2n)]^{j_1}\ldots [S_{b_q}(2n)]^{j_q} \;.
\label{binsum}
\end{eqnarray}
Cases when there are no sums of the type
$S_{a}(n-1)$ or $S_{b}(2n-1)$ on the r.h.s. of~(\ref{binsum}) we characterize
by a ``$-$''--sign replacing indices $(a,i)$ or $(b,j)$ of
$I,J$, respectively.

With this notation we may write the following relations:
\begin{eqnarray}
&&
\frac{1}{2x} I^{j;-}_{a;-;c}(x) =
\left[ 2 J^{j;-}_{a;-;c}(x) - J^{j;-}_{a;-;c+1}(x) \right] ,
\nonumber \\ &&
\frac{1}{2x} I^{j;1}_{a;1;c}(x)
= 2 J^{j;1}_{a;1;c}(x) - J^{j;1}_{a;1;c+1}(x) + J^{j;-}_{a;-;c+1}(x) ,
\nonumber \\ &&
\frac{1}{2x} \left[ I^{-;1}_{-;2;c}(x)-I^{-;2}_{-;1;c}(x) \right] =
2 J^{-;1}_{-;2;c}(x) - 2 J^{-;2}_{-;1;c}(x)
- J^{-;1}_{-;2;c+1}(x) + J^{-;2}_{-;1;c+1}(x)
- 2 J^{-;1}_{-;1;c+1} ,
\nonumber \\ &&
\frac{1}{2x} \sum_{n=1}^\infty \left( 2n \atop n\right) \frac{x^n}{n^c} V_a(n-1)
= \sum_{n=1}^\infty \left( 2n \atop n\right) x^n
\left[ \frac{2 V_a(n)}{(n+1)^c}  - \frac{V_a(n)}{(n+1)^{c+1}}
\right]
\end{eqnarray}

\section{Fermion corrections to the pole masses at one-loop}
\label{Fcone}
\setcounter{equation}{0}
In this Appendix we present, for completeness, the well know~\cite{FJ}
one-loop relations between pole- and ${\overline{\rm MS}}$-masses of
the gauge-bosons.  We divide all corrections into bosonic (diagrams
without any fermions) and fermionic (diagrams exhibiting a fermion
loop) one's: $X^{(1)}_V = X^{(1)}_{V,\mbox{boson}} +
X^{(1)}_{V,\mbox{fermion}}$ where the purely bosonic contributions are
given in Appendix B of I.  Using the notation
\begin{equation}
\frac{M_V^2}{m_V^2} =  1 + \left( \frac{e^2}{16\pi^2\sin^2 \theta_W} \right) X^{(1)}_V
\label{X:def}
\end{equation}
we may write the fermion corrections in the following
form\footnote{For simplicity we assume a
diagonal Cabibbo-Kobayashi-Maskawa matrix.}

\begin{eqnarray}
X^{(1)}_{W,\mbox{fermion}} & = &
\frac{1}{3} n_F \Biggl[ \frac{1}{3} + \frac{1}{3} N_c \Biggr]
+ \sum_{lepton} \Biggl[
\frac{1}{6} B_0(0,m_l^2;m_W^2) \left( 2 - \frac{m_l^2}{m_W^2}
-  \frac{m_l^4}{m_W^4} \right)
\nonumber \\ &&
- \frac{1}{6} \frac{m_l^4}{m_W^4} \left( \ln \frac{m_l^2}{\mu^2}
- 1 \right) \left( 1 + 12 \frac{m_W^2}{m_H^2} \right)
- \frac{1}{3} \frac{m_l^2}{m_W^2} \ln \frac{m_l^2}{\mu^2}
              \Biggr]
\nonumber \\ &&
+ N_c \sum_{\{u,d\}} \Biggl[
- \frac{1}{3} \frac{m_u^2}{m_W^2} \ln \frac{m_u^2}{\mu^2}
- \frac{1}{3} \frac{m_d^2}{m_W^2} \ln \frac{m_d^2}{\mu^2}
\nonumber \\ &&
+ \frac{1}{6}
\left(
2 + 2 \frac{m_u^2 m_d^2}{m_W^4}
-  \frac{m_u^4}{m_W^4}
-  \frac{m_d^4}{m_W^4}
-  \frac{m_u^2}{m_W^2}
-  \frac{m_d^2}{m_W^2}
\right)
B_0(m_d^2,m_u^2;m_W^2)
\nonumber \\ &&
+
\left(
  \frac{1}{6} \frac{m_u^2 m_d^2}{m_W^4}
- \frac{1}{6} \frac{m_u^4}{m_W^4}
- 2 \frac{m_u^4}{m_W^2 m_H^2}
\right)
\left( 1 - \ln \frac{m_u^2}{\mu^2} \right)
\nonumber \\ &&
+
\left(
  \frac{1}{6} \frac{m_u^2 m_d^2}{m_W^4}
- \frac{1}{6} \frac{m_d^4}{m_W^4}
- 2 \frac{m_d^4}{m_W^2 m_H^2}
\right)
\left( 1 - \ln \frac{m_d^2}{\mu^2} \right)
\Biggr]
\label{X1W}
\end{eqnarray}
\begin{eqnarray}
X^{(1)}_{Z,\mbox{fermion}} & = &
\frac{1}{3} n_F \Biggl[
- 2 + \frac{m_Z^2}{m_W^2}
+ \frac{1}{2} \frac{m_Z^2}{m_W^2} B_0(0,0;m_Z^2)
+ \frac{4}{3} \frac{m_W^2}{m_Z^2}
\nonumber \\ && \hspace{10mm}
+ N_c \left( \frac{11}{27}  \frac{m_Z^2}{m_W^2}
+ \frac{20}{27} \frac{m_W^2}{m_Z^2}
- \frac{22}{27}
      \right)
\Biggr]
\nonumber \\ &&
+ \sum_{lepton} \Biggl[
4 \frac{m_l^2}{m_Z^2} \left( 1 - \frac{2}{3}  \frac{m_W^2}{m_Z^2} \right)
                       \left(\ln \frac{m_l^2}{\mu^2} - B_0(m_l^2,m_l^2;m_Z^2) \right)
\nonumber \\ &&
- \frac{m_l^2}{m_W^2} \left( \frac{5}{3} \ln \frac{m_l^2}{\mu^2}
- \frac{7}{6} B_0(m_l^2,m_l^2;m_Z^2) \right)
- 2 \frac{m_l^4}{m_H^2 m_W^2}  \left( 1 -  \ln \frac{m_l^2}{\mu^2} \right)
\nonumber \\ &&
+ B_0(m_l^2,m_l^2;m_Z^2) \left(\frac{5}{6} \frac{m_Z^2}{m_W^2}
+ \frac{4}{3} \frac{m_W^2}{m_Z^2} - 2  \right)
\Biggr]
\nonumber \\ &&
+  N_c \sum_u \Biggl[
2 \frac{m_u^4}{m_H^2 m_W^2} \left( \ln \frac{m_u^2}{\mu^2} - 1\right)
\nonumber \\ && ~~~
+ \left(  \frac{17}{54} \frac{m_Z^2}{m_W^2} + \frac{16}{27} \frac{m_W^2}{m_Z^2}
- \frac{20}{27} \right) B_0(m_u^2,m_u^2;m_Z^2)
\nonumber \\ &&
-   \frac{m_u^2}{m_W^2} \left( \frac{17}{27} \ln \frac{m_u^2}{\mu^2}
- \frac{7}{54} B_0(m_u^2,m_u^2;m_Z^2)
                                   \right)
\nonumber \\ &&
+
  \left( \frac{40}{27} \frac{m_u^2}{m_Z^2} - \frac{32}{27} \frac{m_u^2 m_W^2}{m_Z^4} \right)
  \left( \ln \frac{m_u^2}{\mu^2} - B_0(m_u^2,m_u^2;m_Z^2) \right)
\Biggr]
\nonumber \\ &&
+  N_c \sum_d \Biggl[
   2 \frac{m_d^4}{m_H^2 m_W^2} \left( \ln \frac{m_d^2}{\mu^2} - 1 \right)
\nonumber \\ && ~~~
+ \left(\frac{5}{54}  \frac{m_Z^2}{m_W^2} +  \frac{4}{27} \frac{m_W^2}{m_Z^2}
- \frac{2}{27} \right) B_0(m_d^2,m_d^2;m_Z^2)
\nonumber \\ &&
-   \frac{m_d^2}{m_W^2}  \left( \frac{5}{27} \ln \frac{m_d^2}{\mu^2}
+ \frac{17}{54} B_0(m_d^2,m_d^2;m_Z^2)
                                    \right)
\nonumber \\ &&
+  \left( \frac{4}{27} \frac{m_d^2}{m_Z^2} -  \frac{8}{27} \frac{m_d^2 m_W^2}{m_Z^4} \right)
   \left( \ln \frac{m_d^2}{\mu^2} - B_0(m_d^2,m_d^2;m_Z^2) \right)
\label{X1Z}
\end{eqnarray}
\begin{eqnarray}
X^{(1)}_{H,\mbox{fermion}} & = &
\frac{1}{2} \frac{m_l^2}{m_W^2} \sum_{lepton} \Biggl[ B_0(m_l^2,m_l^2;m_H^2)
 \left( 1 - 4 \frac{m_l^2}{m_H^2} \right)
- 4 \frac{m_l^2}{m_H^2} \left(1 -  \ln \frac{m_l^2}{\mu^2} \right)
               \Biggr]
\nonumber \\
& + &  \frac{1}{2} \frac{m_q^2}{m_W^2} N_c\sum_{quark} \Biggl[
B_0(m_q^2,m_q^2;m_H^2 ) \left( 1 - 4 \frac{m_q^2}{m_H^2} \right)
-  4 \frac{m_q^2}{m_H^2} \left(1 -  \ln \frac{m_q^2}{\mu^2} \right)
                                                    \Biggr]
\nonumber \\
\end{eqnarray}
\begin{eqnarray}
X^{(1)}_{top} & = &
\left(
  \frac{20}{9}
- \frac{7}{36}  \frac{m_Z^2}{m_W^2}
- \frac{17}{36} \frac{m_Z^4}{m_W^2 m_u^2}
+ \frac{10}{9}  \frac{m_Z^2}{m_u^2}
- \frac{8}{9}   \frac{m_W^2}{m_u^2}
- \frac{16}{9}  \frac{m_W^2}{m_Z^2}
\right) B_0(m_Z^2,m_u^2;m_u^2)
\nonumber \\ &&
+ \left(
  \frac{1}{4}
+ \frac{1}{4} \frac{m_d^4}{m_W^2 m_u^2}
+ \frac{1}{4} \frac{m_d^2}{m_u^2}
+ \frac{1}{4} \frac{m_u^2}{m_W^2}
- \frac{1}{2} \frac{m_d^2}{m_W^2}
- \frac{1}{2} \frac{m_W^2}{m_u^2}
\right) B_0(m_W^2,m_d^2;m_u^2)
\nonumber \\ &&
+ \left(
\frac{m_u^2}{m_W^2} - \frac{1}{4} \frac{m_H^2}{m_W^2}
\right) B_0(m_H^2,m_u^2;m_u^2)
+  \frac{155}{36} - \frac{1}{4} \ln \frac{m_W^2}{\mu^2} - \frac{14}{9} \ln \frac{m_u^2}{\mu^2}
\nonumber \\ &&
+ \frac{m_Z^2}{m_W^2} \left( \frac{1}{2} - \frac{17}{36} \ln \frac{m_u^2}{\mu^2} \right)
+ \frac{1}{2} \frac{m_Z^4}{m_W^2 m_H^2} \left( 1 - 3 \ln \frac{m_Z^2}{\mu^2} \right)
\nonumber \\ &&
+ \frac{m_Z^2}{m_u^2} \left( \frac{10}{9} - \frac{17}{36} \frac{m_Z^2}{m_W^2} \right)
\left( 1 - \ln \frac{m_Z^2}{\mu^2} \right)
+ \frac{1}{2} \frac{m_H^2}{m_W^2} \left( 1 - \ln \frac{m_H^2}{\mu^2} \right)
\nonumber \\ &&
+ \frac{1}{4} \frac{m_d^2}{m_W^2} \left( 1 +\frac{m_d^2}{m_u^2} \right) \left( 1 -\ln \frac{m_d^2}{\mu^2} \right)
+ \frac{m_W^2}{m_H^2} \left( 1- 3 \ln \frac{m_W^2}{\mu^2} \right)
\nonumber \\ &&
+ \frac{1}{4} \frac{m_d^2}{m_u^2} \left(1 +  \ln \frac{m_W^2}{\mu^2} - 2 \ln \frac{m_d^2}{\mu^2} \right)
+ \frac{1}{2} \frac{m_u^2}{m_W^2} \left( 1 - \ln \frac{m_u^2}{\mu^2} \right)
\nonumber \\ &&
+ \frac{m_W^2}{m_u^2} \left( -\frac{25}{18} + \frac{1}{2} \ln \frac{m_W^2}{\mu^2}  + \frac{8}{9} \ln \frac{m_Z^2}{\mu^2} \right)
- \frac{16}{9} \frac{m_W^2}{m_Z^2} \left( 2 - \ln \frac{m_u^2}{\mu^2} \right)
\nonumber \\ &&
- 2 N_c \sum_{quark} \frac{m_q^4}{m_W^2 m_H^2} \left( 1 - \ln \frac{m_q^2}{\mu^2} \right)
- 2 \sum_{lepton} \frac{m_l^4}{m_W^2 m_H^2} \left( 1 - \ln \frac{m_l^2}{\mu^2} \right)
\end{eqnarray}
where

$$
B_0(m_1^2,m_2^2;p^2)
= \int\limits_0^1 dx \ln \Biggl( \frac{m_1^2}{\mu^2} x
+ \frac{m_2^2}{\mu^2} (1-x) - \frac{p^2}{\mu^2} x(1-x) - \mbox{i} 0 \Biggr)
$$
denotes a scalar two--point function.
\section{Mass renormalization contributions}
\label{Mrtwo}
\setcounter{equation}{0}

For the parameter renormalization of a two--loop amplitude one
requires the first order derivatives with respect to all relevant
parameters of the one--loop amplitude as may be seen in
(\ref{MS2:subtracted}). Since, we are interested in the on-shell
amplitudes mass derivations implicitly involve a differentiation with
respect to the external momentum, as the ``on--shell momentum'' has
been given the value of a mass. We restrict ourselves to consider the
effect of the mass renormalization contribution, since for the charge
renormalization the derivative with respect to the latter is trivial
and it is included in (\ref{MS2:subtracted}) as a separate term. We
thus consider
\begin{equation}
D_V = \sum_j \delta_j \frac{\partial}{\partial m_j^2} X_{0,V}^{(1)}
\label{DV}
\end{equation}
where the coefficients $\delta_j$ are the one--loop mass
renormalization counter--terms, which depend on the renormalization
scheme. For the \MSb scheme they have to be identified with the
$Z_{m^2_j}^{(1,1)}$ of (\ref{MS2:subtracted}). The explicit
expressions presented below are written down for the third fermion
family $(\nu_\tau,\tau,t,b)$ in the approximation of vanishing
$\tau$-- and $b$--mass. Accordingly in (\ref{DV}), the relevant masses
are indexed by $j=W,Z,H,t$.  For simplicity we give the results
assuming the Cabibbo-Kobayashi-Maskawa matrix to be the unit matrix.
The contribution of a massless family may be obtained by putting $m_t=0
(A_0(m_t)=0)$. The results read
\begin{eqnarray}
&&  D_W =
\delta_{t} \frac{N_c}{d-1} \Biggl\{
 \U \frac{ (d-2) \; m_t^4}{4 \X}
\Biggl[
  \frac{m_W^2}{m_t^2} (2 d -5 ) - \frac{m_t^2}{m_W^2} -
  \frac{m_W^4}{m_t^4} (d-2) - (d-4)
\Biggr]
\nonumber \\ &&
+ \Y
\frac{(d-3) \; m_t^4}{2 \X}
\Biggl[
\frac{m_t^2}{m_W^2}
- \frac{m_W^2}{m_t^2} (2 d - 5)
+ \frac{m_W^4}{m_t^2} (d-2)
+                     (d-4)
  \Biggr]
\nonumber \\ &&
+ \Y \left( \frac{3-d}{2} - \frac{m_t^2}{m_W^2}  \right)
+ \U \Biggl[ \frac{m_t^2}{m_W^2} \frac{d}{4}
+ \frac{m_t^2}{m_H^2} d (d-1)
- \frac{(d-2)^2}{4} \Biggr]
\Biggr \}
\nonumber \\ &&
+ \delta_{H} \Biggl\{
\frac{ (d-2) \; m_H^2 m_W^2 }{2 \g} \Biggl[
         \h
       - \J  \frac{d-3}{d-2}
       - \frac{1}{2} \W
                 \Biggr]
\nonumber \\ && \hspace{10mm}
       + \J \Biggl[ \frac{m_H^2}{m_W^2}  \frac{1}{4}
                    +  \frac{m_W^2}{m_H^2}   \frac{d-3}{2}
                    - \frac{1}{2}
            \Biggr]
\nonumber \\ &&  \hspace{10mm}
       - 2 \U N_c \frac{m_t^4}{m_H^4}
       + \Z \frac{m_Z^4}{m_H^4}  \frac{d-1}{2}
\nonumber \\ &&  \hspace{10mm}
       + \W \Biggl[ \frac{m_W^2}{m_H^2}    \frac{2-d}{4}
                   + \frac{m_W^4}{m_H^4}  (d-1)
                   + \frac{1}{4}
            \Biggr]
       + \h  \Biggl[ \frac{2-d}{4}
                     - \frac{m_H^2}{m_W^2}  \frac{1}{4}
             \Biggr]
\Biggr\}
\nonumber \\ &&
+ \delta_{Z} \frac{1}{d-1} \Biggl\{
       \K \Biggl[\frac{m_Z^2}{m_W^2}   \frac{d-1}{4}
       -  \frac{m_W^2}{m_Z^2}   ( 3 d -5 ) (d-3)
\nonumber \\ && \hspace{55mm}
       -   \frac{m_W^4}{m_Z^4}  2 (d-1) (d-5)
       +  \frac{4 d^2 - 15 d + 13}{2}
             \Biggr]
\nonumber \\ &&
       + \W \Biggl[ \frac{m_W^2}{m_Z^2}  (2 d -3 )(d-2)
       + \frac{m_W^4}{m_Z^4} \frac{(d-1)(d-2)(d-5)}{d-3}
       + \frac{d-1}{4}
             \Biggr]
\nonumber \\ &&
       + \Z \Biggl[ \frac{m_Z^2}{m_W^2}   \frac{1-d}{4}
                   + \frac{m_W^2}{m_Z^2}  \frac{(d-5) (d-2) (d-1)}{2}
                   - \frac{m_Z^2}{m_H^2}  \frac{d (d-1)^2 }{4}
       - \frac{( 3 d-5) ( d-2)}{2}
             \Biggr]
\Biggr \}
\nonumber \\ &&
+ \delta_{W} \frac{1}{d-1} \Biggl\{
\U
\frac{(d-2) \; m_t^4}{\X} \frac{N_c}{4} \Biggl[
          \frac{m_t^4}{m_W^4}
       +  \frac{m_t^2}{m_W^2} (d-4)
       +  \frac{m_W^2}{m_t^2} (d-2)
       -  (2 d -5 )
                   \Biggr]
\nonumber \\ && \hspace{10mm}
+ \Y
\frac{(d-3) \;m_t^4}{\X} \frac{N_c}{4} \Biggl[
          (d-1)
       -  \frac{m_t^4}{m_W^4}
       -  \frac{m_W^4}{m_t^4}  (d-2)
\nonumber \\ && \hspace{70mm}
+ \left( \frac{m_W^2}{m_t^2}  - \frac{m_t^2}{m_W^2} \right) (d-3)
                   \Biggr]
\nonumber \\ &&
+ \frac{ (d-1) \; m_H^2 m_W^2 }{\g} \Biggl[
          \J 2 (d-3)
        + \left( \W - 2 \h \right) (d - 2)
                                        \Biggl]
\nonumber \\ &&
+ \Y N_c \Biggl[
\frac{3}{4} \frac{m_t^4}{m_W^4}
      + \frac{m_t^2}{m_W^2} \frac{d-3}{4}
      + \frac{d-2}{4}
\Biggr]
- \M \frac{(d-2)(d-4)}{4}
\nonumber \\ &&
- \J \Biggl[
 \frac{d}{8} \frac{m_H^4}{m_W^4}
- \frac{1}{2} \frac{m_H^2}{m_W^2}
+ \frac{ (d-1) (d-4) }{2}
\Biggr]
- \frac{1}{2} N_c \U \frac{m_t^4}{m_W^4}
\nonumber \\ &&
- \K \Biggl[
 \frac{d}{8} \frac{m_Z^4}{m_W^4}
+ \frac{m_Z^2}{m_W^2} \frac{ (2 d - 5) (d-1) }{2}
+ 6 (d-1) \frac{m_W^2}{m_Z^2}
+ \frac{d^2 + 3 d - 8}{2}
\Biggr]
\nonumber \\ &&
- \W \Biggl[
  \frac{m_Z^2}{m_W^2} \frac{d}{8}
+ \frac{m_H^2}{m_W^2} \frac{d}{8}
+ \frac{m_W^2}{m_H^2} \frac{d(d-1)^2}{2}
- \frac{m_W^2}{m_Z^2} 3 \frac{(d-1)(d-2)}{d-3}
\nonumber \\ && \hspace{20mm}
- \frac{d-2}{d-3} \frac{d^3 - 13 d^2 + 39 d -31}{2}
\Biggr]
+ \h \Biggl[
\frac{m_H^4}{m_W^4} \frac{d}{8}
+ \frac{(d-1)(d-2)}{2}
\Biggr]
\nonumber \\ &&
+ \Z \Biggl[
\frac{m_Z^4}{m_W^4} \frac{d}{8}
+ \frac{m_Z^2}{m_W^2} \frac{ (2 d - 3) (d-2)}{2}
+ \frac{3 (d-1) (d-2) }{2}
\Biggr ]
\Biggr\} \\&&
\nonumber \\
&&  D_Z =
\delta_{t} \frac{N_c}{d-1} \Biggl\{
\U \frac{m_t^4 (d-2)}{\R} \Biggl[
  \frac{m_Z^4}{m_W^2 m_t^2} \frac{9 d - 43}{18}
- \frac{m_Z^6}{m_W^2 m_t^4} \frac{17(d-2)}{36}
\nonumber \\ && \hspace{30mm}
- \frac{m_W^2 m_Z^2}{ m_t^4} \frac{8 (d-2) }{9}
- \frac{m_W^2}{m_t^2} \frac{32 }{9}
+ \frac{m_Z^2}{m_t^2} \frac{40 }{9}
+ \frac{m_Z^4}{m_t^4} \frac{10 (d-2)}{9}
\Biggr]
\nonumber \\ &&
+ \T \frac{m_t^4 (d-3) }{\R} \Biggl[
  \frac{m_Z^6}{m_W^2 m_t^4} \frac{17 (d-2) }{18}
- \frac{m_Z^4}{m_W^2 m_t^2} \frac{(9 d - 43)}{9}
\nonumber \\ && \hspace{30mm}
+ \frac{m_Z^2 m_W^2}{m_t^4}  \frac{16 (d-2) }{9}
+ \frac{m_W^2}{m_t^2} \frac{64 }{9}
- \frac{m_Z^2}{m_t^2} \frac{80 }{9}
- \frac{m_Z^4}{m_t^4} \frac{20 (d-2)}{9}
\Biggr]
\nonumber \\ &&
+ \U \Biggl[
  \frac{m_Z^2 m_t^2}{m_W^2 m_H^2} d (d-1)
- \frac{m_Z^2}{m_W^2} \frac{17(d-2)^2}{36}
- \frac{m_W^2}{m_Z^2} \frac{8(d-2)^2}{9}
+ \frac{10(d-2)^2}{9}
\Biggr]
\nonumber \\ &&
- \T \Biggl[
  \frac{m_Z^2}{m_W^2} \frac{9 d - 43}{18}
- \frac{32}{9} \frac{m_W^2}{m_Z^2}
+ \frac{40}{9}
\Biggr]
\Biggr \}
\nonumber \\ &&
+ \delta_{H} \Biggl\{
\D \frac{m_Z^4}{2 \Q} \frac{m_H^2}{m_W^2} (3-d)
- 2 \U N_c \frac{m_t^4 m_Z^2}{m_H^4 m_W^2}
\nonumber \\ &&
- \D \Biggl[
  \frac{m_Z^2}{m_W^2} \frac{1}{2}
+ \frac{m_Z^4}{m_W^2 m_H^2} \frac{3-d}{2}
- \frac{1}{4} \frac{m_H^2}{m_W^2}
\Biggr]
+ \W \frac{m_W^2 m_Z^2}{m_H^4} (d-1)
\nonumber \\ &&
+ \h \frac{m_Z^4}{\Q} \frac{m_H^2}{m_W^2} \frac{d-2}{2}
- \Z \frac{m_Z^4}{\Q} \frac{m_H^2}{m_W^2} \frac{d-2}{4}
\nonumber \\ &&
+ \Z \Biggl[
\frac{1}{4} \frac{m_Z^2}{m_W^2}
+ \frac{d-1}{2} \frac{m_Z^6}{m_H^4 m_W^2}
- \frac{d-2}{4} \frac{m_Z^4}{m_W^2 m_H^2}
\Biggr]
- \h \Biggl[  \frac{m_Z^2}{m_W^2} \frac{d-2}{4}
+ \frac{1}{4} \frac{m_H^2}{m_W^2}
\Biggr]
\Biggr \}
\nonumber \\ &&
+ \delta_{W} \frac{1}{d-1}\Biggl\{
\D \Biggl[
\frac{m_Z^2 m_H^2}{m_W^4}
- (d-1) \frac{m_Z^4}{m_W^4}
- \frac{1}{4} \frac{m_H^4}{m_W^4}
\Biggr]
\nonumber \\ &&
+ \I (d-2) \Biggl[
  \frac{2}{9} N_c
- \frac{5}{36} N_c \frac{m_Z^4}{m_W^4}
- \frac{3}{2} \frac{m_Z^4}{m_W^4}
+ 2
\Biggr]
\nonumber \\ &&
- \U N_c  \Biggl[
 2(d-1) \frac{m_t^4 m_Z^2 }{m_W^4 m_H^2}
- \frac{17}{18} (d-2) \frac{m_Z^2 m_t^2}{m_W^4}
+  \frac{m_t^2}{m_Z^2} \frac{16 (d-2)}{9}
\Biggr]
\nonumber \\ &&
+ \T N_c \Biggl[
\frac{m_Z^2 m_t^2}{m_W^4} \frac{9 d -43}{18}
- \frac{17(d-2)}{36} \frac{m_Z^4}{m_W^4}
+ \frac{32}{9} \frac{m_t^2}{m_Z^2}
+ \frac{8(d-2)}{9}
\Biggr]
\nonumber \\ &&
+ \W \Biggl[
\frac{1}{2} \frac{m_Z^2}{m_W^2}
+ \frac{m_W^2}{m_Z^2} (d-1)(d-2)(d+1)
- \frac{m_Z^2}{m_H^2} \frac{(d-1)^2(d-2)}{2}
+ (d-1)(d-2)
\Biggr]
\nonumber \\ &&
+ \Z \Biggl[
  \frac{1}{2} \frac{m_Z^4}{m_W^4}
- \frac{1}{4} \frac{m_Z^2 m_H^2}{m_W^4}
+ \frac{m_Z^6}{m_H^2 m_W^4} \frac{(d-1)^2}{2}
\Biggl]
+ \h \Biggl[
\frac{1}{4} \frac{m_H^4}{m_W^4}
+ \frac{(d-1)}{2} \frac{m_Z^2 m_H^2}{m_W^4}
\Biggl]
\nonumber \\ &&
- B_0(m_W^2,m_W^2;m_Z^2) \Biggl[
  \frac{1}{4} \frac{m_Z^4}{m_W^4}
+ \frac{d-3}{2} \frac{m_Z^2}{m_W^2}
+ 2(d^2-1) \frac{m_W^2}{m_Z^2}
+ (4 d -7 ) (d-1)
\Biggr]
\Biggr \}
\nonumber \\ &&
+ \delta_{Z} \frac{1}{d-1}\Biggl\{
\frac{m_Z^4(d -1) (d-3) }{\Q} \frac{m_H^2}{m_W^2 }
\Biggl[2 \D  + \Z - 2 \h \Biggr]
\nonumber \\ &&
- \D \Biggl[ \frac{m_H^4}{m_W^2 m_Z^2} \frac{d-2}{8}
+ \frac{m_Z^2}{m_W^2} \frac{(d-1)(d-6)}{2}
+ \frac{1}{2} \frac{m_H^2}{m_W^2}
\Biggr]
\nonumber \\ &&
+ \I \Biggl[
  \frac{m_Z^2}{m_W^2} N_c\frac{5 d (d-2)}{72}
+ \frac{m_W^2}{m_Z^2} N_c \frac{(d-2)(d-4)}{9}
- N_c \frac{(d-2)^2}{18}
\nonumber \\ && \hspace{30mm}
+ \frac{m_Z^2}{m_W^2} \frac{3 d (d-2)}{4}
+ \frac{m_W^2}{m_Z^2} (d-2) (d-4)
- \frac{ 3 (d-2)^2}{2}
\Biggr]
\nonumber \\ &&
- \U \frac{N_c m_t^4}{\R}
\Biggl[
 \frac{m_Z^4}{m_W^2 m_t^2} \frac{ (d-2) (9 d - 43) }{72}
- \frac{m_Z^6}{m_W^2 m_t^4} \frac{17 (d-2)^2}{144}
\nonumber \\ && \hspace{20mm}
- \frac{m_W^2 m_Z^2}{m_t^4} \frac{2(d-2)^2}{9}
- \frac{m_W^2}{m_t^2} \frac{8(d-2)}{9}
+ \frac{m_Z^2}{m_t^2}  \frac{10(d-2)}{9}
+ \frac{m_Z^4}{m_t^4} \frac{5 (d-2)^2}{18}
\Biggr]
\nonumber \\ &&
+ \T \frac{N_c m_t^4}{\R}
\Biggl[
\frac{m_Z^4}{m_W^2 m_t^2} \frac{ (d-3) (9 d -43)}{36}
- \frac{m_Z^6}{m_W^2 m_t^4} \frac{17(d-2)(d-3)}{72}
\nonumber \\ &&
- \frac{m_W^2 m_Z^2}{m_t^4} \frac{4(d-2)(d-3)}{9}
- \frac{m_W^2}{m_t^2} \frac{16(d-3)}{9}
+ \frac{m_Z^2}{m_t^2} \frac{20(d-3)}{9}
+ \frac{m_Z^4}{m_t^4} \frac{5(d-2)(d-3)}{9}
\Biggr]
\nonumber \\ &&
+ \U N_c
\Biggl[
\frac{m_Z^2}{m_W^2} \frac{17 (d-2)^2 }{144}
+ \frac{m_t^4}{m_W^2 m_H^2} 2(d-1)
- \frac{m_t^2}{m_W^2}  \frac{ (d-2) (9 d + 25)}{72}
\nonumber \\ && \hspace{20mm}
+ \frac{m_W^2 m_t^2}{m_Z^4} \frac{8(d-2)}{3}
+ \frac{m_W^2}{m_Z^2} \frac{2 (d-2)^2}{9}
- \frac{m_t^2}{m_Z^2}  \frac{10(d-2)}{9}
- \frac{5(d-2)^2}{18}
\Biggr]
\nonumber \\ &&
+  \T N_c
\Biggl[
\frac{m_Z^2}{m_W^2} \frac{17(d-2)}{24}
- \frac{m_t^2}{m_W^2} \frac{9 d - 43}{36}
- \frac{16}{3} \frac{m_W^2 m_t^2}{m_Z^4}
- \frac{m_W^2}{m_Z^2} \frac{4(d-2)}{9}
\nonumber \\ && \hspace{40mm}
+ \frac{20}{9} \frac{m_t^2}{m_Z^2}
- \frac{5(d-2)}{9}
\Biggr]
\nonumber \\ &&
-  \W
\Biggl[
\frac{d}{4}
+ \frac{m_W^2}{m_Z^2} (2 d - 3) (d-2)
+ \frac{m_W^2}{m_H^2} (d-1)^2
+ \frac{m_W^4}{m_Z^4} 3 ( d -1) (d-2)
\Biggr]
\nonumber \\ &&
-  \Z
\Biggl[
\frac{m_Z^2}{m_W^2} \frac{d}{4}
+ \frac{m_Z^4}{m_W^2 m_H^2} \frac{(d-1)^2 (d+2)}{4}
+ \frac{m_H^2}{m_W^2} \frac{d-2}{8}
\Biggr]
\nonumber \\ &&
+  \h
\Biggl[
\frac{m_H^4}{m_W^2 m_Z^2} \frac{d-2}{8}
+ \frac{m_Z^2}{m_W^2} \frac{(d-1)(d-2)}{2}
- \frac{m_H^2}{m_W^2} \frac{d-1}{2}
\Biggr]
\nonumber \\ &&
+  B_0(m_W^2,m_W^2;m_Z^2)
\Biggl[
  \frac{m_Z^2}{m_W^2} \frac{d}{8}
+ \frac{m_W^2}{m_Z^2} \frac{d^2 + 3 d - 8}{2}
+ 6 (d-1) \frac{m_W^4}{m_Z^4}
+ \frac{ (2 d - 5) (d-1)}{2}
\Biggr]
\Biggr \}
\end{eqnarray}
where
$$
\Delta(m_1,m_2,m_3) = 2 m_1^2 m_2 ^2 + 2 m_1^2 m_3^2 + 2 m_2^2 m_3^3
- m_1^4 - m_2^4 -m_3^4 \;.$$
and
\begin{eqnarray}
A_0(m_1^2) & = &  - \frac{1}{m_1^2}
\int \frac{\mbox{d}^d q}{\pi^{d/2} \Gamma(1+\ep)}
\frac{1}{ \bigl( q^2+m_1^2 \bigr)} \; ,
\nonumber \\
B_0(m_1^2, m_2^2; p^2) & = &
\int \frac{\mbox{d}^d q}{\pi^{d/2} \Gamma(1+\ep)}
\frac{1}{ \bigl( q^2+m_1^2 \bigr) \bigl( (p-q)^2+m_2^2 \bigr)} \; .
\end{eqnarray}

\section{Bare two--loop contribution of the massless fermions}
\label{Fctwo}
\setcounter{equation}{0}

As one of our results we present the exact analytic two--loop
contribution to the on--shell self--energies of the gauge bosons.  After
reduction of the set of basic integrals to a minimal set of
master--integrals by means of Tarasov's recurrence
relations~\cite{tarasov-propagator} we obtain the following expressions:

\begin{eqnarray}
&&
-\Biggl( \Pi_W^{(2)} + \Pi_W^{(1)} \Pi_W^{(1)}{}' \Biggr) =
\nonumber \\ &&
V_{00W}
\Biggl\{
{ \frac {\left (d-2\right )}{4 {c}^{2}d\left (d-4\right )\left (d-3\right )
\left (d-1\right )^{2}}}
\Biggl[
8{c}^{2}{d}^{6}
+12{c}^{4}{d}^{5}
-120{c}^{2}{d}^{5}
+596{c}^{2}{d}^{4}
\nonumber \\ &&
+{d}^{4}
-96{c}^{4}{d}^{4}
-1136{c}^{2}{d}^{3}
+268{c}^{4}{d}^{3}
-11{d}^{3}
+508{c}^{2}{d}^{2}
-312{c}^{4}{d}^{2}
+40{d}^{2}-48d
\nonumber \\ &&
+528{c}^{2}d
+128{c}^{4}d
-384{c}^{2} \
\Biggr]
+
\frac {\left (d-2\right )}{4 m_H^{2}m_W^{2}\left (m_H^2-4 m_W^4\right )\left (d-1\right )^{2}}
\Biggl[
32{d}^{3}m_W^{6}
\nonumber \\ &&
-8{d}^{3}m_W^{4}m_H^{2}
+28{d}^{2}m_W^{4}m_H^{2}
-128{d}^{2}m_W^{6}
+m_H^{6}d
-4m_H^{4}dm_W^{2}
-28dm_W^{4}m_H^{2}
\nonumber \\ &&
+160dm_W^{6}
+16m_H^{4}m_W^{2}
-4m_H^{6}
-64 m_W^{6}
+8 m_W^{4}m_H^{2}
\Biggr]
\Biggr\}
\nonumber \\ &&
+ V_{00Z}
\Biggl\{
{\frac {\left (d-2\right )}{12 \left (d-1\right )^{2}{c}^{4}d}}
\Biggl[
-80{c}^{6}{d}^{4}
+64{c}^{8}{d}^{4}
+40{c}^{4}{d}^{4}
-88{c}^{2}{d}^{3}
+4{d}^{3}
+520{c}^{6}{d}^{3}
\nonumber \\ &&
-480{c}^{8}{d}^{3}
-124{c}^{4}{d}^{3}
+434{c}^{2}{d}^{2}
-236{c}^{4}{d}^{2}
-872{c}^{6}{d}^{2}
-51{d}^{2}
+992{c}^{8}{d}^{2}
+78d
\nonumber \\ &&
-508{c}^{2}d
-576{c}^{8}d
+632{c}^{4}d
+272{c}^{6}d
+32{c}^{2}
-16-64{c}^{4}
\Biggr]
\nonumber \\ &&
-{\frac {\left (d-2\right )^{2}\left (5{ m_Z^2}+8{c}^{2}{m_W^2}-10{c}^{2}{ m_Z^2}\right )}{3 {c}^{2}{ m_H^2}}}
\Biggr\}
\nonumber \\ &&
- F_{0000Z}
  {\frac {m_W^4 \left (d-2\right )\left (-1+2{c}^{2}+2{c}^{4}\right
)\left (2{c}^{4}-{c}^{2}d+8{c}^{2}+2\right )}
{6 \left (d-1\right ){c}^{6}}}
\nonumber \\ &&
- F_{Z0W00} 8{\frac {m_W^4 \left ({c}^{2}+2\right )\left (d-2\right )}{d-1}}
\nonumber \\ &&
+ V_{W00Z}
{\frac {m_W^2 \left (d-2\right )}{12 \left (d-1\right )^{2}{c}^{6}}}
\Biggl[
-120{c}^{2}{d}^{2}
+500{c}^{4}{d}^{2}
-552{c}^{6}{d}^{2}
+96{c}^{8}{d}^{2}
+256{c}^{10}{d}^{2}
\nonumber \\ &&
-1408{c}^{10}d
-15d
-288{c}^{8}d
+2856{c}^{6}d
-2444{c}^{4}d
+570{c}^{2}d
-640{c}^{2}
+30
+2712{c}^{4}
\nonumber \\ &&
-3456{c}^{6}
+832{c}^{8}
+1152{c}^{10}
\Biggr]
\nonumber \\ &&
+ V_{Z00W}
{\frac {m_W^2 \left (d-2\right )}{4 \left (d-1\right )^{2}{c}^{4}}}
\Biggl[
24{c}^{6}{d}^{2}
-8{c}^{2}{d}^{2}
+44{c}^{4}{d}^{2}
-d-80{c}^{6}d
+42{c}^{2}d
-204{c}^{4}d
\nonumber \\ &&
+4
-58{c}^{2}
+208{c}^{4}
+56{c}^{6}
\Biggr]
\nonumber \\ &&
+ V_{H00W} \Biggl\{
- {m_W^2 \frac {\left (d-2\right )^{2}}{d-1}}
- {\frac {\left (d-2\right )}{4 m_W^{2}\left (m_H^2-4m_W^2\right ) \left (d-1\right )^{2}}}
\Biggl[
-8{d}^{2}m_W^{6}
+32dm_W^{6}
+m_H^{6}d
\nonumber \\ &&
-10 m_H^{4}dm_W^{2}
+24dm_W^{4}m_H^{2}
-4m_H^{6}
-24m_W^{6}
-72 m_W^{4}m_H^{2}
+34m_H^{4}m_W^{2}
\Biggr]
\Biggr\}
\nonumber \\ &&
- J_{00Z}
\frac {1}{12 \left (d-1\right )^{2}{c}^{4}d}
\Biggl[
256{c}^{2}d
-302{c}^{2}{d}^{2}
-432d
-64{c}^{2}
+432{d}^{2}
+128{c}^{4}
-20{c}^{4}d
\nonumber \\ &&
-206{c}^{4}{d}^{2}
+128-8{c}^{2}{d}^{4}
+103{c}^{2}{d}^{3}
-56{c}^{4}{d}^{4}
+220{c}^{4}{d}^{3}
+12{d}^{4}
-140{d}^{3}
+52{c}^{6}{d}^{4}
\nonumber \\ &&
-612{c}^{6}{d}^{3}
-2240{c}^{6}d
+2008{c}^{6}{d}^{2}
+768{c}^{6}
\Biggr]
\nonumber \\ &&
-J_{00Z}'
{\frac { m_W^2 \left (c-1\right )\left (c+1\right )}{3 d\left (d-1\right )^{2}{c}^{6}}}
\Biggl[
-72{c}^{6}d
+4{c}^{6}{d}^{3}
+36{c}^{6}{d}^{2}
+32{c}^{6}+24{c}^{4}d
+20{c}^{4}{d}^{3}
+16{c}^{4}
\nonumber \\ &&
-48{c}^{4}{d}^{2}
-15{c}^{2}{d}^{2}
-8{c}^{2}
+2{c}^{2}{d}^{3}
+18{c}^{2}d
+18{d}^{2}
-24d
-2{d}^{3}
+8
\Biggr]
\nonumber \\ &&
+ J_{00H} \Biggl\{
{\frac {d-2}{2 \left (d-1\right )^{2}}}
+ {\frac {\left (d-2\right )
\left (3d-8\right )\left (4dm_W^{4}-4m_W^{2}m_H^{2}
-4m_W^{4}+m_H^{4}\right )
}{4 m_W^{2}\left (m_H^2-4 m_W^2 \right )\left (d-1\right )^{2}}}
\Biggr\}
\nonumber \\ &&
+ J_{00H}' \Biggl\{
4 m_W^2 {\frac {d-2}{d-1}}
+ \frac {\left (d-2\right )}{m_W^{2}\left (m_H^2 -4 m_W^2 \right ) \left (d-1\right )^{2}}
\Biggl[
12dm_W^{6}
-12m_W^{6}
+m_H^{6}
\nonumber \\ &&
-5m_H^{4}m_W^{2}
+4m_W^{4}m_H^{2}
\Biggr]
\Biggr\}
\nonumber \\ &&
+ J_{00W}
\frac {1}{12{c}^{4}d\left (2d-7\right )\left (d-4\right )\left (d-1\right )^{2}}
\Biggl[
-30240{c}^{2}d
+61000{c}^{2}{d}^{2}
+1680d
-2580{d}^{2}
\nonumber \\ &&
-53760{c}^{4}
+249600{c}^{4}d
-361120{c}^{4}{d}^{2}
-3340{c}^{2}{d}^{5}
+18110{c}^{2}{d}^{4}
-47720{c}^{2}{d}^{3}
\nonumber \\ &&
-744{c}^{4}{d}^{6}
+11988{c}^{4}{d}^{5}
-75452{c}^{4}{d}^{4}
+233208{c}^{4}{d}^{3}
+240{c}^{2}{d}^{6}
-345{d}^{4}
+1440{d}^{3}
\nonumber \\ &&
+30{d}^{5}
-18432{c}^{8}d
+10840{c}^{6}{d}^{4}
-29472{c}^{6}{d}^{3}
-1992{c}^{6}{d}^{5}
-36928{c}^{8}{d}^{3}
+43264{c}^{8}{d}^{2}
\nonumber \\ &&
+14624{c}^{8}{d}^{4}
-2720{c}^{8}{d}^{5}
-22272{c}^{6}d
+40352{c}^{6}{d}^{2}
+144{c}^{6}{d}^{6}
+192{c}^{8}{d}^{6}
\Biggr]
\nonumber \\ &&
- V_{000}
{\frac {2 \left (d-2\right )
\left({d}^{5}-8{d}^{4}+27{d}^{3}-38{d}^{2}-24d+96\right )
\left (c-1\right )\left (c+1\right )
}{3 \left (d-4\right )^{2}\left (d-3\right )\left (d-1\right )^{2}}}
\nonumber \\ &&
+ J_W J_{00}\Biggl\{
{\frac {\left (d-2\right )}{2 d\left (d-4\right )\left (d-1\right )^{2}{c}^{2}}}
\Biggl[
2{c}^{2}{d}^{5}
-24{c}^{2}{d}^{4}
+2{c}^{4}{d}^{4}
-6{c}^{4}{d}^{3}
+60{c}^{2}{d}^{3}
+4{c}^{4}{d}^{2}
\nonumber \\ &&
+34{c}^{2}{d}^{2}
-{d}^{2}
-136{c}^{2}d
+4d
+64{c}^{2}
\Biggr ]
- {\frac {\left (d-2\right )}{2 m_H^{2}m_W^{2}\left (m_H^2-4m_W^2\right ) \left (d-1\right )^{2}}}
\Biggl[
2{d}^{3}m_W^{4}m_H^{2}
\nonumber \\ &&
-8{d}^{3}m_W^{6}
+32{d}^{2}m_W^{6}
-6{d}^{2}m_W^{4}m_H^{2}
-40dm_W^{6}
+4dm_W^{4}m_H^{2}
+m_H^{6}
+16m_W^{6}
-4m_H^{4}m_W^{2}
\Biggr]
\Biggr\}
\nonumber \\ &&
- J_Z J_{00}
\Biggl\{
{\frac {\left (d-2\right )}{6 d\left (d-1\right )^{2}{c}^{2}}}
\Biggl[
-6{c}^{4}{d}^{4}
+4{c}^{6}{d}^{3}
+52{c}^{4}{d}^{3}
-8{c}^{2}{d}^{3}
+36{c}^{6}{d}^{2}
-122{c}^{4}{d}^{2}
+19{c}^{2}{d}^{2}
\nonumber \\ &&
+4{d}^{2}
-7{c}^{2}d
+92{c}^{4}d
-7d
-72{c}^{6}d
+8{c}^{2}
+32{c}^{6}
-16{c}^{4}
\Biggr]
- {\frac {{ m_Z^2}\left (d-2\right )^{2}}{2 { m_H^2}}}
\Biggr\}
\nonumber \\ &&
+ J_H J_{00}
\Biggl\{ \left( 1-\frac{d}{2} \right)
+ {\frac {\left (d-2\right )}{2 m_H^{2}m_W^{2}\left (m_H^2-4m_W^2\right )\left (d-1\right )^{2}}}
\Biggl[
-4{d}^{2}m_W^{6}
+2{d}^{2}m_W^{4}m_H^{2}
+12d{m_W}^{6}
\nonumber \\ &&
-6dm_W^{4}m_H^{2}
+m_H^{6}
-4m_H^{4}m_W^{2}
-8m_W^{6}
+4m_W^{4}m_H^{2}
\Biggr]
\Biggr\}
\nonumber \\ &&
+ J_{00} J_{00}
{\frac {m_W^2 \left (d-2\right )}{6 {c}^{4}\left (d-4\right )\left (d-1\right )^{2}}}
\Biggl[
-16{c}^{4}{d}^{3}
+{c}^{2}{d}^{3}
-8{c}^{2}{d}^{2}
-2{d}^{2}
+8{c}^{6}{d}^{2}
+140{c}^{4}{d}^{2}
+19{c}^{2}d
\nonumber \\ &&
-384{c}^{4}d
+10d
-32{c}^{6}d
-12{c}^{2}
-8
+296{c}^{4}
+24{c}^{6}
\Biggr]
\nonumber \\ &&
+ J_{00} J_{ZW}
{\frac {m_W^2 \left (d-2\right )}{2 {c}^{4}\left (d-1\right )^{2}}}
\Biggl[
4{c}^{6}{d}^{2}
+8{c}^{4}{d}^{2}
-52{c}^{4}d
+5{c}^{2}d
-16{c}^{6}d
+60{c}^{4}
+1
-13{c}^{2}
+12{c}^{6}
\Biggr]
\nonumber  \\ &&
+ J_{00} J_{HW}
{\frac {\left (d-2\right )}{2 \left (m_H^2-4m_W^2\right )\left (d-1\right )^{2}m_W^{2}}}
\Biggl[
4{d}^{2}m_W^{6}
-16dm_W^{6}
-4dm_W^{4}m_H^{2}
+m_H^{4}dm_W^{2}
+m_H^{6}
\nonumber  \\ &&
-9m_H^{4}m_W^{2}
+12m_W^{6}
+20 m_W^{4}m_H^{2}
\Biggr] \;,
\end{eqnarray}


\begin{eqnarray}
&&
- \Biggl( \Pi_Z^{(2)} + \Pi_Z^{(1)} \Pi_Z^{(1)}{}' \Biggr) =
\nonumber \\ &&
- V_{00W}
\Biggl\{
{\frac {\left (d-2\right )}{6 {c}^{2}d\left (d-1\right )^{2}}}
\Biggl[
-24{c}^{4}{d}^{4}
+164{c}^{4}{d}^{3}
-4{c}^{2}{d}^{3}
+8{c}^{6}{d}^{3}
-340{c}^{4}{d}^{2}
-10{c}^{2}{d}^{2}
+72{c}^{6}{d}^{2}
\nonumber \\ &&
+11{d}^{2}
+232{c}^{4}d
-20d
+34{c}^{2}d
-144{c}^{6}d
+16{c}^{2}
+64{c}^{6}
-32{c}^{4}
\Biggr]
+ 2{\frac {{m_Z}^{2}\left (d-2\right )^{2}}{{m_H}^{2}}}
\Biggr\}
\nonumber \\ &&
+ V_{00Z}
\Biggl\{
{\frac {1}{54}}{\frac {\left (d-2\right )}{{c}^{4}d\left (d-1\right )^{2}}}
\Biggl[
-976{c}^{6}{d}^{3}
-760{c}^{2}{d}^{3}
+190{d}^{3}
+352{c}^{8}{d}^{3}
+1248{c}^{4}{d}^{3}
-3872{c}^{8}{d}^{2}
\nonumber \\ &&
+10736{c}^{6}{d}^{2}
-13872{c}^{4}{d}^{2}
-2180{d}^{2}
+8540{c}^{2}{d}^{2}
-11090{c}^{2}d
+4928{c}^{8}d
+17832{c}^{4}d
\nonumber \\ &&
-13664{c}^{6}d
+2885d
+3904{c}^{6}
-4992{c}^{4}
-760
+3040{c}^{2}
-1408{c}^{8}
\Biggr]
\nonumber \\ &&
+
{\frac {\left (d-2\right )\left (5+8{c}^{4}-10{c}^{2}\right )}
{12 {m_H}^{2}{c}^{4}{m_Z}^{2}\left (m_H^2-4m_Z^2\right )\left (d-1\right )^{2}}}
\Biggl[
16{d}^{3}{m_Z}^{6}
-4{d}^{3}{m_Z}^{4}{m_H}^{2}
+12{d}^{2}{m_Z}^{4}{m_H}^{2}
\nonumber \\ &&
-64{d}^{2}{m_Z}^{6}
+80d{m_Z}^{6}
+{m_H}^{6}d-4{m_H}^{4}d{m_Z}^{2}
-8d{m_Z}^{4}{m_H}^{2}
+16{m_H}^{4}{m_Z}^{2}
\nonumber \\ &&
-4{m_H}^{6}
-32{m_Z}^{6}
\Biggr]
\Biggr\}
\nonumber \\ &&
- F_{0000W}
{\frac {m_Z^4\left (d-2\right )\left (2{c}^{2}+2{c}^{4}-1\right )
\left (2{c}^{4}-{c}^{2}d+8{c}^{2}+2\right )}{3 \left (d-1\right ){c}^{2}}}
\nonumber \\ &&
- 8 F_{W0W00}{\frac {m_Z^4 {c}^{4}\left (2+{c}^{2}\right )\left (d-2\right )}{d-1}}
- V_{H00Z}
\Biggl\{
{\frac {m_Z^2 \left (d-2\right )^{2}\left (-10{c}^{2}+5+8{c}^{4}\right )}{3 {c}^{4}\left (d-1\right )}}
\nonumber \\ &&
+
{\frac {\left (d-2\right )\left (5+8{c}^{4}-10{c}^{2}\right )}{12 {m_Z}^{2}{c}^{4}
\left (m_H^2-4m_Z^2\right )\left (d-1\right )^{2}}}
\Biggl[
-8{d}^{2}{m_Z}^{6}
+{m_H}^{6}d
+24d{m_Z}^{4}{m_H}^{2}
-10{m_H}^{4}d{m_Z}^{2}
\nonumber \\ &&
+32d{m_Z}^{6}
-4{m_H}^{6}
+34{m_H}^{4}{m_Z}^{2}
-24{m_Z}^{6}
-72{m_Z}^{4}{m_H}^{2}
\Biggr]
\Biggr\}
\nonumber \\ &&
+ V_{W00W}
{\frac {m_Z^2 \left (d-2\right )}{2 \left (d-1\right )^{2}{c}^{2}}}
\Biggl[
44{c}^{4}{d}^{2}
+24{c}^{6}{d}^{2}
-8{c}^{2}{d}^{2}
-80{c}^{6}d
-204{c}^{4}d
-d
+42{c}^{2}d
\nonumber \\ &&
+4
-58{c}^{2}
+56{c}^{6}
+208{c}^{4}
\Biggr]
\nonumber \\ &&
+ J_{00W}
\frac{1}{6 d\left (d-1\right )^{2}{c}^{2}}
\Biggl[
-192
-1008{c}^{6}{d}^{2}
-28{c}^{2}d
-358{c}^{2}{d}^{2}
+608{c}^{4}d
+512d
+256{c}^{2}
\nonumber \\ &&
-512{c}^{4}
-512{c}^{6}
-88{c}^{4}{d}^{2}
-478{d}^{2}
+1344{c}^{6}d
+159{d}^{3}
-148{c}^{4}{d}^{3}
+152{c}^{6}{d}^{3}
+266{c}^{2}{d}^{3}
\nonumber \\ &&
+24{c}^{6}{d}^{4}
-52{c}^{2}{d}^{4}
+44{c}^{4}{d}^{4}
-16{d}^{4}
\Biggr]
\nonumber \\ &&
+ J_{00W}'
{\frac { m_Z^2 2 \left (c-1\right )\left (c+1\right )}{3 d\left (d-1\right )^{2}{c}^{2}}}
\Biggl[
32{c}^{6}
+36{c}^{6}{d}^{2}
-72{c}^{6}d
+4{c}^{6}{d}^{3}
+16{c}^{4}
-48{c}^{4}{d}^{2}
\nonumber \\ &&
+20{c}^{4}{d}^{3}
+24{c}^{4}d
+18{c}^{2}d
-15{c}^{2}{d}^{2}
-8{c}^{2}
+2{c}^{2}{d}^{3}
-24d
-2{d}^{3}
+8
+18{d}^{2}
\Biggr]
\nonumber \\ &&
+ J_{00H}
{\frac {\left (8{c}^{4}+5-10{c}^{2}\right )\left (d-2\right )}{6 {c}^{4}\left (d-1\right )^{2}}}
+ J_{00H}'
{ m_Z^2 4 \frac {\left (8{c}^{4}+5-10{c}^{2}\right )\left (d-2\right )}{3 \left (d-1\right ){c}^{4}}}
\nonumber \\ &&
+ J_{00H}
{\frac {\left (d-2\right )\left (5+8{c}^{4}-10{c}^{2}\right )
\left (3d-8\right )
\left (4d{m_Z}^{4}
-4{m_H}^{2}{m_Z}^{2}
+{m_H}^{4}-4{m_Z}^{4}\right )}{12 {c}^{4}{m_Z}^{2}\left (m_H^2-4m_Z^2\right )\left (d-1\right )^{2}}}
\nonumber \\ &&
+ J_{00H}'
{\frac {\left (d-2\right )
\left (
12d{m_Z}^{6}
+4{m_Z}^{4}{m_H}^{2}
+{m_H}^{6}
-5{m_H}^{4}{m_Z}^{2}
-12{m_Z}^{6}
\right )
\left (5+8{c}^{4}-10{c}^{2}\right )}{3 {c}^{4}{m_Z}^{2}\left (m_H^2-4m_Z^2\right )\left (d-1\right )^{2}}}
\nonumber \\ &&
- J_{00Z}
{\frac {\left (7{d}^{3}+168d-70{d}^{2}-80\right )
\left (-380{c}^{2}+95+624{c}^{4}-488{c}^{6}+176{c}^{8}\right )
}{27 {c}^{4}d\left (d-1\right )}}
\nonumber \\ &&
- V_{000}
{\frac {8}{27}}{\frac {\left (d-2\right )\left (d-3\right )\left ({
d}^{2}-4d+8\right )\left (c-1\right )\left (c+1\right )\left (44{c
}^{4}-61{c}^{2}+26\right )}{{c}^{2}\left (d-4\right )^{2}\left (d-1
\right )}}
\nonumber \\ &&
- F_{0000Z}{\frac {m_Z^4}{54}}{\frac {\left (d-2\right )\left (d-12\right )\left (
-380{c}^{2}+95+624{c}^{4}-488{c}^{6}+176{c}^{8}\right )}{
\left (d-1\right ){c}^{4}}}
\nonumber \\ &&
+ J_W J_{00}
\Biggl\{
{\frac {\left (d-2\right )}{3 \left (d-1\right )^{2}{c}^{4}d}}
\Biggl[
16{c}^{8}{d}^{4}
-20{c}^{6}{d}^{4}
+10{c}^{4}{d}^{4}
-34{c}^{2}{d}^{3}
+176{c}^{6}{d}^{3}
-32{c}^{4}{d}^{3}
\nonumber \\ &&
+2{d}^{3}
-160{c}^{8}{d}^{3}
+368{c}^{8}{d}^{2}
-23{d}^{2}
-114{c}^{4}{d}^{2}
+172{c}^{2}{d}^{2}
-340{c}^{6}{d}^{2}
+120{c}^{6}d
\nonumber \\ &&
+268{c}^{4}d
+34d
-204{c}^{2}d
-224{c}^{8}d
-8
-32{c}^{4}
+16{c}^{2}
\Biggr]
- {\frac {{m_Z}^{2}\left (d-2\right )^{2}\left (5+8{c}^{4}-10{c}^{2}\right )}{3 {c}^{2}{m_H}^{2}}}
\Biggr\}
\nonumber \\ &&
+ J_Z J_{00}
\Biggl\{
{\frac {\left (d-2\right )}{54 \left (d-1\right )^{2}{c}^{4}d}}
\Biggl[
-760{c}^{2}{d}^{3}
-976{c}^{6}{d}^{3}
+352{c}^{8}{d}^{3}
+1248{c}^{4}{d}^{3}
+190{d}^{3}
\nonumber \\ &&
-13800{c}^{4}{d}^{2}
-2135{d}^{2}
-3872{c}^{8}{d}^{2}
+10736{c}^{6}{d}^{2}
+8450{c}^{2}{d}^{2}
-13664{c}^{6}d
+2795d
\nonumber \\ &&
+17688{c}^{4}d
-10910{c}^{2}d
+4928{c}^{8}d
-760
-1408{c}^{8}
+3040{c}^{2}
-4992{c}^{4}
+3904{c}^{6}
\Biggr]
\nonumber \\ &&
-{\frac {\left (d-2\right )
\left (5+8{c}^{4}-10{c}^{2}\right )}{6 {c}^{4}{m_H}^{2}\left (m_H^2-4m_Z^2\right )
\left (d-1\right )^{2}{m_Z}^{2}}}
\Biggl[
{d}^{3}{m_Z}^{4}{m_H}^{2}
-4{d}^{3}{m_Z}^{6}
-2{d}^{2}{m_Z}^{4}{m_H}^{2}
+16{d}^{2}{m_Z}^{6}
\nonumber \\ &&
-20d{m_Z}^{6}
-d{m_Z}^{4}{m_H}^{2}
+2{m_Z}^{4}{m_H}^{2}
-4{m_H}^{4}{m_Z}^{2}
+{m_H}^{6}
+8{m_Z}^{6}
\Biggr]
\Biggr\}
\nonumber \\ &&
- J_H J_{00}
\Biggl\{
{\frac {\left (8{c}^{4}+5-10{c}^{2}\right )\left (d-2\right )}{6 {c}^{4}}}
- {\frac {\left (d-2\right )\left (5+8{c}^{4}-10{c}^{2}\right )}
{6 {c}^{4}{m_H}^{2}\left (m_H^2-4 m_Z^2 \right ) \left (d-1\right )^{2}{m_Z}^{2}}}
\Biggl[
2{d}^{2}{m_Z}^{4}{m_H}^{2}
\nonumber \\ &&
-4{d}^{2}{m_Z}^{6}
+12d{m_Z}^{6}
-6d{m_Z}^{4}{m_H}^{2}
+{m_H}^{6}
-4{m_H}^{4}{m_Z}^{2}
+4{m_Z}^{4}{m_H}^{2}
-8{m_Z}^{6}
\Biggr]
\Biggr\}
\nonumber \\ &&
-J_{00}J_{00}
{\frac {m_Z^2}{54}}{\frac {\left (d-2\right )}{{c}^{4}\left (d-4\right )\left (d-1\right )^{2}}}
\Biggl[
1920{c}^{4}{d}^{3}
-1816{c}^{6}{d}^{3}
+768{c}^{8}{d}^{3}
+205{d}^{3}
-942{c}^{2}{d}^{3}
\nonumber \\ &&
-14244{c}^{4}{d}^{2}
-1070{d}^{2}
-6696{c}^{8}{d}^{2}
+14840{c}^{6}{d}^{2}
+5928{c}^{2}{d}^{2}
-43120{c}^{6}d
+1345d
\nonumber \\ &&
+36972{c}^{4}d
-12074{c}^{2}d
+20360{c}^{8}d
-1380
-21344{c}^{8}
+12488{c}^{2}
+46224{c}^{6}
-38688{c}^{4}
\Biggr]
\nonumber \\ &&
+ J_{00}J_{WW}
{\frac {m_Z^2\left (d-2\right )}{6 \left (d-1\right )^{2}{c}^{4}}}
\Biggl[
-40{c}^{2}{d}^{2}
-144{c}^{6}{d}^{2}
+16{c}^{8}{d}^{2}
+140{c}^{4}{d}^{2}
+64{c}^{10}{d}^{2}
+920{c}^{6}d
\nonumber \\ &&
-808{c}^{4}d
-5d
+200{c}^{2}d
+16{c}^{8}d
-512{c}^{10}d
+10-230{c}^{2}
+448{c}^{10}
+224{c}^{8}
\nonumber \\ &&
-1240{c}^{6}
+968{c}^{4}
\Biggr]
\nonumber \\ &&
+ J_{00}J_{HZ}
{\frac {\left (d-2\right )\left (5+8{c}^{4}-10{c}^{2}\right )}
{6 {c}^{4}\left (m_H^2-4m_Z^2\right ) \left (d-1\right )^{2}{m_Z}^{2}}}
\Biggl[
4{d}^{2}{m_Z}^{6}
+{m_H}^{4}d{m_Z}^{2}
-16d{m_Z}^{6}
-4d{m_Z}^{4}{m_H}^{2}
\nonumber \\ &&
+{m_H}^{6}
+12{m_Z}^{6}
+20{m_Z}^{4}{m_H}^{2}
-9{m_H}^{4}{m_Z}^{2}
\Biggr]  \;,
\end{eqnarray}
where

\begin{eqnarray}
F_{ABCDE} & = &
\int \frac{d^d k_1}{\Gamma(1+\ep)} \frac{d^d k_2}{\Gamma(1+\ep)}
P^{(1)}(k_1,m_A) P^{(1)}(k_2,m_B)
\nonumber \\ &&
P^{(1)}(k_1-p,m_C)
P^{(1)}(k_2-p,m_D)
P^{(1)}(k_1-k_2,m_E) \;,
\nonumber \\
V_{ABCD} & = &
\int \frac{d^d k_1}{\Gamma(1+\ep)} \frac{d^d k_2}{\Gamma(1+\ep)}
P^{(1)}(k_1-p,m_A)
P^{(1)}(k_1-k_2,m_B)
\nonumber \\ &&
P^{(1)}(k_2,m_C)
P^{(1)}(k_1,m_D) \;,
\nonumber \\
J_{00A} & = &
\int \frac{d^d k_1}{\Gamma(1+\ep)} \frac{d^d k_2}{\Gamma(1+\ep)}
P^{(1)}(k_1-p,0)
P^{(1)}(k_1-k_2,0)
P^{(1)}(k_2,m_A) \;,
\nonumber \\
J_{00A}' & = &
\int \frac{d^n k_1}{\Gamma(1+\ep)} \frac{d^n k_2}{\Gamma(1+\ep)}
P^{(1)}(k_1-p,0)
P^{(1)}(k_1-k_2,0)
P^{(2)}(k_2,m_A) \;,
\nonumber \\
V_{ABC} & = &
\int \frac{d^d k_1}{\Gamma(1+\ep)} \frac{d^d k_2}{\Gamma(1+\ep)}
P^{(1)}(k_1,m_A)
P^{(1)}(k_1-k_2,m_B)
P^{(1)}(k_2,m_C) \;,
\nonumber \\
J_{AB} & = &
\int \frac{d^d k_1}{\Gamma(1+\ep)}
P^{(1)}(k_1-p,m_A) P^{(1)}(k_1,m_B) \;,
\nonumber \\
J_{A} & = &
\int \frac{d^d k_1}{\Gamma(1+\ep)} P^{(1)}(k_1,m_A) \;,
\end{eqnarray}
with $c=m_W/m_Z=\cos \Theta_W$ and $P^{(\sigma)}(p,m) =
1/(p^2+m^2)^\sigma$ and set $\{A,B,C,D,E\}$ denotes the particles.  The
external momentum belongs to the mass shell.

\section{$\overline{{\rm MS}}$ vs. pole masses at two-loop order}
\setcounter{equation}{0}

After expansion of the diagrams with respect to $\sin^2 \theta_W$ we
get rid of one of the boson masses and write the functions $X^{(2)}_V$
introduced in (\ref{result}) in the form
\begin{equation}
  X^{(2)}_V = \frac{m_H^4}{m_W^4}\sum_{k=0}^{4} \sin^{2k}\theta_W\, A^V_k.
\label{X2V}
\end{equation}
In particular for the $Z$ boson
propagator we eliminate $m_W$ and vice versa.  Consequently, the
coefficients $A_i^V$ in the above formula are functions of the Higgs and top
masses and one of the boson masses.  We expand this function with respect to
$m_V^2/m_{H,t}^2$
$$
A_i^V = \sum_{j=0}^5 B_{i,j}^V(m_H^2,m_t^2) (m_V^2)^j
$$
and calculate analytically the first sixth coefficients~\footnote{
In parameterization (\ref{X2V}) $B_{j,0} = 0$ for $j > 0$.}.
This is not a naive Taylor expansion. The general
rules for asymptotic expansions \cite{asymptotic} allow us to extract
also logarithmic dependences, or in other words, to preserve all
analytical properties of the original diagrams.
In the result of the asymptotic expansion all propagator
diagrams are reduced to single scale massive diagrams
(including the two-loop bubbles).
In the finite part, we meet the following constants and functions:
\begin{eqnarray}
S_0 & = & \frac{\pi}{\sqrt{3}} \sim 1.813799365...,
\hspace{3cm}
S_1 = \frac{\pi}{\sqrt{3}} \ln 3 \sim 1.992662272...,
\nonumber \\
S_2 & = &  \frac{4}{9} \frac{\Cl{2}{\tfrac{\pi}{3}}}{\sqrt{3}}
\sim 0.260434137632162...,
~~~~~
S_3 = \pi \Cl{2}{\tfrac{\pi}{3}} \sim 3.188533097...
\end{eqnarray}
Furthermore, $\ln(m_V^2)$ denotes
$\ln\left( m_V^2/\mu^2 \right)$ where $\mu$ is the 't Hooft
scale. We also introduce the notation

\begin{eqnarray}
&&
z_H = \frac{m_Z^2}{m_H^2},
\quad
z_t = \frac{m_Z^2}{m_t^2},
\quad
\omega_H = \frac{m_W^2}{m_H^2},
\quad
\omega_t = \frac{m_W^2}{m_t^2},
\nonumber \\ &&
t_H = \frac{m_t^2}{m_H^2},
\quad
y = \frac{m_t^2}{m_H^2-m_t^2} ,
\quad
r = \frac{m_t^2}{m_H^2-4m_t^2} ,
\nonumber \\ &&
\mbox{L}_2 = \Li{2}{\frac{m_H^2}{m_t^2}} ,
\quad
\mbox{L}_1 = \ln \left( 1- \frac{m_H^2}{m_t^2} \right) ,
\quad
F = \Phi\left(\frac{m_H^2}{4m_t^2} \right) ,
\end{eqnarray}
where
$N_m$ is a number of massless fermion families.
and
$\Phi(z)$ is the finite part of two-loop bubble master integrals
defined in  \cite{DT93}. We rewrite is as follow (see \cite{bastei_ep}-\cite{DK2001}):
\begin{eqnarray}
\Phi(z) =
2 \frac{1-\eta}{1+\eta} \left[ \Li{2}{\eta}-\Li{2}{\frac{1}{\eta}} \right] \;, \quad
\eta = \frac{1-\sqrt{\frac{z}{z-1}}}{1+\sqrt{\frac{z}{z-1}}}.
\end{eqnarray}
Below, we present the coefficients for ${\rm Re}\; A^V_{i,fermion}$
including fermionic contribution only, which are defined as $$
A^V_{i,full} = A^V_{i,boson} + A^V_{i,fermion} $$ and $ A^V_{i,boson}$
given in Appendix~D of \cite{I}. In the following we present the
coefficients in the form $A^V_{i,j}=B_{i,j}^V(m_H^2,m_t^2)\;(m_V^2)^j$.
\subsection{The expansion coefficients for the $W$}
\setcounter{equation}{0}

\begin{eqnarray}
&&
A^W_{0,0} =
{\frac {9}{8}} \ln (\omega_H)\ln (m_W^2)t_H
-{\frac {9}{8}}\ln (\omega_t)\ln (m_W^2)t_H
+3/8{\pi }^{2}t_H
-{\frac {11}{16}}{\pi }^{2}{t_H}^{2}
+1/2{\pi }^{2}{t_H}^{3}
\nonumber \\ &&
-{\frac {9}{16}}\mbox{L}_1\ln (\omega_H)
+{\frac {9}{16}}\mbox{L}_1\ln (\omega_t)
-{\frac {57}{16}}\ln (\omega_H)t_H
+{\frac {57}{8}}\ln (\omega_H){t_H}^{2}
+{\frac {9}{16}}\ln (\omega_t) t_H
\nonumber \\ &&
-{\frac {57}{4}}\ln (\omega_t){t_H}^{2}
+15\ln (\omega_t){t_H}^{3}
+36\ln (\omega_t){t_H}^{4}
+{\frac {9}{8}}\left (\ln (\omega_t)\right )^{2}t_H
-{\frac {117}{16}}\left (\ln (\omega_t)\right )^{2}{t_H}^{2}
\nonumber \\ &&
-9/2\left (\ln (\omega_t)\right )^{2}{t_H}^{3}
+36\left (\ln (\omega_t)\right )^{2}{t_H}^{4}
+3\ln (m_W^2)t_H
+{\frac {57}{8}}\ln (m_W^2){t_H}^{2}
-15\ln (m_W^2){t_H}^{3}
\nonumber \\ &&
-36\ln (m_W^2){t_H}^{4}
-{\frac {117}{16}}\left (\ln (m_W^2)\right )^{2}{t_H}^{2}
-9/2\left (\ln (m_W^2)\right )^{2}{t_H}^{3}
+36\left (\ln (m_W^2)\right )^{2}{t_H}^{4}
\nonumber \\ &&
+{\frac {9}{16}}Ft_H
+{\frac {45}{8}}F{t_H}^{2}
-{\frac {27}{2}}F{t_H}^{3}
-9/4\mbox{L}_2t_H
+{\frac {45}{16}}\mbox{L}_2{t_H}^{2}
-{\frac {9}{8}}\mbox{L}_2{t_H}^{3}
-9/4\mbox{L}_1\ln (\omega_t)t_H
\nonumber \\ &&
-{\frac {81}{16}}t_H
+{\frac {9}{16}}\mbox{L}_2
+9/4\mbox{L}_1\ln (\omega_H)t_H
-{\frac {45}{16}}\mbox{L}_1\ln (\omega_H){t_H}^{2}
+{\frac {9}{8}}\mbox{L}_1\ln (\omega_H){t_H}^{3}
\nonumber \\ &&
+{\frac {45}{16}}\mbox{L}_1\ln (\omega_t){t_H}^{2}
-{\frac {9}{8}}\mbox{L}_1\ln (\omega_t){t_H}^{3}
-{\frac {9}{8}}\ln (\omega_h)\ln (\omega_t)t_H
+{\frac {117}{8}}\ln (\omega_t)\ln (m_W^2){t_H}^{2}
\nonumber \\ &&
+9\ln (\omega_t)\ln (m_W^2){t_H}^{3}
-72\ln (\omega_t)\ln (m_W^2){t_H}^{4}
+27{t_H}^{3}
-{\frac {3}{32}}{\pi }^{2}
+{\frac {213}{64}}{t_H}^{2}
-{\frac {9}{32}}F
\end{eqnarray}
\begin{eqnarray}
&&
A^W_{0,1} =
1/4\mbox{L}_2\omega_H{t_H}^{2}
+{\frac {177}{2}}\ln (m_W^2)\omega_H{t_H}^{2}
-{\frac {339}{16}}\ln (m_W^2)\omega_ht_H
-5F\omega_H{t_H}^{3}
\nonumber \\ &&
+{\frac {925}{48}}\ln (\omega_t)\omega_Ht_H
-3/4\ln (\omega_H)\ln (\omega_t)\omega_H
-2/3\ln (\omega_H)\omega_HN_m
+3/4\ln (\omega_t)\ln (m_W^2)\omega_H
\nonumber \\ &&
+{\frac {113}{16}}\left (\ln (m_W^2)\right )^{2}\omega_ht_H
+16/3\omega_H{t_H}^{2}N_m-1/2\mbox{L}_1\ln (\omega_H)\omega_Ht_H
-1/4\mbox{L}_1\ln (\omega_H)\omega_H{t_H}^{2}
\nonumber \\ &&
+{\frac {37}{24}}\mbox{L}_1\ln (\omega_H)\omega_H{t_H}^{3}
+1/2\mbox{L}_1\ln (\omega_t)\omega_Ht_H
+1/4\mbox{L}_1\ln (\omega_t)\omega_H{t_H}^{2}
-{\frac {37}{24}}\mbox{L}_1\ln (\omega_t)\omega_H{t_H}^{3}
\nonumber \\ &&
+3/8\mbox{L}_1\ln (\omega_h)\omega_t
-{\frac {79}{48}}\ln (\omega_H)\omega_Ht_H
+{\frac {11}{12}}\mbox{L}_1\ln (\omega_t)\omega_H
-{\frac {11}{72}}{\pi }^{2}\omega_H
+1/16{\pi }^{2}\omega_t
+{\frac {977}{192}}\omega_ht_H
\nonumber \\ &&
-{\frac {1451}{24}}\omega_H{t_H}^{2}
-{\frac {7}{9}}\omega_HN_m
-{\frac {1621}{24}}\ln (\omega_t)\omega_H{t_H}^{2}
+{\frac {11}{6}}\ln (m_W^2)\omega_HN_m
+6\left (\ln (\omega_t)\right )^{2}\omega_H{t_H}^{2}
\nonumber \\ &&
-{\frac {33}{16}}\omega_Ht_HS_0
-{\frac {11}{12}}\mbox{L}_1\ln (\omega_h)\omega_H
+{\frac {41}{12}}F\omega_H{t_H}^{2}
+{\frac {23}{36}}\omega_H
-{\frac {43}{2}}\left (\ln (m_W^2)\right )^{2}\omega_H{t_H}^{2}
\nonumber \\ &&
+\ln (\omega_H)\ln (m_W^2)\omega_H
-\left (\ln (m_W^2)\right )^{2}\omega_H
-{\frac {7}{48}}{\pi }^{2}\omega_ht_H
-2/3\omega_Ht_HN_m
+{\frac {99}{2}}\omega_h{t_H}^{2}S_0
\nonumber \\ &&
-3/8\mbox{L}_1\ln (\omega_t)\omega_t
+{\frac {277}{24}}\ln (\omega_H)\omega_H{t_H}^{2}
+3/16\left (\ln (\omega_t)\right )^{2}\omega_Ht_H
-{\frac {5}{24}}F\omega_ht_H
+1/2\mbox{L}_2\omega_Ht_H
\nonumber \\ &&
-{\frac {37}{24}}\mbox{L}_2\omega_H{t_H}^{3}
+\ln (\omega_H)\ln (m_W^2)\omega_HN_m
-\left (\ln (m_W^2)\right )^{2}\omega_HN_m
-1/24\ln (\omega_H)\omega_H
\nonumber \\ &&
-1/2\ln (\omega_t)\omega_H
+5/6\ln (m_W^2)\omega_H
-5/6F\omega_H
+3/16F\omega_t
+{\frac {11}{12}}\mbox{L}_2\omega_H
-3/8\mbox{L}_2\omega_t
\nonumber \\ &&
-5/4\ln (\omega_H)\ln (\omega_t)\omega_ht_H
+5\ln (\omega_H)\ln (\omega_t)\omega_H{t_H}^{2}
+5/4\ln (\omega_H)\ln (m_W^2)\omega_Ht_H
\nonumber \\ &&
-5\ln (\omega_h)\ln (m_W^2)\omega_H{t_H}^{2}
-{\frac {99}{8}}\ln (\omega_t)\omega_Ht_HS_0
-4/3\ln (\omega_t)\omega_Ht_HN_m
+{\frac {99}{2}}\ln (\omega_t)\omega_H{t_H}^{2}S_0
\nonumber \\ &&
+16/3\ln (\omega_t)\omega_H{t_H}^{2}N_m
-{\frac {29}{4}}\ln (\omega_t)\ln (m_W^2)\omega_Ht_H
+{\frac {31}{2}}\ln (\omega_t)\ln (m_W^2)\omega_H{t_H}^{2}
\nonumber \\ &&
+{\frac {99}{8}}\ln (m_W^2)\omega_Ht_HS_0
+7/3\ln (m_W^2)\omega_Ht_HN_m
-{\frac {99}{2}}\ln (m_W^2)\omega_H{t_H}^{2}S_0
\nonumber \\ &&
-{\frac {40}{3}}\ln (m_W^2)\omega_h{t_H}^{2}N_m
-2\left (\ln (m_W^2)\right )^{2}\omega_ht_HN_m
+8\left (\ln (m_W^2)\right )^{2}\omega_h{t_H}^{2}N_m
\nonumber \\ &&
+2\ln (\omega_t)\ln (m_W^2)\omega_Ht_HN_m
-8\ln (\omega_t)\ln (m_W^2)\omega_H{t_H}^{2}N_m
\end{eqnarray}
\begin{eqnarray}
&&
A^W_{0,2} =
-{\frac {1}{10368}}\omega_H
\Biggl(
-12960\mbox{L}_2\omega_H{t_H}^{2}
-95904\ln (m_W^2)\omega_H{t_H}^{2}
-98172\ln (m_W^2)\omega_Ht_H
\nonumber \\ &&
+139968F\omega_H{t_H}^{3}
+162\omega_Hy
-5292\ln (\omega_t)\omega_Ht_H
-17280\ln (\omega_h)\ln (\omega_t)\omega_H
\nonumber \\ &&
-26880\ln (\omega_H)\omega_HN_m
-97632\ln (\omega_t)\ln (m_W^2)\omega_H
-139968\left (\ln (m_W^2)\right )^{2}\omega_Ht_H
\nonumber \\ &&
-6480\mbox{L}_1\ln (\omega_H)\omega_Ht_H
+12960\mbox{L}_1\ln (\omega_H)\omega_H{t_H}^{2}
-28080\mbox{L}_1\ln (\omega_H)\omega_H{t_H}^{3}
\nonumber \\ &&
+6480\mbox{L}_1\ln (\omega_t)\omega_Ht_H
-12960\mbox{L}_1\ln (\omega_t)\omega_H{t_H}^{2}
+28080\mbox{L}_1\ln (\omega_t)\omega_H{t_H}^{3}
\nonumber \\ &&
+1296\mbox{L}_1\ln (\omega_H)\omega_t
+39312\ln (\omega_h)\omega_Ht_H
+4320\mbox{L}_1\ln (\omega_t)\omega_H
-12024{\pi }^{2}\omega_H
\nonumber \\ &&
+216{\pi }^{2}\omega_t
+276426\omega_Ht_H
-516672\omega_H{t_H}^{2}
+1144080\omega_HN_m
-866052\omega_HS_2
\nonumber \\ &&
-15984\ln (\omega_t)\omega_H{t_H}^{2}
-666816\ln (m_W^2)\omega_HN_m
-4320\mbox{L}_1\ln (\omega_h)\omega_H
-69552F\omega_H{t_H}^{2}
\nonumber \\ &&
+886115\omega_H
+3888\omega_t
+559872\left (\ln (m_W^2)\right )^{2}\omega_H{t_H}^{2}
-116424\omega_HS_0
\nonumber \\ &&
+23040\ln (\omega_H)\ln (m_W^2)\omega_H
+133056\left (\ln (m_W^2)\right )^{2}\omega_H
+15552{\pi }^{2}\omega_Ht_H
-1296\mbox{L}_1\ln (\omega_t)\omega_t
\nonumber \\ &&
-214704\ln (\omega_H)\omega_H{t_H}^{2}
+69984\left (\ln (\omega_t)\right )^{2}\omega_Ht_H
+27648F\omega_Ht_H
+6480\mbox{L}_2\omega_Ht_H
\nonumber \\ &&
+28080\mbox{L}_2\omega_H{t_H}^{3}
+23040\ln (\omega_H)\ln (m_W^2)\omega_HN_m
+114624\left (\ln (m_W^2)\right )^{2}\omega_HN_m
\nonumber \\ &&
-20742\ln (\omega_H)\omega_H
+287154\ln (\omega_t)\omega_H
-608340\ln (m_W^2)\omega_H
-7344F\omega_H
+648F\omega_t
\nonumber \\ &&
+4320\mbox{L}_2\omega_h
-1296\mbox{L}_2\omega_t
+34992\ln (\omega_H)\ln (\omega_t)\omega_ht_H
-93312\ln (\omega_H)\ln (\omega_t)\omega_H{t_H}^{2}
\nonumber \\ &&
-34992\ln (\omega_H)\ln (m_W^2)\omega_Ht_H
+93312\ln (\omega_H)\ln (m_W^2)\omega_H{t_H}^{2}
+69984\ln (\omega_t)\ln (m_W^2)\omega_Ht_H
\nonumber \\ &&
-559872\ln (\omega_t)\ln (m_W^2)\omega_H{t_H}^{2}
-55296\omega_HS_3
+43008\omega_h{N_m}^{2}\ln (m_W^2)
\nonumber \\ &&
-18432\omega_H{N_m}^{2}\left (\ln (m_W^2)\right )^{2}
-6156\omega_H\ln (\omega_t)y
+103680\omega_H\zeta (3)
-1296\omega_H\left (\ln (\omega_H)\right )^{2}
\nonumber \\ &&
+23760\omega_H\left (\ln (\omega_t)\right )^{2}
-20480\omega_H{N_m}^{2}
+2916\ln (\omega_H)\omega_t-2916\ln (m_W^2)\omega_t
\nonumber \\ &&
+228096\omega_HN_mS_0\ln (m_W^2)
+27648\omega_H\ln (\omega_t)\ln (m_W^2)N_m
-99792\omega_h\ln (\omega_t)S_0
\nonumber \\ &&
-32256\omega_H\ln (\omega_t)N_m
-42624\omega_HN_m{\pi }^{2}
-5184\omega_H\left (\ln (\omega_h)\right )^{2}N_m
+6156\omega_H\ln (\omega_H)y
\nonumber \\ &&
+162\omega_h\ln (\omega_H){y}^{2}
+18432\omega_H{N_m}^{2}{\pi }^{2}
+414720\omega_HN_m\zeta (3)
-162\omega_H\ln (\omega_t){y}^{2}
\nonumber \\ &&
+228096\omega_HS_0\ln (m_W^2)
-38016\omega_HN_mS_0
-221184\omega_HN_mS_3
-1539648\omega_hN_mS_2
\Biggr )
\end{eqnarray}
\begin{eqnarray}
&&
A^W_{1,1} =
{\frac {1}{576}}t_H\omega_H
\Biggl (
-476{\pi }^{2}-972S_0
+8259
-9504t_HS_0
-3552t_H+2376\ln (\omega_t)S_0
\nonumber \\ &&
+336\ln (\omega_t)
-9504\ln (\omega_t)t_HS_0
-3552\ln (\omega_t)t_H
+612\left (\ln (\omega_t)\right )^{2}
-2304\left (\ln (\omega_t)\right )^{2}t_H
\nonumber \\ &&
-576\ln (\omega_t)\ln (m_W^2)
+2016\ln (\omega_t)\ln (m_W^2)t_H
-2376\ln (m_W^2)S_0
-660\ln (m_W^2)
\nonumber \\ &&
+9504\ln (m_W^2)t_HS_0+960\ln (m_W^2)t_H
-36\left (\ln (m_W^2)\right )^{2}
+288\left (\ln (m_W^2)\right )^{2}t_H
\Biggr )
\end{eqnarray}

\begin{eqnarray}
&&
A^W_{1,2} =
-{\frac {1}{1296}}{\omega_H}^{2}
\Biggl (
-1128{\pi }^{2}N_m
+576\ln (m_W^2)N_m
-1296\left (\ln (m_W^2)\right )^{2}N_m
+10368\zeta (3)N_m
\nonumber \\ &&
+5832\ln (\omega_t)\ln (m_W^2)t_H
-24300\ln (\omega_t)t_H
-31104\ln (\omega_t){t_H}^{2}
+5832\left (\ln (\omega_t)\right )^{2}t_H
\nonumber \\ &&
+15552\ln (m_W^2)t_H
+7776\ln (m_W^2){t_H}^{2}
+46656\left (\ln (m_W^2)\right )^{2}{t_H}^{2}
+32238t_H+5364S_0
\nonumber \\ &&
+1452N_m
-46656\ln (\omega_t)\ln (m_W^2){t_H}^{2}
+3969\ln (\omega_t)
+1116\ln (m_W^2)
-899{\pi }^{2}
-27216{t_H}^{2}
\nonumber \\ &&
+1458S_2
+2304S_3
+1728\zeta (3)
-342\left (\ln (\omega_t)\right )^{2}
-1296\left (\ln (m_W^2)\right )^{2}
-11664\left (\ln (m_W^2)\right )^{2}t_H
\nonumber \\ &&
+972\ln (\omega_t)\ln (m_W^2)
-6336S_0N_m
-17496S_2N_m
+9216S_3N_m
-8029
\nonumber \\ &&
-4752\ln (m_W^2)S_0
-4752\ln (m_W^2)S_0N_m
+2700\ln (\omega_t)S_0
\Biggr )
\end{eqnarray}
\begin{eqnarray}
&&
A^W_{2,1} =
-{\frac {1}{576}}t_H\omega_H
\Biggl (
476{\pi }^{2}
+1872S_0
-8619+8640t_HS_0
-6528t_H-2160\ln (\omega_t)S_0
\nonumber \\ &&
+2184\ln (\omega_t)
+8640\ln (\omega_t)t_HS_0
-6528\ln (\omega_t)t_H
-612\left (\ln (\omega_t)\right )^{2}
+2304\left (\ln (\omega_t)\right )^{2}t_H
\nonumber \\ &&
+576\ln (\omega_t)\ln (m_W^2)
-2016\ln (\omega_t)\ln (m_W^2)t_H
+2160\ln (m_W^2)S_0
-1860\ln (m_W^2)
\nonumber \\ &&
-8640\ln (m_W^2)t_HS_0
xsy+9120\ln (m_W^2)t_H
+36\left (\ln (m_W^2)\right )^{2}
-288\left (\ln (m_W^2)\right )^{2}t_H
\Biggr )
\end{eqnarray}

\begin{eqnarray}
&&
A^W_{2,2} =
{\frac {1}{2592}}{\omega_H}^{2}
\Biggl (
-17496\ln (\omega_t)\ln (m_W^2)t_H
+3564\ln (\omega_t)t_H
+139968\ln (\omega_t){t_H}^{2}
\nonumber \\ &&
-17496\left (\ln (\omega_t)\right )^{2}t_H
+22680\ln (m_W^2)t_H
-69984\ln (m_W^2){t_H}^{2}
-139968\left (\ln (m_W^2)\right )^{2}{t_H}^{2}
\nonumber \\ &&
-70686t_H
-63114S_0
-13708N_m
+139968\ln (\omega_t)\ln (m_W^2){t_H}^{2}
-9561\ln (\omega_t)
\nonumber \\ &&
-10632\ln (m_W^2)
+4991{\pi }^{2}
+104976{t_H}^{2}
-113238S_2
+3888\zeta (3)
+90\left (\ln (\omega_t)\right )^{2}
\nonumber \\ &&
+4752\left (\ln (m_W^2)\right )^{2}
+34992\left (\ln (m_W^2)\right )^{2}t_H
-7488\ln (\omega_t)S_0
-1944\ln (\omega_t)\ln (m_W^2)
\nonumber \\ &&
+25344\ln (m_W^2)S_0
-103152S_0N_m
-141912S_2N_m
+4752\left (\ln (m_W^2)\right )^{2}N_m
+5768{\pi }^{2}N_m
\nonumber \\ &&
+25344\ln (m_W^2)S_0N_m
+1728\zeta (3)N_m
-9552\ln (m_W^2)N_m
-228
\Biggr )
\end{eqnarray}

\subsection{The expansion coefficients for the $Z$}

\begin{eqnarray}
&&
A_{0,0}^Z =
-{\frac {9}{8}}\ln (z_t)\ln (z_h)t_h
+{\frac {9}{8}}\ln (z_h)\ln (m_Z^2)t_h
-{\frac {9}{8}}\ln (z_t)\ln (m_Z^2)t_h
-{\frac {5}{16}}{\pi }^{2}{t_h}^{2}
\nonumber \\ &&
+1/2{\pi }^{2}{t_h}^{3}
-{\frac {75}{16}}\ln (z_h)t_h
+6\ln (z_h){t_h}^{2}
+{\frac {57}{16}}\ln (m_Z^2)t_h
+{\frac {105}{16}}\ln (m_Z^2){t_h}^{2}
\nonumber \\ &&
-{\frac {39}{2}}\ln (m_Z^2){t_h}^{3}
-36\ln (m_Z^2){t_h}^{4}
-{\frac {117}{16}}\left (\ln (m_Z^2)\right )^{2}{t_h}^{2}
-9/2\left (\ln (m_Z^2)\right )^{2}{t_h}^{3}
\nonumber \\ &&
+36\left (\ln (m_Z^2)\right )^{2}{t_h}^{4}
+{\frac {9}{8}}\ln (z_t)t_h
-{\frac {201}{16}}\ln (z_t){t_h}^{2}
+{\frac {39}{2}}\ln (z_t){t_h}^{3}
+36\ln (z_t){t_h}^{4}
\nonumber \\ &&
+{\frac {9}{8}}\left (\ln (z_t)\right )^{2}t_h
-{\frac {117}{16}}\left (\ln (z_t)\right )^{2}{t_h}^{2}
-9/2\left (\ln (z_t)\right )^{2}{t_h}^{3}
+36\left (\ln (z_t)\right )^{2}{t_h}^{4}
-{\frac {9}{8}}Ft_h
\nonumber \\ &&
+{\frac {135}{16}}F{t_h}^{2}
-{\frac {27}{2}}F{t_h}^{3}
-{\frac {81}{16}}t_h
+{\frac {117}{8}}\ln (z_t)\ln (m_Z^2){t_h}^{2}
\nonumber \\ &&
+9\ln (z_t)\ln (m_Z^2){t_h}^{3}
-72\ln (z_t)\ln (m_Z^2){t_h}^{4}
+{\frac {9}{64}}{t_h}^{2}
+{\frac {63}{2}}{t_h}^{3}
\end{eqnarray}

\begin{eqnarray}
&&
A_{0,1}^Z =
-3/8\ln (z_t)\ln (z_h)z_h
+{\frac {5}{48}}t_hz_hr
-1/2Ft_hz_h
-{\frac {49}{48}}\ln (z_h)t_hz_h
+10\ln (z_h){t_h}^{2}z_h
\nonumber \\ &&
+5/8\ln (z_h)t_hz_h{r}^{2}
+5/4\ln (z_h)\ln (m_Z^2)t_hz_h
-5\ln (z_h)\ln (m_Z^2){t_h}^{2}z_h
+{\frac {99}{8}}\ln (m_Z^2)t_hz_hS_0
\nonumber \\ &&
+4/3\ln (m_Z^2)t_hz_hN_m
+{\frac {579}{64}}t_hz_h
-{\frac {121}{2}}{t_h}^{2}z_h
-{\frac {7}{9}}z_hN_m
+{\frac {7}{48}}\ln (z_h)z_h
+{\frac {49}{48}}\ln (m_Z^2)z_h
\nonumber \\ &&
-{\frac {7}{16}}\ln (z_t)z_h
+{\frac {409}{128}}F{t_h}^{2}z_h
+{\frac {31}{288}}z_h
+\ln (z_h)\ln (m_Z^2)z_h
-\left (\ln (m_Z^2)\right )^{2}z_h
+3/8{\pi }^{2}t_hz_h
\nonumber \\ &&
-{\frac {33}{4}}t_hz_hS_0
+{\frac {99}{2}}{t_h}^{2}z_hS_0
+\ln (z_h)\ln (m_Z^2)z_hN_m
-\left (\ln (m_Z^2)\right )^{2}z_hN_m
+16/3{t_h}^{2}z_hN_m
\nonumber \\ &&
-{\frac {55}{48}}\ln (z_h)t_hz_hr
-2/3\ln (z_h)z_hN_m
-{\frac {281}{16}}\ln (m_Z^2)t_hz_h
+87\ln (m_Z^2){t_h}^{2}z_h
\nonumber \\ &&
+{\frac {11}{6}}\ln (m_Z^2)z_hN_m
+{\frac {113}{16}}\left (\ln (m_Z^2)\right )^{2}t_hz_h
-{\frac {43}{2}}\left (\ln (m_Z^2)\right )^{2}{t_h}^{2}z_h
+{\frac {205}{12}}\ln (z_t)t_hz_h
\nonumber \\ &&
-{\frac {135}{2}}\ln (z_t){t_h}^{2}z_h
+3/8\ln (z_t)\ln (m_Z^2)z_h
+{\frac {15}{16}}\left (\ln (z_t)\right )^{2}t_hz_h
+3\left (\ln (z_t)\right )^{2}{t_h}^{2}z_h
\nonumber \\ &&
-5F{t_h}^{3}z_h
-{\frac {99}{2}}\ln (m_Z^2){t_h}^{2}z_hS_0
-{\frac {40}{3}}\ln (m_Z^2){t_h}^{2}z_hN_m
-2\left (\ln (m_Z^2)\right )^{2}t_hz_hN_m
\nonumber \\ &&
+8\left (\ln (m_Z^2)\right )^{2}{t_h}^{2}z_hN_m
+{\frac {55}{48}}\ln (z_t)t_hz_hr
-5/8\ln (z_t)t_hz_h{r}^{2}
-{\frac {99}{8}}\ln (z_t)t_hz_hS_0
\nonumber \\ &&
-4/3\ln (z_t)t_hz_hN_m
+{\frac {99}{2}}\ln (z_t){t_h}^{2}z_hS_0
+16/3\ln (z_t){t_h}^{2}z_hN_m
-5/4\ln (z_t)\ln (z_h)t_hz_h
\nonumber \\ &&
+5\ln (z_t)\ln (z_h){t_h}^{2}z_h
-8\ln (z_t)\ln (m_Z^2)t_hz_h
+{\frac {37}{2}}\ln (z_t)\ln (m_Z^2){t_h}^{2}z_h
-{\frac {115}{384}}Ft_hz_hr
\nonumber \\ &&
+{\frac {5}{32}}Ft_hz_h{r}^{2}
+2\ln (z_t)\ln (m_Z^2)t_hz_hN_m
-8\ln (z_t)\ln (m_Z^2){t_h}^{2}z_hN_m
\end{eqnarray}

\begin{eqnarray}
&&
A_{0,2}^Z =
-{\frac {16}{9}}{\pi }^{2}{z_h}^{2}{N_m}^{2}
+11/3{z_h}^{2}S_0N_m
+{\frac {49}{128}}F{z_h}^{2}{r}^{4}
+{\frac {18911}{23040}}\ln (z_h){z_h}^{2}{r}^{2}
\nonumber \\ &&
-{\frac {112}{27}}\ln (m_Z^2){z_h}^{2}{N_m}^{2}
-3/2{\pi }^{2}t_h{z_h}^{2}
+1/2\left (\ln (z_h)\right )^{2}{z_h}^{2}N_m
+{\frac {5673}{10240}}\ln (z_h){z_h}^{2}r
\nonumber \\ &&
-{\frac {27}{4}}\left (\ln (z_t)\right )^{2}t_h{z_h}^{2}
-{\frac {1865}{768}}Ft_h{z_h}^{2}
+{\frac {25979}{30720}}\ln (z_t)t_h{z_h}^{2}
+{\frac {297}{2}}{z_h}^{2}S_2N_m
\nonumber \\ &&
+18\ln (z_h){t_h}^{2}{z_h}^{2}
+{\frac {121}{144}}{\pi }^{2}{z_h}^{2}
-{\frac {487163}{15360}}t_h{z_h}^{2}
+{\frac {377}{8}}{t_h}^{2}{z_h}^{2}
+{\frac {7729}{46080}}{z_h}^{2}r
-{\frac {2557}{11520}}{z_h}^{2}{r}^{2}
\nonumber \\ &&
-22\ln (m_Z^2){z_h}^{2}S_0
+{\frac {64}{3}}S_3{z_h}^{2}N_m
+{\frac {25}{9}}{\pi }^{2}{z_h}^{2}N_m
-40\zeta (3){z_h}^{2}N_m
-{\frac {26047}{6144}}\ln (z_h)t_h{z_h}^{2}
\nonumber \\ &&
-{\frac {433}{384}}\ln (z_h){z_h}^{2}{r}^{3}
+{\frac {49}{32}}\ln (z_h){z_h}^{2}{r}^{4}
+{\frac {70}{27}}\ln (z_h){z_h}^{2}N_m
-{\frac {20}{9}}\ln (z_h)\ln (m_Z^2){z_h}^{2}
\nonumber \\ &&
-{\frac {9}{160}}\ln (z_h)z_tz_h
+{\frac {2343}{160}}\ln (m_Z^2)t_h{z_h}^{2}
+{\frac {37}{4}}\ln (m_Z^2){t_h}^{2}{z_h}^{2}
+{\frac {3473}{54}}\ln (m_Z^2){z_h}^{2}N_m
\nonumber \\ &&
+{\frac {27}{2}}\left (\ln (m_Z^2)\right )^{2}t_h{z_h}^{2}
-54\left (\ln (m_Z^2)\right )^{2}{t_h}^{2}{z_h}^{2}
-{\frac {199}{18}}\left (\ln (m_Z^2)\right )^{2}{z_h}^{2}N_m
\nonumber \\ &&
+{\frac {16}{9}}\left (\ln (m_Z^2)\right )^{2}{z_h}^{2}{N_m}^{2}
+{\frac {9}{160}}\ln (m_Z^2)z_tz_h
+{\frac {27}{8}}\ln (z_h)\ln (m_Z^2)t_h{z_h}^{2}
\nonumber \\ &&
-9\ln (z_h)\ln (m_Z^2){t_h}^{2}{z_h}^{2}
-{\frac {20}{9}}\ln (z_h)\ln (m_Z^2){z_h}^{2}N_m
-22\ln (m_Z^2){z_h}^{2}S_0N_m
\nonumber \\ &&
-{\frac {27}{8}}\ln (z_t)\ln (z_h)t_h{z_h}^{2}
+9\ln (z_t)\ln (z_h){t_h}^{2}{z_h}^{2}
-{\frac {27}{4}}\ln (z_t)\ln (m_Z^2)t_h{z_h}^{2}
\nonumber \\ &&
+54\ln (z_t)\ln (m_Z^2){t_h}^{2}{z_h}^{2}
-4/3\ln (z_t)\ln (m_Z^2){z_h}^{2}N_m
+16/3S_3{z_h}^{2}
-{\frac {2746861}{51840}}{z_h}^{2}
\nonumber \\ &&
+{\frac {49}{192}}{z_h}^{2}{r}^{3}
+{\frac {297}{8}}{z_h}^{2}S_2
+{\frac {55}{6}}{z_h}^{2}S_0
-{\frac {7969}{72}}{z_h}^{2}N_m
+{\frac {160}{81}}{z_h}^{2}{N_m}^{2}
-7\zeta (3){z_h}^{2}
\nonumber \\ &&
+{\frac {8413}{4320}}\ln (z_h){z_h}^{2}
+{\frac {5}{16}}\left (\ln (z_h)\right )^{2}{z_h}^{2}
+{\frac {50209}{864}}\ln (m_Z^2){z_h}^{2}
-{\frac {77}{6}}\left (\ln (m_Z^2)\right )^{2}{z_h}^{2}
\nonumber \\ &&
-{\frac {38411}{1440}}\ln (z_t){z_h}^{2}
+{\frac {35}{48}}\left (\ln (z_t)\right )^{2}{z_h}^{2}
-{\frac {3}{40}}z_tz_h
+3/16F{z_h}^{2}
+{\frac {17}{4}}\ln (z_t){t_h}^{2}{z_h}^{2}
\nonumber \\ &&
-{\frac {5673}{10240}}\ln (z_t){z_h}^{2}r
-{\frac {18911}{23040}}\ln (z_t){z_h}^{2}{r}^{2}
+{\frac {433}{384}}\ln (z_t){z_h}^{2}{r}^{3}
-{\frac {49}{32}}\ln (z_t){z_h}^{2}{r}^{4}
\nonumber \\ &&
+{\frac {55}{4}}\ln (z_t){z_h}^{2}S_0
+{\frac {14}{9}}\ln (z_t){z_h}^{2}N_m
+5/6\ln (z_t)\ln (z_h){z_h}^{2}
+{\frac {113}{24}}\ln (z_t)\ln (m_Z^2){z_h}^{2}
\nonumber \\ &&
+{\frac {53695}{6144}}F{t_h}^{2}{z_h}^{2}
-{\frac {27}{2}}F{t_h}^{3}{z_h}^{2}
+{\frac {89}{768}}F{z_h}^{2}r
+{\frac {1441}{6144}}F{z_h}^{2}{r}^{2}
-{\frac {241}{768}}F{z_h}^{2}{r}^{3}
\end{eqnarray}

\begin{eqnarray}
&&
A_{1,1}^Z =
\ln (z_t)\ln (z_h)z_h
+1/2\ln (z_h)t_hz_h
-{\frac {165}{4}}\ln (m_Z^2)t_hz_hS_0
-8/3\ln (m_Z^2)t_hz_hN_m
\nonumber \\ &&
-{\frac {9919}{288}}t_hz_h
+{\frac {607}{3}}{t_h}^{2}z_h
+{\frac {14}{9}}z_hN_m
-1/3\ln (z_h)z_h
-2\ln (m_Z^2)z_h
+7/6\ln (z_t)z_h
\nonumber \\ &&
+1/4F{t_h}^{2}z_h
-2/9z_h
-2\ln (z_h)\ln (m_Z^2)z_h
+2\left (\ln (m_Z^2)\right )^{2}z_h
-3/8{\pi }^{2}t_hz_h
+{\frac {201}{8}}t_hz_hS_0
\nonumber \\ &&
-165{t_h}^{2}z_hS_0
-2\ln (z_h)\ln (m_Z^2)z_hN_m
+2\left (\ln (m_Z^2)\right )^{2}z_hN_m
-{\frac {32}{3}}{t_h}^{2}z_hN_m
\nonumber \\ &&
+3\ln (z_h)t_hz_hr
+4/3\ln (z_h)z_hN_m
+{\frac {1267}{24}}\ln (m_Z^2)t_hz_h
-{\frac {790}{3}}\ln (m_Z^2){t_h}^{2}z_h
\nonumber \\ &&
-11/3\ln (m_Z^2)z_hN_m
-{\frac {127}{8}}\left (\ln (m_Z^2)\right )^{2}t_hz_h
+57\left (\ln (m_Z^2)\right )^{2}{t_h}^{2}z_h
-{\frac {625}{12}}\ln (z_t)t_hz_h
\nonumber \\ &&
+{\frac {583}{3}}\ln (z_t){t_h}^{2}z_h
-\ln (z_t)\ln (m_Z^2)z_h
+{\frac {11}{8}}\left (\ln (z_t)\right )^{2}t_hz_h
-12\left (\ln (z_t)\right )^{2}{t_h}^{2}z_h
\nonumber \\ &&
+165\ln (m_Z^2){t_h}^{2}z_hS_0
+{\frac {80}{3}}\ln (m_Z^2){t_h}^{2}z_hN_m
+4\left (\ln (m_Z^2)\right )^{2}t_hz_hN_m
\nonumber \\ &&
-16\left (\ln (m_Z^2)\right )^{2}{t_h}^{2}z_hN_m
-3\ln (z_t)t_hz_hr
+{\frac {165}{4}}\ln (z_t)t_hz_hS_0
+8/3\ln (z_t)t_hz_hN_m
\nonumber \\ &&
-165\ln (z_t){t_h}^{2}z_hS_0
-{\frac {32}{3}}\ln (z_t){t_h}^{2}z_hN_m
+{\frac {29}{2}}\ln (z_t)\ln (m_Z^2)t_hz_h
-45\ln (z_t)\ln (m_Z^2){t_h}^{2}z_h
\nonumber \\ &&
+3/4Ft_hz_hr
-4\ln (z_t)\ln (m_Z^2)t_hz_hN_m
+16\ln (z_t)\ln (m_Z^2){t_h}^{2}z_hN_m
\end{eqnarray}

\begin{eqnarray}
&&
A_{1,2}^Z =
16/3{\pi }^{2}{z_h}^{2}{N_m}^{2}
+{\frac {374}{9}}{z_h}^{2}S_0N_m
-{\frac {103}{54}}\ln (z_h){z_h}^{2}{r}^{2}
+{\frac {32}{3}}\ln (m_Z^2){z_h}^{2}{N_m}^{2}
\nonumber \\ &&
+3{\pi }^{2}t_h{z_h}^{2}
-\left (\ln (z_h)\right )^{2}{z_h}^{2}N_m
-{\frac {1321}{1080}}\ln (z_h){z_h}^{2}r
+9\left (\ln (z_t)\right )^{2}t_h{z_h}^{2}
+{\frac {391}{288}}Ft_h{z_h}^{2}
\nonumber \\ &&
+{\frac {17969}{360}}\ln (z_t)t_h{z_h}^{2}
-720{z_h}^{2}S_2N_m
-{\frac {167}{54}}{\pi }^{2}{z_h}^{2}
-{\frac {1913}{180}}t_h{z_h}^{2}
-42{t_h}^{2}{z_h}^{2}
-{\frac {233}{540}}{z_h}^{2}r
\nonumber \\ &&
+{\frac {23}{27}}{z_h}^{2}{r}^{2}
+{\frac {242}{3}}\ln (m_Z^2){z_h}^{2}S_0
-{\frac {832}{9}}S_3{z_h}^{2}N_m
-{\frac {107}{9}}{\pi }^{2}{z_h}^{2}N_m
+176\zeta (3){z_h}^{2}N_m
\nonumber \\ &&
-{\frac {577}{72}}\ln (z_h)t_h{z_h}^{2}
+{\frac {46}{9}}\ln (z_h){z_h}^{2}{r}^{3}
-{\frac {140}{27}}\ln (z_h){z_h}^{2}N_m
+{\frac {40}{9}}\ln (z_h)\ln (m_Z^2){z_h}^{2}
\nonumber \\ &&
+1/5\ln (z_h)z_tz_h
-{\frac {509}{10}}\ln (m_Z^2)t_h{z_h}^{2}
+12\ln (m_Z^2){t_h}^{2}{z_h}^{2}
-{\frac {595}{3}}\ln (m_Z^2){z_h}^{2}N_m
\nonumber \\ &&
-18\left (\ln (m_Z^2)\right )^{2}t_h{z_h}^{2}
+72\left (\ln (m_Z^2)\right )^{2}{t_h}^{2}{z_h}^{2}
+{\frac {107}{3}}\left (\ln (m_Z^2)\right )^{2}{z_h}^{2}N_m
\nonumber \\ &&
-16/3\left (\ln (m_Z^2)\right )^{2}{z_h}^{2}{N_m}^{2}
-1/5\ln (m_Z^2)z_tz_h
+{\frac {40}{9}}\ln (z_h)\ln (m_Z^2){z_h}^{2}N_m
\nonumber \\ &&
+{\frac {242}{3}}\ln (m_Z^2){z_h}^{2}S_0N_m
+9\ln (z_t)\ln (m_Z^2)t_h{z_h}^{2}
-72\ln (z_t)\ln (m_Z^2){t_h}^{2}{z_h}^{2}
\nonumber \\ &&
+{\frac {40}{9}}\ln (z_t)\ln (m_Z^2){z_h}^{2}N_m
-{\frac {208}{9}}S_3{z_h}^{2}
+{\frac {475837}{3240}}{z_h}^{2}-180{z_h}^{2}S_2
-{\frac {289}{18}}{z_h}^{2}S_0
\nonumber \\ &&
+{\frac {4091}{12}}{z_h}^{2}N_m
-{\frac {80}{27}}{z_h}^{2}{N_m}^{2}
+{\frac {112}{3}}\zeta (3){z_h}^{2}
-{\frac {548}{135}}\ln (z_h){z_h}^{2}
-1/2\left (\ln (z_h)\right )^{2}{z_h}^{2}
\nonumber \\ &&
-{\frac {39517}{216}}\ln (m_Z^2){z_h}^{2}
+41\left (\ln (m_Z^2)\right )^{2}{z_h}^{2}
+{\frac {51127}{540}}\ln (z_t){z_h}^{2}
-7/6\left (\ln (z_t)\right )^{2}{z_h}^{2}
\nonumber \\ &&
+{\frac {4}{15}}z_tz_h
-1/2F{z_h}^{2}
-48\ln (z_t){t_h}^{2}{z_h}^{2}
+{\frac {1321}{1080}}\ln (z_t){z_h}^{2}r
+{\frac {103}{54}}\ln (z_t){z_h}^{2}{r}^{2}
\nonumber \\ &&
-{\frac {46}{9}}\ln (z_t){z_h}^{2}{r}^{3}
-{\frac {619}{12}}\ln (z_t){z_h}^{2}S_0
-{\frac {116}{27}}\ln (z_t){z_h}^{2}N_m
-{\frac {20}{9}}\ln (z_t)\ln (z_h){z_h}^{2}
\nonumber \\ &&
-{\frac {565}{36}}\ln (z_t)\ln (m_Z^2){z_h}^{2}
+{\frac {577}{72}}F{t_h}^{2}{z_h}^{2}
-{\frac {71}{288}}F{z_h}^{2}r
-{\frac {7}{12}}F{z_h}^{2}{r}^{2}
+{\frac {23}{18}}F{z_h}^{2}{r}^{3}
\end{eqnarray}

\begin{eqnarray}
&&
A_{2,1}^Z =
-4/3\ln (z_t)\ln (z_h)z_h
-2/3\ln (z_h)t_hz_h
+{\frac {183}{4}}\ln (m_Z^2)t_hz_hS_0
+{\frac {32}{9}}\ln (m_Z^2)t_hz_hN_m
\nonumber \\ &&
+{\frac {1567}{36}}t_hz_h
-{\frac {2138}{9}}{t_h}^{2}z_h
-{\frac {56}{27}}z_hN_m
+4/9\ln (z_h)z_h
+8/3\ln (m_Z^2)z_h
-{\frac {14}{9}}\ln (z_t)z_h
\nonumber \\ &&
-1/3F{t_h}^{2}z_h
+{\frac {8}{27}}z_h
+8/3\ln (z_h)\ln (m_Z^2)z_h
-8/3\left (\ln (m_Z^2)\right )^{2}z_h
-{\frac {109}{4}}t_hz_hS_0
\nonumber \\ &&
+183{t_h}^{2}z_hS_0
+8/3\ln (z_h)\ln (m_Z^2)z_hN_m
-8/3\left (\ln (m_Z^2)\right )^{2}z_hN_m
+{\frac {128}{9}}{t_h}^{2}z_hN_m
\nonumber \\ &&
-4\ln (z_h)t_hz_hr
-{\frac {16}{9}}\ln (z_h)z_hN_m
-{\frac {518}{9}}\ln (m_Z^2)t_hz_h
+{\frac {2324}{9}}\ln (m_Z^2){t_h}^{2}z_h
\nonumber \\ &&
+{\frac {44}{9}}\ln (m_Z^2)z_hN_m
+{\frac {31}{6}}\left (\ln (m_Z^2)\right )^{2}t_hz_h
-{\frac {62}{3}}\left (\ln (m_Z^2)\right )^{2}{t_h}^{2}z_h
+{\frac {524}{9}}\ln (z_t)t_hz_h
\nonumber \\ &&
-{\frac {2042}{9}}\ln (z_t){t_h}^{2}z_h
+4/3\ln (z_t)\ln (m_Z^2)z_h
-8/3\left (\ln (z_t)\right )^{2}t_hz_h
+{\frac {32}{3}}\left (\ln (z_t)\right )^{2}{t_h}^{2}z_h
\nonumber \\ &&
-183\ln (m_Z^2){t_h}^{2}z_hS_0
-{\frac {320}{9}}\ln (m_Z^2){t_h}^{2}z_hN_m
-16/3\left (\ln (m_Z^2)\right )^{2}t_hz_hN_m
\nonumber \\ &&
+{\frac {64}{3}}\left (\ln (m_Z^2)\right )^{2}{t_h}^{2}z_hN_m
+4\ln (z_t)t_hz_hr
-{\frac {183}{4}}\ln (z_t)t_hz_hS_0
-{\frac {32}{9}}\ln (z_t)t_hz_hN_m
\nonumber \\ &&
+183\ln (z_t){t_h}^{2}z_hS_0
+{\frac {128}{9}}\ln (z_t){t_h}^{2}z_hN_m
-5/2\ln (z_t)\ln (m_Z^2)t_hz_h
+10\ln (z_t)\ln (m_Z^2){t_h}^{2}z_h
\nonumber \\ &&
-Ft_hz_hr
+16/3\ln (z_t)\ln (m_Z^2)t_hz_hN_m
-{\frac {64}{3}}\ln (z_t)\ln (m_Z^2){t_h}^{2}z_hN_m
\end{eqnarray}

\begin{eqnarray}
&&
A_{2,2}^Z =
-16/3{\pi }^{2}{z_h}^{2}{N_m}^{2}
-{\frac {7295}{27}}{z_h}^{2}S_0N_m
+{\frac {206}{81}}\ln (z_h){z_h}^{2}{r}^{2}
-{\frac {32}{27}}\ln (m_Z^2){z_h}^{2}{N_m}^{2}
\nonumber \\ &&
-3/2{\pi }^{2}t_h{z_h}^{2}
+4/3\left (\ln (z_h)\right )^{2}{z_h}^{2}N_m
+{\frac {1321}{810}}\ln (z_h){z_h}^{2}r
-9/2\left (\ln (z_t)\right )^{2}t_h{z_h}^{2}
-{\frac {391}{216}}Ft_h{z_h}^{2}
\nonumber \\ &&
-{\frac {34453}{540}}\ln (z_t)t_h{z_h}^{2}
+{\frac {2655}{2}}{z_h}^{2}S_2N_m
+{\frac {995}{216}}{\pi }^{2}{z_h}^{2}
+{\frac {4208}{135}}t_h{z_h}^{2}
+39{t_h}^{2}{z_h}^{2}
+{\frac {233}{405}}{z_h}^{2}r
\nonumber \\ &&
-{\frac {92}{81}}{z_h}^{2}{r}^{2}
-{\frac {638}{9}}\ln (m_Z^2){z_h}^{2}S_0
+{\frac {1408}{9}}S_3{z_h}^{2}N_m
+{\frac {1025}{54}}{\pi }^{2}{z_h}^{2}N_m
-{\frac {860}{3}}\zeta (3){z_h}^{2}N_m
\nonumber \\ &&
+{\frac {577}{54}}\ln (z_h)t_h{z_h}^{2}
-{\frac {184}{27}}\ln (z_h){z_h}^{2}{r}^{3}
+{\frac {560}{81}}\ln (z_h){z_h}^{2}N_m
-{\frac {160}{27}}\ln (z_h)\ln (m_Z^2){z_h}^{2}
\nonumber \\ &&
-{\frac {4}{15}}\ln (z_h)z_tz_h
+{\frac {3457}{60}}\ln (m_Z^2)t_h{z_h}^{2}
-42\ln (m_Z^2){t_h}^{2}{z_h}^{2}
+{\frac {3583}{27}}\ln (m_Z^2){z_h}^{2}N_m
\nonumber \\ &&
+9\left (\ln (m_Z^2)\right )^{2}t_h{z_h}^{2}
-36\left (\ln (m_Z^2)\right )^{2}{t_h}^{2}{z_h}^{2}
-{\frac {97}{3}}\left (\ln (m_Z^2)\right )^{2}{z_h}^{2}N_m
\nonumber \\ &&
+16/3\left (\ln (m_Z^2)\right )^{2}{z_h}^{2}{N_m}^{2}
+{\frac {4}{15}}\ln (m_Z^2)z_tz_h
-{\frac {160}{27}}\ln (z_h)\ln (m_Z^2){z_h}^{2}N_m
\nonumber \\ &&
-{\frac {638}{9}}\ln (m_Z^2){z_h}^{2}S_0N_m
-9/2\ln (z_t)\ln (m_Z^2)t_h{z_h}^{2}
+36\ln (z_t)\ln (m_Z^2){t_h}^{2}{z_h}^{2}
\nonumber \\ &&
-{\frac {112}{27}}\ln (z_t)\ln (m_Z^2){z_h}^{2}N_m
+{\frac {352}{9}}S_3{z_h}^{2}
-{\frac {1126103}{19440}}{z_h}^{2}
+{\frac {2655}{8}}{z_h}^{2}S_2
-{\frac {1744}{27}}{z_h}^{2}S_0
\nonumber \\ &&
-{\frac {68035}{324}}{z_h}^{2}N_m
-{\frac {1040}{81}}{z_h}^{2}{N_m}^{2}
-{\frac {533}{9}}\zeta (3){z_h}^{2}
+{\frac {2192}{405}}\ln (z_h){z_h}^{2}
+2/3\left (\ln (z_h)\right )^{2}{z_h}^{2}
\nonumber \\ &&
+{\frac {3458}{27}}\ln (m_Z^2){z_h}^{2}
-{\frac {113}{3}}\left (\ln (m_Z^2)\right )^{2}{z_h}^{2}
-{\frac {206869}{2160}}\ln (z_t){z_h}^{2}
+{\frac {23}{108}}\left (\ln (z_t)\right )^{2}{z_h}^{2}
\nonumber \\ &&
-{\frac {16}{45}}z_tz_h
+2/3F{z_h}^{2}
+60\ln (z_t){t_h}^{2}{z_h}^{2}
-{\frac {1321}{810}}\ln (z_t){z_h}^{2}r
-{\frac {206}{81}}\ln (z_t){z_h}^{2}{r}^{2}
\nonumber \\ &&
+{\frac {184}{27}}\ln (z_t){z_h}^{2}{r}^{3}
+{\frac {983}{18}}\ln (z_t){z_h}^{2}S_0
-{\frac {64}{81}}\ln (z_t){z_h}^{2}N_m
+{\frac {80}{27}}\ln (z_t)\ln (z_h){z_h}^{2}
\nonumber \\ &&
+{\frac {437}{27}}\ln (z_t)\ln (m_Z^2){z_h}^{2}
-{\frac {577}{54}}F{t_h}^{2}{z_h}^{2}
+{\frac {71}{216}}F{z_h}^{2}r
+{\frac {7}{9}}F{z_h}^{2}{r}^{2}
-{\frac {46}{27}}F{z_h}^{2}{r}^{3}
\end{eqnarray}



\begin{figure}
\begin{minipage}{13.0cm}
\centerline{\vbox{\epsfysize=80mm \epsfbox{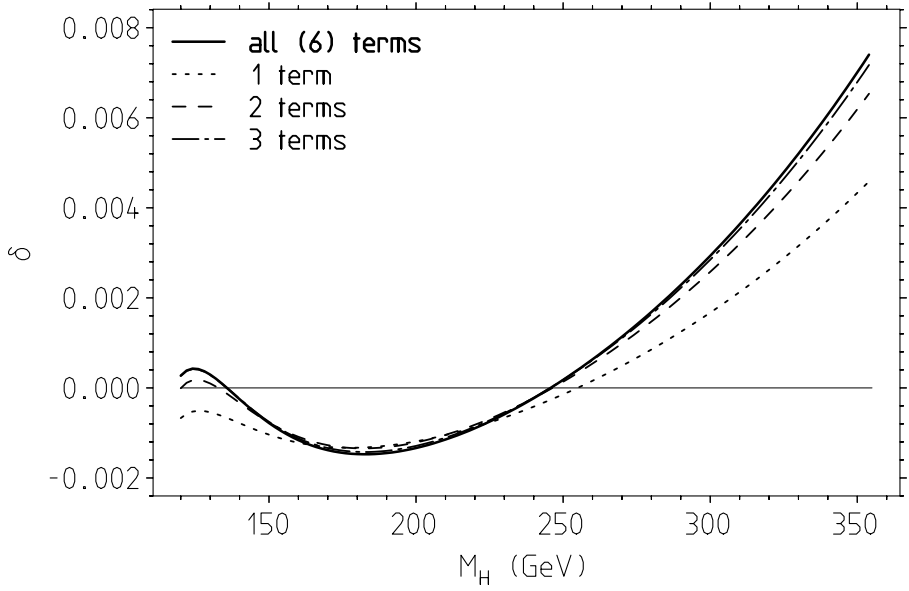}~~~~~~}}
\end{minipage}

\begin{minipage}{13.0cm}
\centerline{\vbox{\epsfysize=80mm \epsfbox{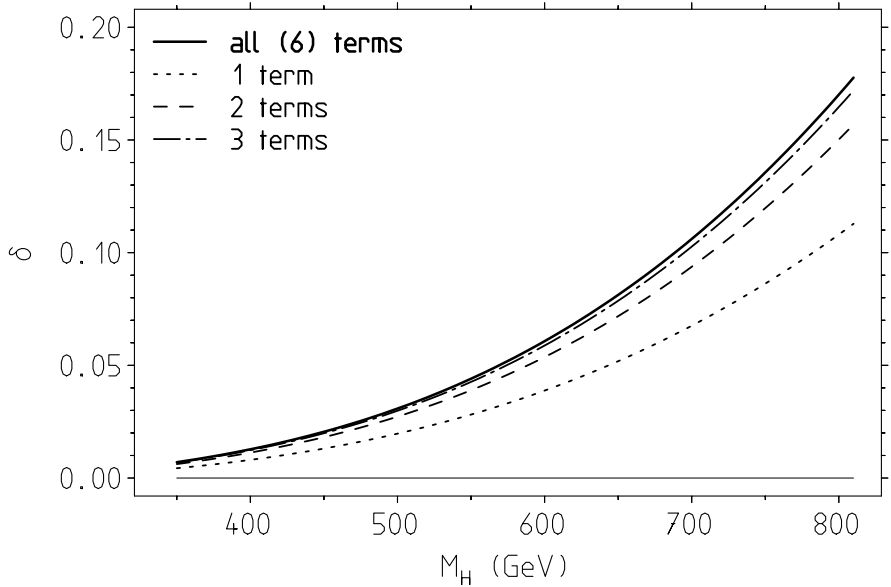}}}
\end{minipage}
\caption{\it
The dependence on the number of coefficients of the expansion with
respect to $\sinW$ of the two--loop corrections $\delta \equiv
\Biggl \{ \Pi_Z^{(2)} + \Pi_Z^{(1)} \Pi_Z^{(1)}{}' \Biggr\}_{\msb}
$
(see \ref{MS2:subtracted}) as a
function of the Higgs mass. The dotted, dashed, dot-dashed and full lines show
results obtained with the first one, two, three and all calculated (six)
coefficients, respectively.
Upper plot: for intermediate Higgs masses.
Lower plot: for heavy Higgs masses.}
\label{f:sex}
\end{figure}

\begin{figure}
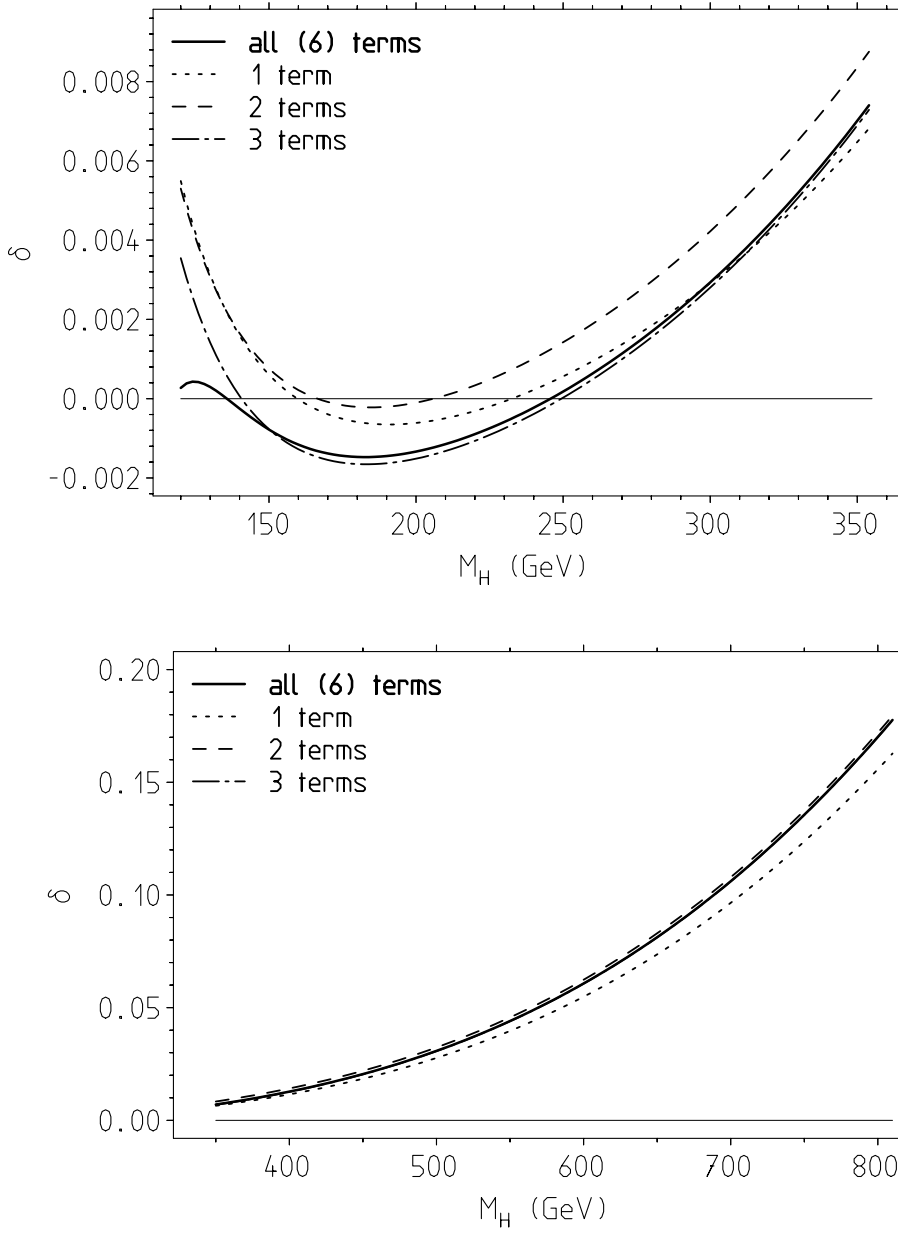

\begin{minipage}{13.0cm}
\centerline{\vbox{\epsfysize=80mm \epsfbox{fig.5a}~~~~~~}}
\end{minipage}

\begin{minipage}{13.0cm}
\centerline{\vbox{\epsfysize=80mm \epsfbox{fig.5b}}}
\end{minipage}
\caption{\it
The dependence on the number of coefficients of the expansion with
respect to $m^2_Z/m^2_i$ ($i=H,t$) used for the evaluation of the
two-loop corrections.  We show $\delta \equiv
\Biggl \{ \Pi_Z^{(2)} + \Pi_Z^{(1)} \Pi_Z^{(1)}{}' \Biggr\}_{\msb}$
(see \ref{MS2:subtracted}) as a function of the Higgs mass. The dotted,
dashed, dot-dashed and full lines show results obtained with the first
one, two, three and all calculated (six) coefficients, respectively.
Upper plot: for intermediate Higgs masses.  Lower plot: for heavy
Higgs masses.}
\label{f:zex}
\end{figure}

\begin{figure}
\begin{minipage}{13.0cm}
\centerline{\vbox{\epsfysize=80mm \epsfbox{fig.6a}}}
\end{minipage}

\begin{minipage}{13.0cm}
\centerline{\vbox{\epsfysize=80mm ~~\epsfbox{fig.6b}}}
\end{minipage}
\caption{\it
Corrections to the relation
$\Delta_W \equiv M^2_W/m^2_W(M_W)-1 $
as a function of the Higgs mass $M_H$ for intermediate Higgs masses.
Upper plot: the various two-loop corrections.
Lower plot: the complete one- and two-loop correction.}
\label{f:wl}
\end{figure}

\begin{figure}
\begin{minipage}{13.0cm}
\centerline{\vbox{\epsfysize=80mm \epsfbox{fig.7a}}}
\end{minipage}

\begin{minipage}{13.0cm}
\centerline{\vbox{\epsfysize=80mm ~~\epsfbox{fig.7b}}}
\end{minipage}
\caption{\it
Corrections to the relation
$\Delta_Z \equiv M^2_Z/m^2_Z(M_Z)-1$
as a function of the Higgs mass $M_H$ for intermediate Higgs masses.
Upper plot: the various two-loop corrections.
Lower plot: the complete one- and two-loop correction.}
\label{f:zl}
\end{figure}

\begin{figure}
\begin{minipage}{13.0cm}
\centerline{\vbox{\epsfysize=80mm \epsfbox{fig.8a}}}
\end{minipage}

\begin{minipage}{13.0cm}
\centerline{\vbox{\epsfysize=80mm \epsfbox{fig.8b}}}
\end{minipage}
\caption{\it
Corrections to the relation
$\Delta_W \equiv M^2_W/m^2_W(M_W)-1$
as a function of the Higgs mass $M_H$ for heavy Higgs masses.
Upper plot: the various two-loop corrections.
Lower plot: the complete one- and two-loop correction.}
\label{f:wh}
\end{figure}

\begin{figure}
\begin{minipage}{13.0cm}
\centerline{\vbox{\epsfysize=80mm \epsfbox{fig.9a}}}
\end{minipage}

\begin{minipage}{13.0cm}
\centerline{\vbox{\epsfysize=80mm \epsfbox{fig.9b}}}
\end{minipage}
\caption{\it
Corrections to the relation
$\Delta_Z \equiv M^2_Z/m^2_Z(M_Z)-1 $
as a function of the Higgs mass $M_H$ for heavy Higgs masses.
Upper plot: the various two-loop corrections.
Lower plot: the complete one- and two-loop correction.}
\label{f:zh}
\end{figure}

\begin{figure}
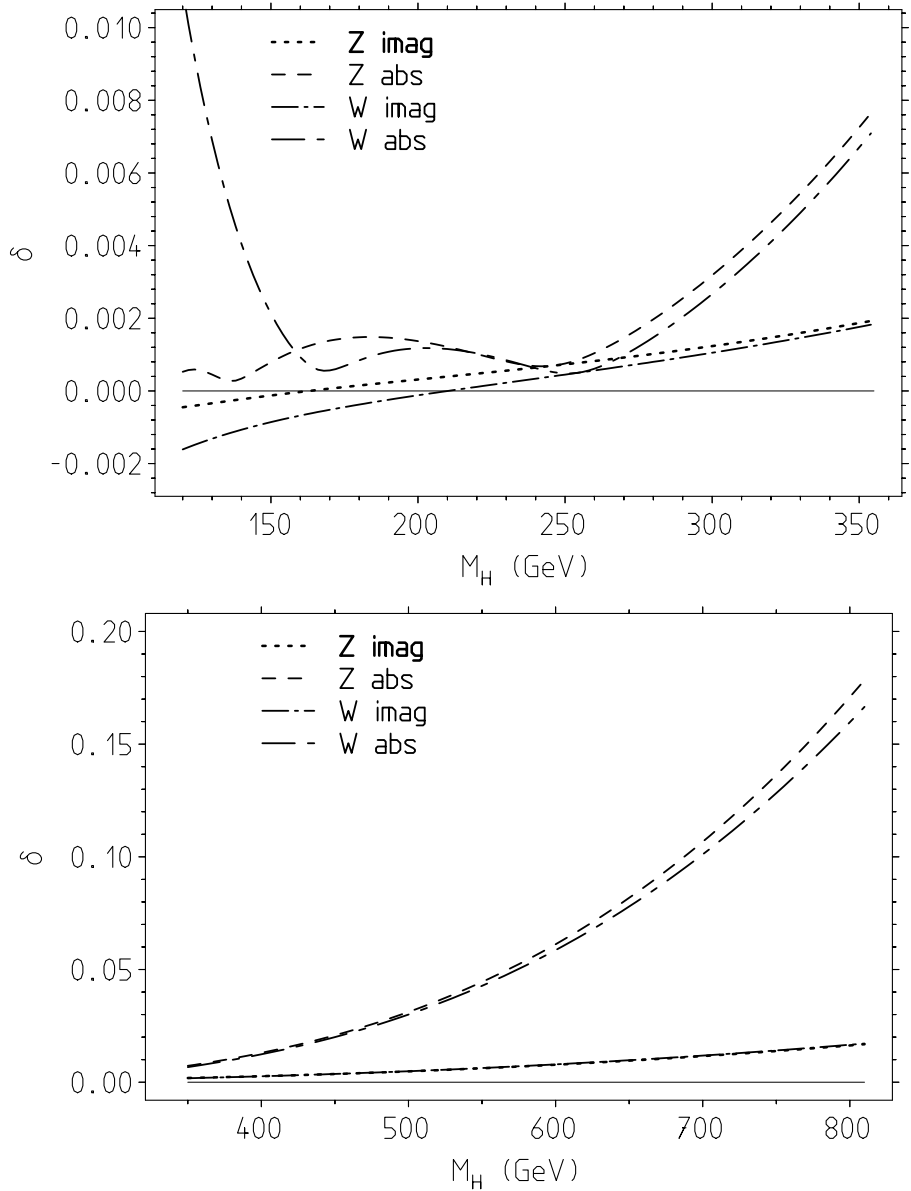

\begin{minipage}{13.0cm}
\centerline{\vbox{\epsfysize=80mm \epsfbox{fig.10a}}}
\end{minipage}

\begin{minipage}{13.0cm}
\centerline{\vbox{\epsfysize=80mm \epsfbox{fig.10b}}}
\end{minipage}
\caption{\it
Two--loop imaginary part $\delta={\rm Im} s_{P,\:V}/m_V^2$ {\tt (imag)} and
absolute value $\delta=|s_{P,\:V}|/m_V^2$ {\tt (abs)} of the pole position
$s_P$ for the $Z$ {\tt (Z)} and the $W$ {\tt (W)} as a function of the
Higgs mass $M_H$.  Upper plot: for intermediate Higgs masses.  Lower
plot: for heavy Higgs masses.}
\label{f:im}
\end{figure}

\begin{figure}
\begin{minipage}{13.0cm}
\centerline{\vbox{\epsfysize=80mm \epsfbox{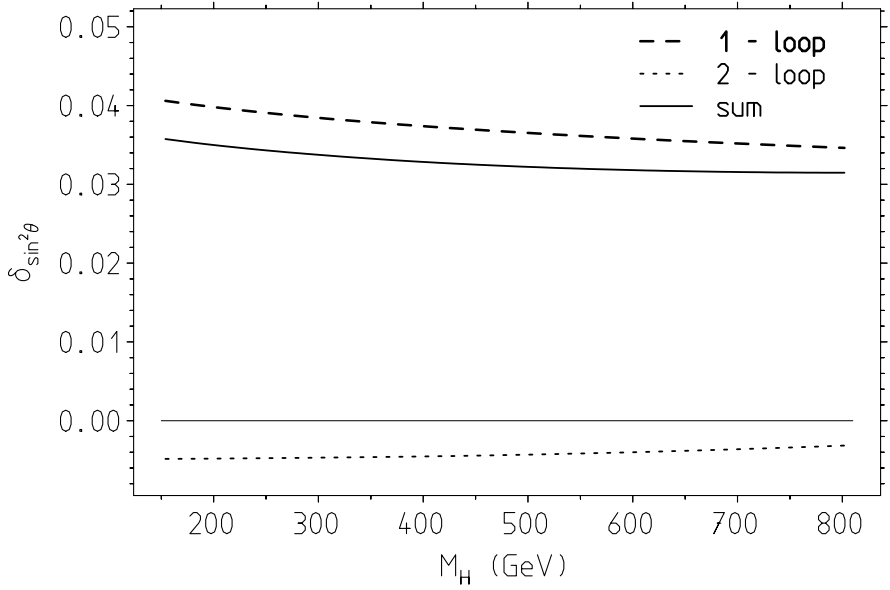}~~~~~~}}
\end{minipage}

\caption{\it
One- and two-loop corrections to $\delta_{\sin^2\Theta}=
\sin^2\theta^{\msb}_W/\sin^2\theta^{\rm OS}_W-1$
(see (\ref{sinus}))
as a function  of the Higgs mass $m_H$ ($\mu=M_Z$).
\label{sinofmh}
}
\end{figure}

\end{document}